\documentclass[trackchanges,twocolumn]{aastex701}

\usepackage{newtxtext,newtxmath}
\usepackage[T1]{fontenc}
\usepackage{lineno}
\usepackage{graphicx}
\usepackage{amsmath}
\usepackage{lipsum}
\usepackage{hyperref}
\usepackage{longtable}
\usepackage[figuresright]{rotating}
\usepackage{multirow}
\usepackage{afterpage} 
\usepackage{enumitem}
\usepackage{placeins}

\begin{document}

\title{Void Galaxies and AGN Activity in ZOBOV-identified \texttt{TNG300} Voids from $z = 3$ to $z = 0$}

\author[0000-0002-0212-4563]{Olivia Curtis}
\affiliation{Department of Astronomy \& Astrophysics, The Pennsylvania State University, 251 Pollock Road, University
Park, PA 16802, USA}\affiliation{Department of Astronomy \& Institute for Astrophysical Research, 725 Commonwealth Ave., Boston University, Boston, MA 02215, USA}
\email[show]{ocurtis@psu.edu}

\author[0000-0001-6928-4345]{Bryanne McDonough}
\affiliation{Physics Department, Adelphi University, 1 South Ave., Garden City, NY 11530, USA}\affiliation{Department of Physics, Northeastern University, 360 Huntington Ave, Boston, MA 02115, USA}\affiliation{Department of Astronomy \& Institute for Astrophysical Research, 725 Commonwealth Ave., Boston University, Boston, MA 02215, USA}
\email{bmcdonough@adelphi.edu}

\author[0000-0001-7917-7623]{Tereasa G. Brainerd}
\affiliation{Department of Astronomy \& Institute for Astrophysical Research, 725 Commonwealth Ave., Boston University, Boston, MA 02215, USA}
\email{brainerd@bu.edu}

\begin{abstract}

We study void galaxies in the TNG300 simulation between redshifts $z=3$ and $z=0$. Cosmic void catalogs were constructed using a watershed-based void-finding algorithm, and we define {four} populations of field galaxies for our investigation: [1] galaxies that are members of a {watershed} void, [2] galaxies that are located within a radius $r \leq 0.8 R_{\rm eff}$ of the center of a void, {[3] galaxies interior to spheres centered on void centers that have underdensity contrasts $<-0.8$,} and [4] non-void galaxies. {We show that population statistics on void galaxy properties can be recovered from watershed-based void catalogs.} {Differences between galaxy populations are most pronounced interior to the shell-crossing surface (i.e., population [3]) where densities are intermediate to high.} Compared to non-void galaxies at all redshifts, the {density} controlled galaxies are {bluer}, {smaller}, more actively star forming, more massive, and less metal enriched. At redshifts $\geq 1$, these differences are less apparent{, likely caused by resolution and selection effects incurred by attempting to define a density-controlled sample from a watershed-based void finding algorithm}. Further, we investigate the fraction of galaxies with Active Galactic Nuclei (AGN) and find that our {density} controlled population has AGN fractions that are significantly higher than those of non-void galaxy population ($79.8 \pm 0.4$\% higher at $z=0.0$ and $61.5\pm 0.7$\% higher at larger redshifts).

\end{abstract}



\section{Introduction}\label{sec:intro}

Cosmic voids are not entirely devoid of luminous material and thus contain a wealth of astrophysical information (see, e.g., \citealt{gregory1978}; \citealt{sheth}; \citealt{hoyle2004}). Due to their rarefied environments, the locations of galaxies within cosmic voids are expected to affect their evolution (see, e.g., \citealt{peebles2001}; \citealt{kreckel2012}; \citealt{curtis2022}; \citealt{cavity1}). In particular, compared to galaxies that are located outside of voids, void galaxies have been observed to be bluer, more actively star forming, and less massive (see, e.g., \citealt{rojas2004}; \citealt{croton2005}; \citealt{hoyle2012}; \citealt{kreckel2012}; \citealt{douglass2017}; \citealt{rodriguez2022}; \citealt{rosasguevara2022}; \citealt{cavity1}; \citealt{cavity2}). 

The general framework is that, due to their relatively isolated environments, void galaxies retain most of their gas since they should experience fewer major interactions than non-void galaxies \citep{kauffmann2004,cen2011}. However, in a recent study of void galaxies in the \texttt{TNG300} suite of simulations \citep{illustris1,illustris2,illustris3,illustris4,illustris5}, \cite{rodriguesmedrano2024} found that void and non-void galaxies experience the same mean number of mergers over the course of the simulation, but void galaxies tend to experience these mergers later than non-void galaxies. This paints a picture in which the assembly of stellar mass in void galaxies is slower than that of non-void galaxies. Indeed, \cite{cavity1} and \cite{cavity2} found that, within the ongoing Calar Alto void integral-field treasury survey (CAVITY) of void galaxies, void galaxies assembled $50\%$ of their stellar mass $1.03\pm0.06$~Gyr later than galaxies in walls and filaments.

Still, there have been results in the literature regarding the differences between void and non-void galaxies {that do not fully align}. For example, observations have found little correlation between factors like star formation rates, chemical abundances, and metallicities between void and non-void galaxies (see, e.g., \citealt{szomoru1996}; \citealt{patiri2006}; 
\citealt{moorman2014};
\citealt{liu2015};
\citealt{douglass2017};
\citealt{douglass2019}; \citealt{wegner2019}; \citealt{dominguezgomez2022}). Therefore, it is important to continue quantifying how the physical properties of void galaxies evolve over cosmic time in order to further test these claims to a high degree of statistical significance.

Another open question involves the degree to which void environments affect the triggering of Active Galactic Nuclei (AGN). That is, some studies have concluded that there is no correlation between AGN activity and location within the cosmic web (e.g., \citealt{carter2001}; \citealt{karhunen2014}; \citealt{sabater2015}; \citealt{amiri2019}; \citealt{habouzit}). Other studies have concluded that there is either a higher fraction of AGN within voids (e.g., \citealt{kauffmann2004}; \citealt{constantin2008}; \citealt{platenPhD}; \citealt{lopes2017}; \citealt{mishra2021}; \citealt{curtis2024}; \citealt{Aradhey2025}), or a lower fraction of AGN within voids (e.g., \citealt{manzer2014}; \citealt{argudofern2018}; \citealt{ceccarelli2021}). In particular, \cite{mishra2021} looked at the AGN fraction as a function of void-centric distance in the Baryon Oscillation Spectroscopic Survey \cite{BOSS} of the Sloan Digital Sky Survey III \citep{SDSSIII} and found that, compared to galaxies in the outer regions of voids, galaxies in the inner regions of voids have AGN fractions that are $\sim45\%$ higher. Beyond redshifts $z \sim 0.7$, the AGN fractions of void vs.\ non-void galaxies is an open question in both observations and simulations.


{One of the driving factors behind these results is the fact that the choice of void-finding algorithm also strongly influences the results of void galaxy surveys \citep{colberg2008, Veyrat2023, Zaidouni2025}. In particular, the region of shell-crossing --- where out-flowing spherical shells of material build up to form a void ridge --- plays an important role in galaxy evolution, with galaxies interior to this surface showing the most significant differences. Many void finding algorithms are based on the \texttt{VoidFinder} algorithm \citep{elad1997, hoyle2004}, which traces the shell-crossing surface but enforces spherical geometry, or watershed-based methods that do not enforce geometry but include much of the surface of the shell-crossing region in their definitions \citep{platen2007, neyrinck2008}.} 

{In practice, there is no unifying void definition, and different algorithms can probe different problems \citep{colberg2008}, with spherical void finders being better suited for the study of void galaxy properties and watershed-based algorithms being appropriate for studies of the morphologies and matter distribution of voids. Another consideration is how the centers of voids are defined --- the centers of \texttt{VoidFinder} voids are often poorly constrained as their centers are defined on a grid of arbitrary resolution \citep{douglass2022}, while the centers of watershed-based voids have natural definitions for how their centers are defined (e.g., a volume-weighted barycenter \citealt{Nadathur2015}). An ancillary goal of this manuscript is to thus show how population statistics of void galaxies can be recovered from watershed-based void catalogs out to higher redshifts than has been done in recent studies (e.g., \citealt{habouzit, Veyrat2023, Zaidouni2025}).}


Throughout this paper, we use the void catalogs that were obtained \cite{curtis2025} for the purposes of studying the evolution of voids in a $\Lambda$ Cold Dark Matter ($\Lambda$CDM) universe. \cite{curtis2025} identified voids in the galaxy field of eleven snapshots of the magnetohydrodynamical (MHD) galaxy formation simulation \texttt{TNG300} using the Voronoi-based ZOnes Bordering on Voidness (ZOBOV; \citealt{neyrinck2008}) voidfinder algorithm. A benefit of using \texttt{TNG300} is that full particle information for each of these eleven snapshots is available, allowing us to perform epoch-by-epoch analyses of the physical properties of void galaxies. 
In this paper, we use the void galaxy catalogs to determine whether these objects comprise a distinct population of galaxies in the context of this particular state-of-the-art cosmological MHD simulation. We also investigate the degree to which the location of a galaxy within the cosmic web influences the triggering of the AGN phenomenon. 

This paper is structured as follows. We discuss \texttt{TNG300} in~\S\ref{sec:TNG300} and the void catalogs in~\S\ref{sec:voidcats}. In \S\ref{sec:ch5:galcuts}, we describe the galaxy populations that we investigate. We also discuss the magnitude, size, and mass cuts that we apply to the galaxy populations. In the subsections of \S\ref{sec:ch5:phyproperties}, we present the following physical properties of the galaxies: $(g-r)$ colors, radial sizes, luminosity functions, mass functions, specific star formation rates, and stellar and gas chemical abundance ratios. In \S\ref{sec:ch5:AGN}, we discuss the differences in the AGN fraction and the supermassive black hole mass vs.\ stellar mass relation for the galaxy populations. We present a discussion of our results and summarize our conclusions in \S\ref{sec:ch5:discussion}. {Unless otherwise stated}, error bars are calculated from 10,000 bootstrap resamplings of the data.

\section{TNG300 Simulation}
\label{sec:TNG300}

The data for our analyses were obtained from the highest-resolution run of the \texttt{TNG300} simulation. \texttt{TNG300} has a box size of $205h^{-1}$Mpc and contains $2500^3$ dark matter particles and $2500^3$ hydrodynamical gas cells. The simulation has a dark matter resolution limit of $m_{\rm dm} = 4.0 \times 10^7 h^{-1} M_\odot$ and a baryonic resolution limit of $m_{\rm b} = 7.5 \times 10^6 h^{-1} M_\odot$. The simulation adopted the cosmological parameters obtained by the \cite{Planck15} (i.e., $\Omega_{\Lambda,0} = 0.6911$,
$\Omega_{m,0} = 0.3089$, $\Omega_{b,0} = 0.0486$, $\sigma_8 = 0.8159$, $n_s = 0.9667$, and $h = 0.6774$.)

We obtain stellar masses, star formation rates, gas and stellar chemical abundance ratios, sizes, black hole masses, and black hole accretion rates from the halo and subhalo catalogs. We use the subhalo magnitudes of \citet{illustris1} to assign luminosities to the galaxies as this catalog includes the effects of dust obscuration on the simulated galaxies. Following \cite{illustris3}, we also apply a correction to the stellar masses from the \texttt{TNG300} catalog (i.e., to account for the fact that the stellar masses in \texttt{TNG300} are not as well converged as in the benchmark \texttt{TNG100} simulation; see Appendix A of \citealt{illustris3}).

\section{Void Catalogs}
\label{sec:voidcats}

\begin{figure*}[!htbp]
    \centering
    \includegraphics[width=0.8\textwidth]{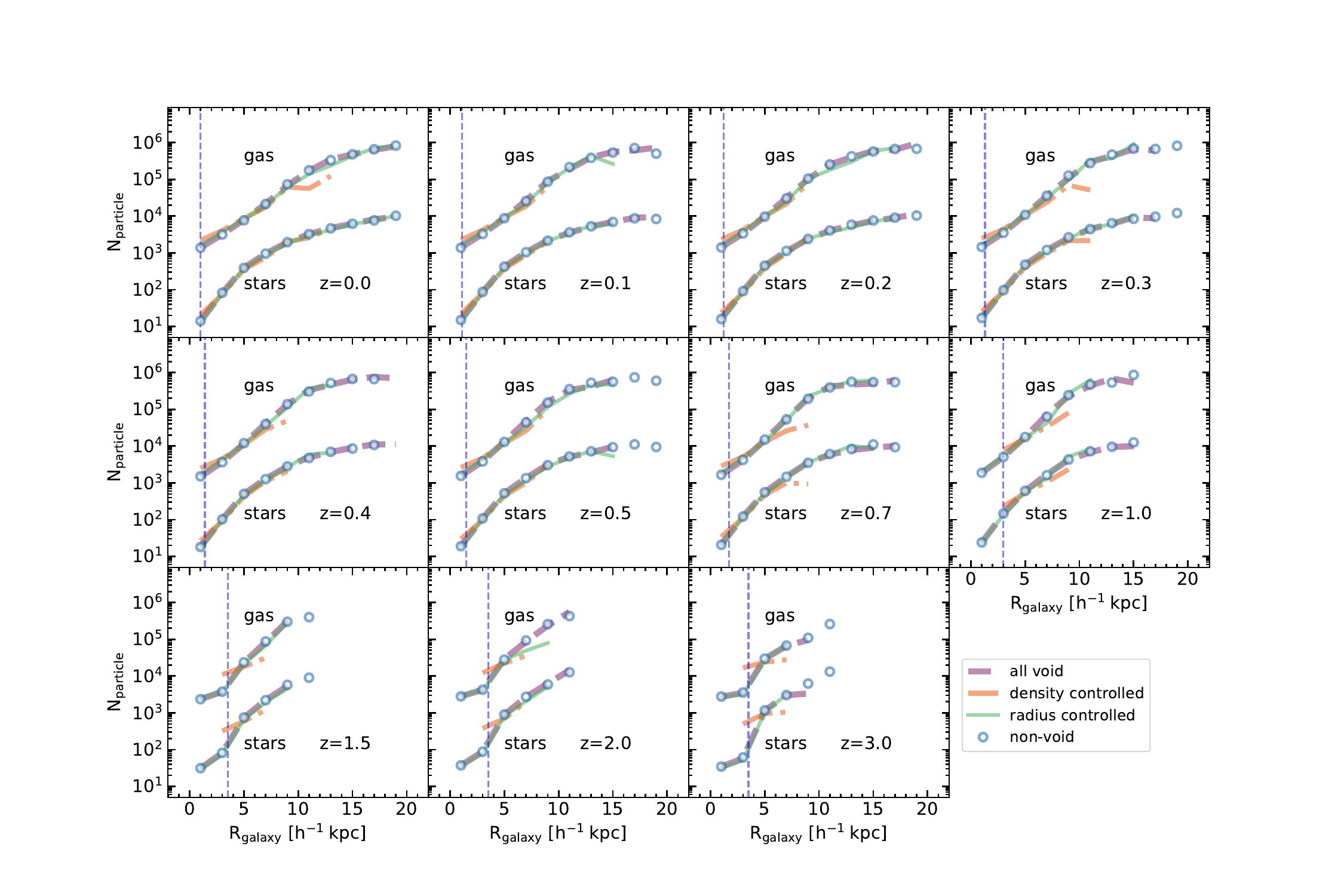}
    \caption{Median number of gas cells {and stellar particles per galaxy as a function of redshift. Each panel shows galaxies from different snapshots. Results are shown for all void galaxies (solid purple), non-void galaxies (circles), radius-controlled galaxies (dashed green), and density-controlled galaxies (dash dotted orange).} The vertical blue lines indicate the minimum galaxy radii cuts (see text). {For visibility, the number of stellar particles have been shifted down by an order of magnitude.}}
    \label{fig:ch5:ngasvsr}
\end{figure*}

To identify void galaxies, we use the \texttt{TNG300} void catalogs of \cite{curtis2025}. A detailed description of the voids and how these catalogs were made can be found in \cite{curtis2025}, but we provide a brief discussion of the catalogs below. In summary, these catalogs were created using the ZOBOV algorithm \citep{neyrinck2008}, as implemented by the publicly available code \texttt{REVOLVER}\footnote{https://github.com/seshnadathur/Revolver} \citep{nadathur2019}.

ZOBOV can be summarized in four steps. First, a 3D Voronoi tesselation on the galaxy field (see, e.g., \citealt{schaap2007}) is performed. The local density of particle $i$ is then estimated as $1/V(i)$, where $V(i)$ is the volume of the Voronoi cell surrounding that particle.
Next, a watershed transformation is used to merge Voronoi cells together to form "zones" (see, e.g., \citealt{platen2007}). The least dense Voronoi cell in the survey forms a "zone center." All denser cells that neighbor the zone center are then added to the zone. This process then repeats for all cells on the boundary of the zone until a cell with a higher density than the previous cell is reached. 
Lastly, ZOBOV merges zones together to form voids. This is done by merging together all neighboring zones that that have a minimum Voronoi cell density that is greater than that of the first zone. That is, a zone annexes all neighboring zones until it reaches a zone that has a zone center with a lower density estimate than it does. 

\begin{deluxetable*}{c c c c c c c c c}
\tabletypesize{\scriptsize}
\tablecaption{Redshift, snapshot number, total number of subhalos in that snapshot, total number of resolved subhalos in that snapshot, total number of galaxies in large groups or clusters, and total number of galaxies in our {four} galaxy populations (see text). Note that $N_{\rm resolved}=N_{\rm cluster}+N_{\rm void}+N_{\rm non-void}$. \label{tab:ch5:galaxysummary}}
\tablehead{
\colhead{Redshift} & \colhead{Snapshot Number} & \colhead{$N_{\rm galaxies}$} & \colhead{$N_{\rm resolved}$} & \colhead{$N_{\rm cluster}$} & \colhead{$N_{\rm all void}$} & \colhead{$N_{\rm non -void}$} & \colhead{$N_{\rm{radius -controlled}}$} & \colhead{$N_{\rm{density -controlled}}$}
}
\startdata
0.0 & 99 & 342635 & 253590 & 27189 & 84453 & 141948 & 16137  & 4031 \\
0.1 & 91 & 338602 & 249806 & 23578 & 79753 & 146475 & 15103 & 3993 \\
0.2 & 84 & 334219 & 244457 & 20457 & 75730 & 148270 & 14872 & 3791 \\
0.3 & 78 & 329227 & 237428 & 16652 & 72489 & 148287 & 14325 & 3633 \\
0.4 & 72 & 325406 & 227652 & 13777 & 67722 & 146153 & 12933 & 3501 \\
0.5 & 67 & 322657 & 223054 & 12378 & 60568 & 150108 & 11526 & 3134 \\
0.7 & 59 & 315329 & 210121 & 7909 & 64395 & 137817 & 12472 & 3198 \\
1.0 & 50 & 295677 & 68437 & 1555 & 25130 & 41752 & 4686 & 1396 \\
1.5 & 40 & 229435 & 40792 & 415 & 11958 & 28419 & 2184 & 758 \\
2.0 & 33 & 170014 & 21558 & 56 & 5763 & 15739 & 1046 & 366 \\
3.0 & 25 & 118841 & 4772 & 0 & 1246 & 3526 & 182 & 134 \\
\enddata
\end{deluxetable*}

This procedure thus creates a hierarchy of voids, ranging from one massive void that encompasses the entire survey to small individual voids. To account for this, \texttt{REVOLVER} does not merge zones together to form voids and instead defines a void to be any single zone that has a volume above the median value of all other zones. \cite{curtis2025} then further pruned their void catalogs by estimating the probability that any individual void arose due to Poisson fluctuations, only keeping those deemed to be significant to a $3\sigma$ confidence level (see, e.g., \citealt{neyrinck2008}).

\cite{curtis2025} ran ZOBOV on eleven \texttt{TNG300} snapshots, corresponding to redshifts $z=0.0,$ $0.1,$ $0.2,$ $0.3,$ $0.4,$ $0.5,$ $0.7$, $1.0$, $1.5,$ $2.0,$ and $3.0$. ZOBOV was run on all galaxies in the snapshot that were identified as being cosmological in origin as selected by the \texttt{TNG} team. 
{This was done by selecting systems that, at their formation time, were satellites within a virial radius of their hosts and those that exhibited anomalously low dark-matter fractions, which are fragments of a larger halo and not an object that formed from structure formation processes. We also require that all galaxies have} absolute $r$-band luminosities $M_r \leq-14.5$ and stellar masses $M_\ast \geq 10^{8}h^{-1}M_\odot$ (with the exception of the $z=3.0$ snapshot, which had a mass cutoff of $10^{7.75}h^{-1}M_\odot$). In general, each snapshot contains $70-178$ significant voids that have median radii $\sim20h^{-1}$Mpc. The density profiles of all voids were consistent with the reverse-spherical top-hat distribution that underdense regions of the universe are expected to exhibit (see, e.g., \citealt{Icke}; \citealt{sheth}), indicating that voids in the catalogs were truly the largest, emptiest regions of the simulation.

\section{Galaxy Selection}
\label{sec:ch5:galcuts}

To obtain our galaxy catalogs, we perform various cuts on the subhalo catalogs. First, within each snapshot, we remove any subhalo that is flagged as being as non-cosmological in origin. Next, since we are only interested in the differences between void galaxies and those in a typical field environment, we remove any galaxy that belongs to a parent halo or cluster more massive than $10^{14}h^{-1}M_{\rm \odot}$. Then, we perform a minimum stellar mass cut by removing any galaxy with a stellar mass $< 10^{8.5}h^{-1}M_{\rm \odot}$, allowing for galaxies with $\gtrsim50$ stellar particles to be present in the final catalogs.

Finally, we implement a resolution criterion for the radius of each subhalo. For this, we define a minimum galaxy radius, $R_{\rm gal,min}$, for each snapshot to be

\begin{equation}
\label{eq:ch5:rgalmin}
R_{\rm gal,min}=
    \begin{cases}
        1.0\times \epsilon & \text{if } z<1.0\\
        1.5\times \epsilon & \text{if } z\geq1.0 
    \end{cases}
    \; .
\end{equation}

\noindent Here, $\epsilon$ is the gravitational softening length used for each snapshot, and $z$ is the corresponding redshift.\texttt{TNG300} uses a static gravitational softening length of $2.0h^{-1}$kpc (in co-moving coordinates) for snapshots corresponding to redshifts $z>1$ while the softening length at lower redshifts is $[1+z]h^{-1}$kpc. We adopt a piece-wise definition for the minimum galaxy radius because of a marked discontinuity in the number of gas cells and stellar particles in small galaxies at large redshifts, and we will discuss this further below.

After implementing these cuts, we define {four} populations of field galaxies using the \texttt{TNG300} void catalogs from \cite{curtis2025}: 

\begin{enumerate}[label={[\arabic*]}]
    \item all void galaxies,
   { \item galaxies within spheres that are centered on the void center and have radii $0.8$ times the effective radius ($R_{\rm eff}$) of the void,}
    {\item galaxies within spheres that are centered on the void center and have underdensity contrasts $<-0.8$, and}
    \item and non-void galaxies.
\end{enumerate}

\noindent Throughout, we will refer to these populations as: [1] ``void galaxies,'' [2] ``radius-controlled galaxies,'' {[3] ``density-controlled galaxies,''} and [{4}] ``non-void galaxies.''


\begin{figure*}[!t]
    \centering
    \includegraphics[width=0.7\textwidth]{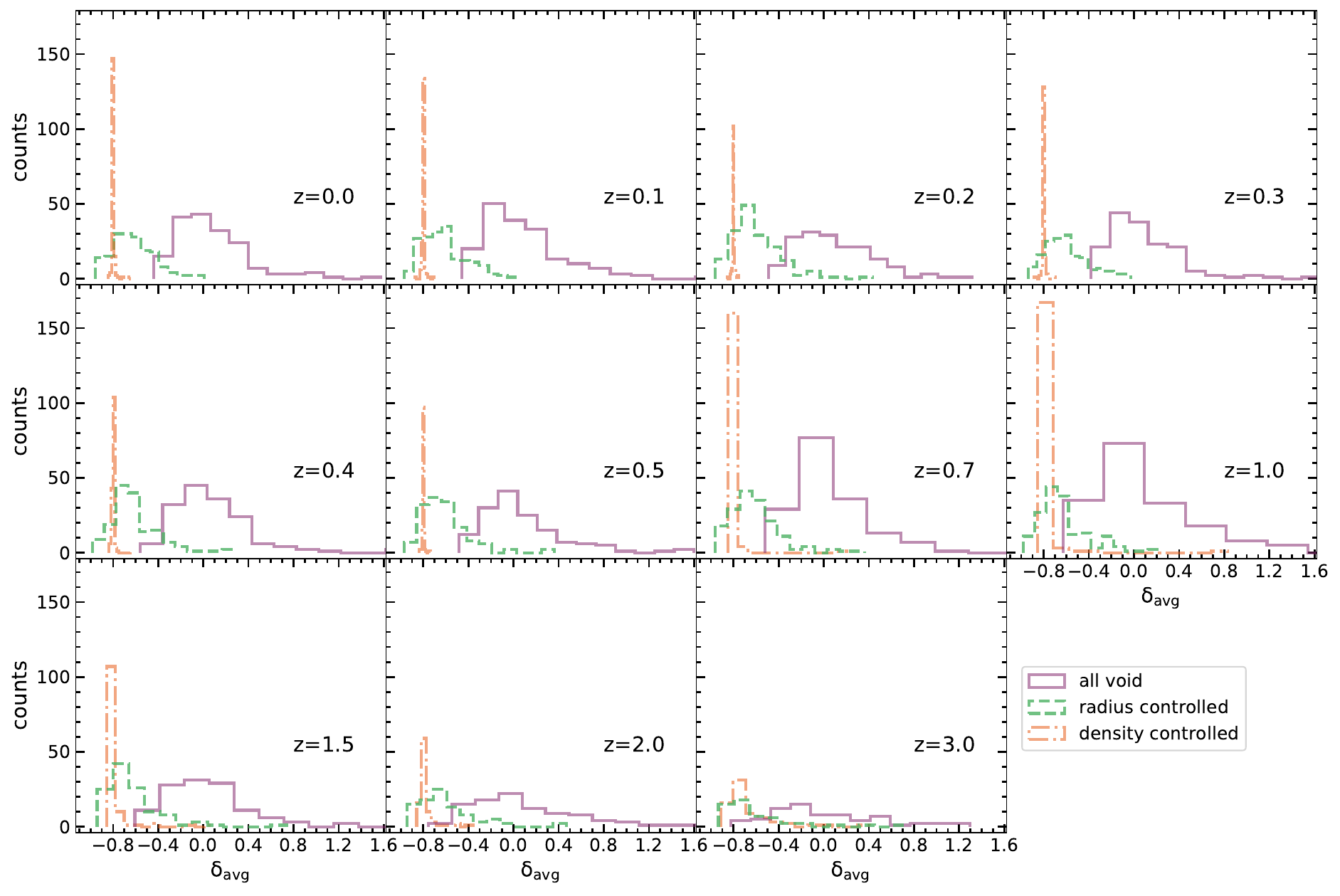}
    \caption{Histograms showing average tracer underdensities for all voids (solid purple), radius-controlled underdensity regions (dashed green), and density-controlled underdensity regions (dash dotted orange).}
    \label{fig:underdensities}
\end{figure*}

The ``void galaxies'' population contains all field galaxies that reside within the bounds of a \texttt{ZOBOV} void which, as mentioned above will contain galaxies that are located in shell-crossing regions.  The ``non-void galaxies'' population consists of any field galaxy that is not a member of one of the voids.

The ``radius-controlled galaxies'' population contains only those void galaxies within a sphere of radius $0.8$ times the effective radius of the void, centered on the weighted center of the void. (Note: for a \texttt{ZOBOV} void, $R_{\rm eff}$ is the radius of a sphere with the same volume as the amorphously shaped void region.) This population consists of a sample of void galaxies that reside within the inner regions of voids. It contains fewer galaxies that reside along the void ridges and results in a sample of objects that are similar to the sample that would be obtained if a {SVF} had been used to create the void catalog. 

{The density-controlled sample creates a sample of interior void galaxies that can be directly compared to void galaxy catalogs attained with SVFs. We construct this sample of galaxies in the following way. First, from a given void center, we order galaxies by increasing void-centric distance and find the first $i-$th galaxy that satisfies}

\begin{equation}
    \delta_{\rm avg} \equiv \frac{n}{\bar{n}}-1=\frac{i}{\frac{4}{3}\pi r_i^3} \geq -0.8.
\end{equation}

\noindent {Here, $\delta_{avg}$ is the average underdensity contrast of the sphere, n is the number density of galaxies within the sphere, $\bar{n}$ is the average number density of the simulation, and $r_i$ is the distance to the $i-$th galaxy. This maps the shell-crossing surface of voids \citep{sheth}, exterior to which mixing between the void environment and surrounding filaments occur that can bias population-level studies of void galaxies \citep{Zaidouni2025}. Except at higher redshifts where tracer sparsity effects take over, this defines spheres with average underdensity contrasts $\sim-0.8$.}

Figure~\ref{fig:ch5:ngasvsr} shows the median numbers of gas cells and stellar particles for all void galaxies {(solid purple)}, non-void galaxies (circles), radius-controlled galaxies ({dashed green}), {and density-controlled galaxies (dash dotted orange)} as a function of galaxy radius for the eleven \texttt{TNG300} snapshots. The vertical blue lines indicate the value of $R_{\rm gal,min}$ for each snapshot. 

The discontinuities in the number of gas cells and stellar particles in small galaxies at large redshifts can be seen in the $z\geq1.0$ panels of Figures~\ref{fig:ch5:ngasvsr}, where, to the left and right of the vertical lines, there are jumps in the function of an order of magnitude. For redshifts $z\leq0.7$, there are no such jumps, which justifies the use of a smaller value of $R_{\rm gal,min}$ at these redshifts. 
As most galaxies in \texttt{TNG300} have radii $\lesssim6h^{-1}$kpc (see~\ref{sec:ch5:sizes}), including these unresolved galaxies would have biased our galaxy populations due to their large number and lack of stellar particles and gas cells. Hence, we have defined $R_{\rm gal,min}$ to remove the unresolved galaxies with radii $\lesssim3.5h^{-1}$kpc at $z\geq1.0$ while keeping as many galaxies as possible at lower redshifts. This choice may account for a number of the differences seen between the $z<1$ and $z\geq1$ populations that are seen below (e.g., the jump that is seen in the top panel of Figure~\ref{fig:ch5:median_Ms_over_z}). However, this choice does affect all galaxy populations equally, so comparisons between galaxy populations can still be made.  

\begin{figure*}[]
    \centering
    \includegraphics[width=0.8\textwidth]{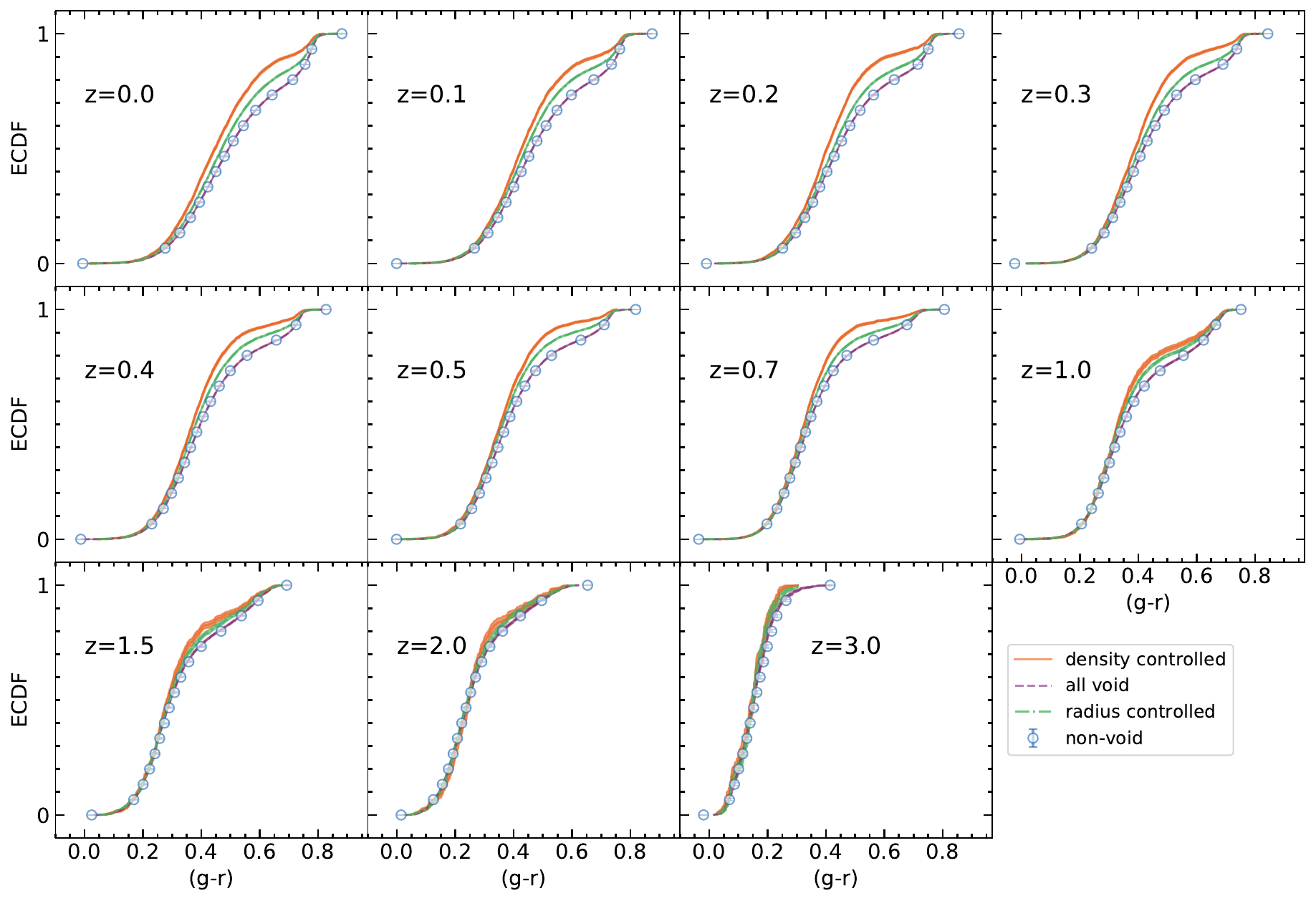}
    \caption{{Empirical} cumulative $(g-r)$ probability distribution functions for all void galaxies (purple), non-void galaxies (circles), radius-controlled galaxies (green){, and density-controlled (orange). Shaded regions show the 1$\sigma$ spread in the data.}}
    \label{fig:ch5:colormag}
\end{figure*}

\begin{figure}[!h]
    \centering
    \includegraphics[width=0.45\textwidth]{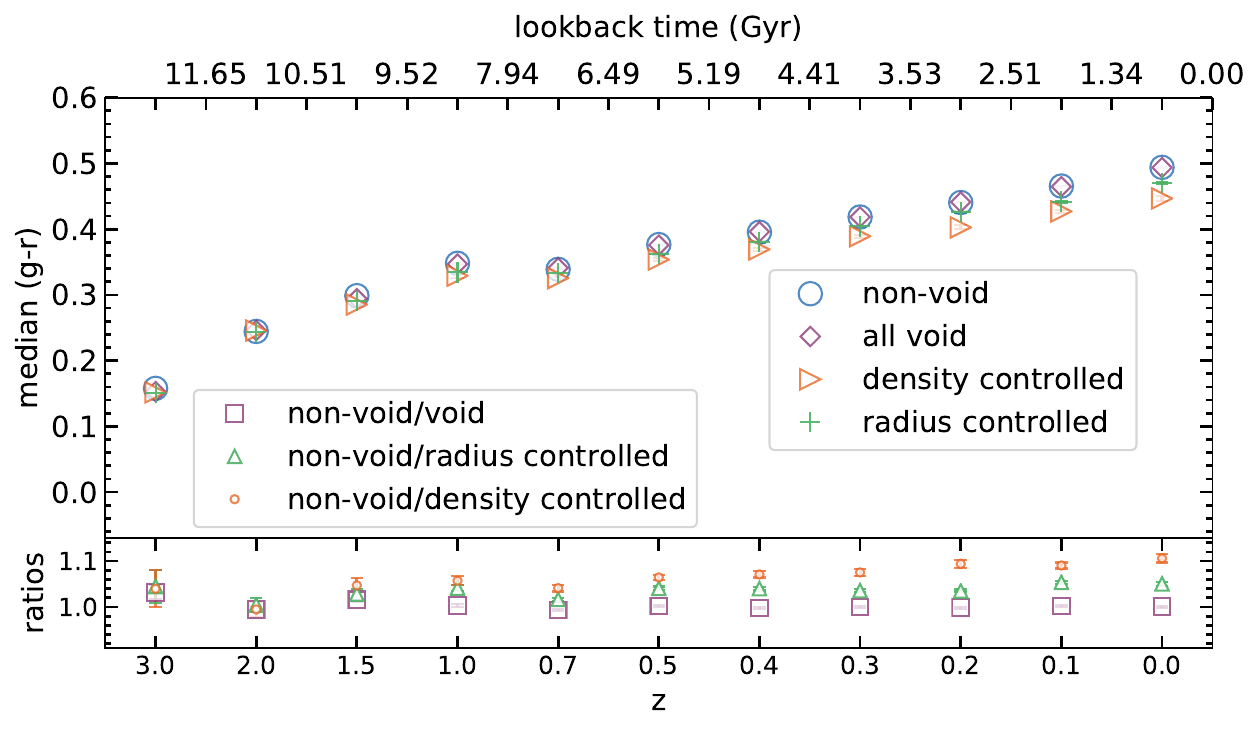}
    \caption{{Top:} Median $(g-r)$ color for all void (diamonds), non-void (circles), radius-controlled (crosses), {and density-controlled galaxies (triangles)} as a function of redshift. {Bottom:} Ratios of the corresponding points for the non-void and void (squares), radius-controlled (triangles){, and density-controlled (circles) galaxies.}}
    \label{fig:ch5:colormagmedians}
\end{figure}

\begin{figure*}[]
    \centering
    \includegraphics[width=0.7\textwidth]{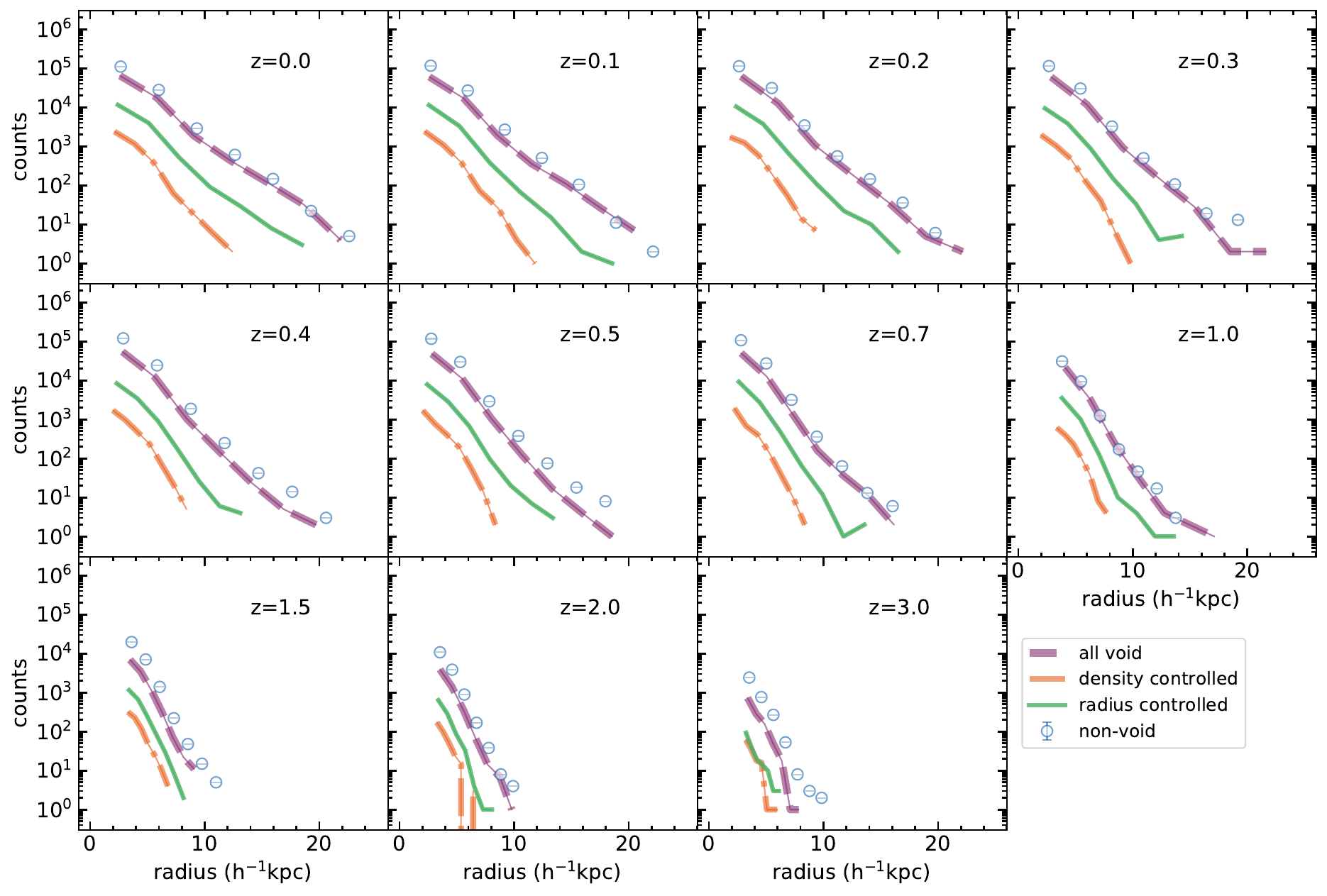}
    \caption{Number of galaxies as a function of radius for non-void (circles), all void ({purple}), radius-controlled ({green}){, and density-controlled galaxies (orange)}.}
    \label{fig:ch5:galradmosaic}
\end{figure*}

We present the total number of galaxies in each population for each redshift snapshot in Table~\ref{tab:ch5:galaxysummary}. The third column shows the total number of subhalos in each snapshot when only the cut to remove subhalos flagged as non-cosmological in origin is applied. Column four shows the total number of resolved subhalos in each snapshot, and column five shows the total number of subhalos in massive groups or clusters. The sixth through {ninth} columns show the total number of void, non-void, radius-controlled, {and density-controlled} galaxies when all quality and resolution cuts are applied. For the most part, the total number of galaxies in each population increases over time, likely due to the fact that more galaxies meet our resolution criteria at lower redshifts. 

{In Figure~\ref{fig:underdensities}, we plot $\delta_{\rm avg}$ for the underdense region defined by the complete void galaxy sample (purple), the radius-controlled sample (green), and the density-controlled sample (orange). With the exception of the higher redshift snapshots, where the density of available tracers limits the available regions that we can define, each density-controlled region has underdensity contrasts $\sim-0.8$ by definition. Because the {ZOBOV} algorithm includes much of the shell-crossing region in its definition of voids, the full void samples extend out to $\delta_{\rm avg}\sim1.6$ for a few voids in each snapshot. Most galaxies in \texttt{ZOBOV} are often located in regions near shell-crossing (i.e., near the edges of voids), where densities are intermediate to high. Truncating the voids to spheres of $0.8R_{\rm eff}$ successfully lowers the average underdensity contrasts of the specified regions to values comparable to those identified with SVFs, but variations in void shapes and sizes still allow galaxies in more overdense environments to contaminate the sample.}

\section{Physical Properties of Void Galaxies over Cosmic Time}
\label{sec:ch5:phyproperties}

\subsection{Color Distributions}
\label{sec:ch5:colormag}

Figure~\ref{fig:ch5:colormag} shows {empirical} cumulative probability distribution functions ({ECDFs}) for {non-void galaxies (circles), all void galaxies (purple line), radius-controlled galaxies (green line), and density-controlled galaxies (orange line). Shaded regions show the $1\sigma$ spread in the data.} for each redshift snapshot. Generally, each population becomes redder with time as the stellar populations within the galaxies mature. Furthermore, the ECDFs of each population evolve similarly over cosmic time.
All galaxies are very blue at high redshifts; i.e., the majority have
$(g-r)$ colors in the range 0.1 -- 0.3.
The ECDFs steadily shift towards $(g-r) >0.4$ at redshifts $z\leq0.5$. Compared to the void and non-void galaxies, there is a {noticeable overabundance of density-controlled galaxies with $(g-r)\sim0.4$}. 

The evolution in galaxy colors is better illustrated by Figure~\ref{fig:ch5:colormagmedians}. The top panel of this Figure shows the median $(g-r)$ colors of each galaxy population as a function of redshift and the bottom panel shows the ratios of the corresponding points between the non-void and void (squares), radius-controlled (triangles), {and density-controlled (circles)} as a function of redshift. From $z=3.0$ to $z=0.7$, the median colors of all {four} populations are consistent within the errorbars, and this remains true for the void and non-void galaxies down to $z=0.0$. Compared to the non-void galaxies, the median colors of the radius-controlled galaxies are consistently $5.0\pm0.3\%$ lower between $z=0.5$ and $z=0.0$. {Compared to the non-void galaxies, those in the density-controlled sample show a more prominent deviation that is growing with time --- the latter being $5.7\pm0.9\%$ bluer at $z=0.7$ and $10.6\pm0.9\%$ bluer at $z=0.0$.} 

\begin{figure}[!h]
    \centering
    \includegraphics[width=0.47\textwidth]{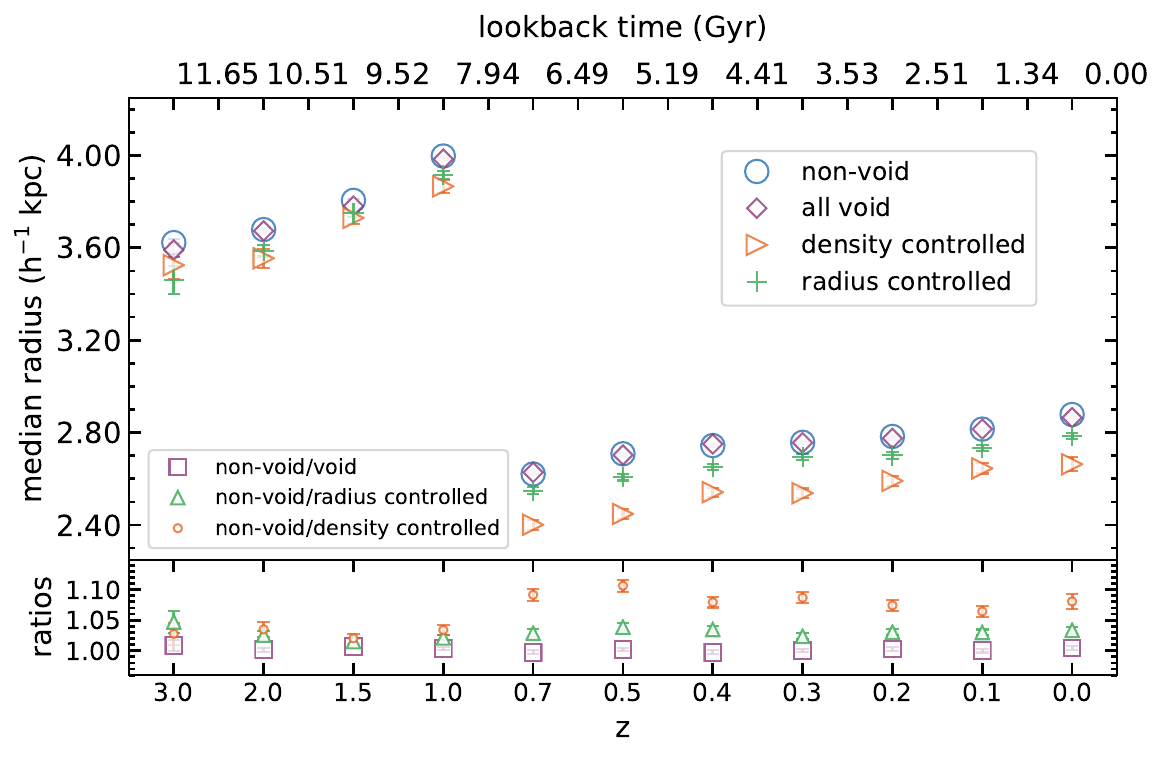}
    \caption{Same as Figure~\ref{fig:ch5:colormagmedians} except for the median galaxy radius as a function of redshift.}
    \label{fig:ch5:galradmedians}
\end{figure}

\begin{figure*}[!t]
    \centering
    \includegraphics[width=0.7\textwidth]{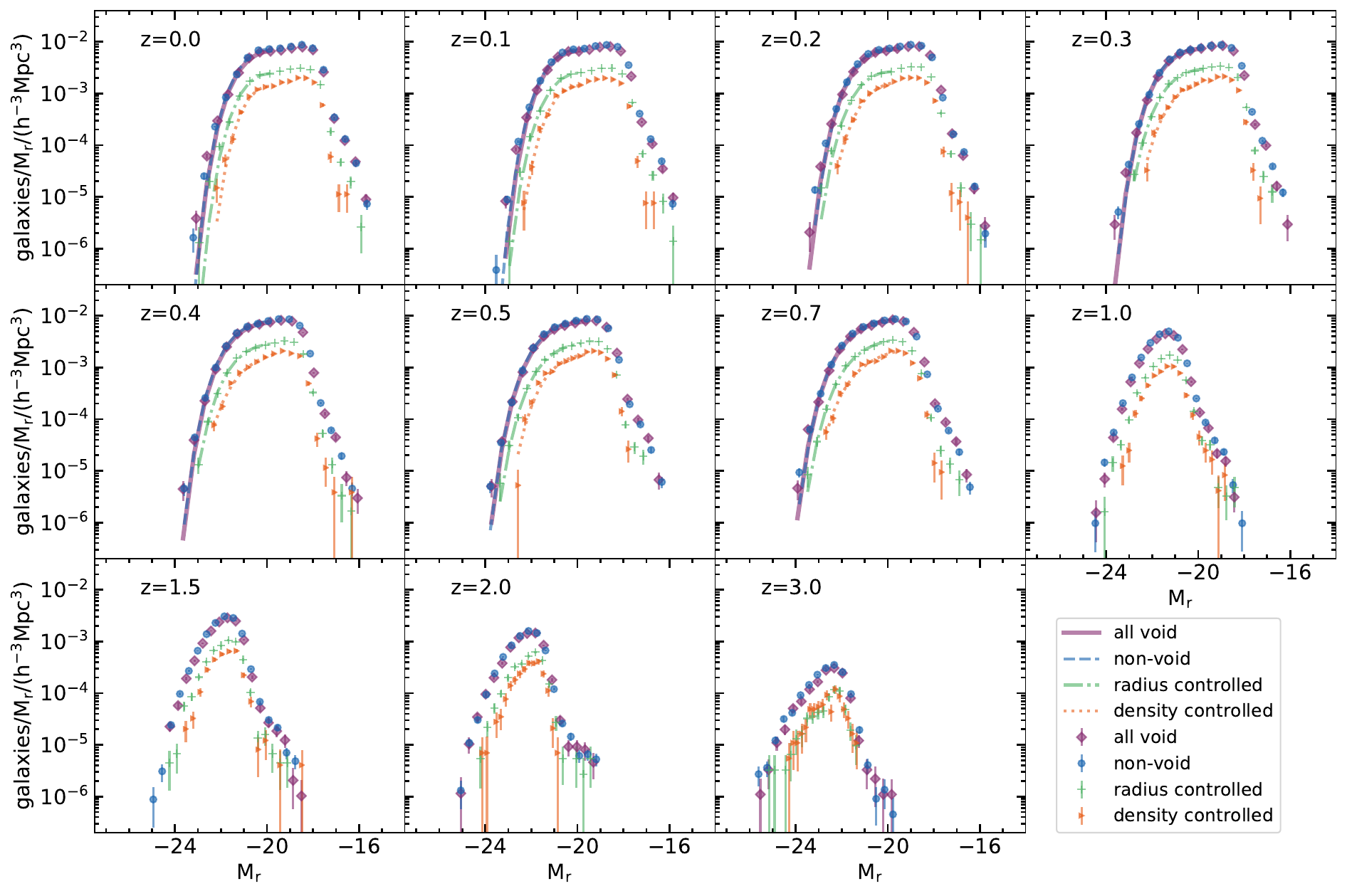}
    \caption{Luminosity functions for {non-void (blue), all void (purple), radius-controlled (green), and density-controlled (orange) galaxies}. Dashed lines show best-fit Schechter functions, which are fit down to the faintest absolute magnitude for which the luminosity functions are still increasing. For $z\geq1.0$, we are unable to fit the luminosity functions by Schechter functions due to the small number of galaxies in these snapshots.}
    \label{fig:ch5:lumfuncvoidgal}
\end{figure*}

\begin{deluxetable*}{c c c c | c c c}
\tablecaption{Best-fit Schechter parameters for non-void, all void, radius-controlled{, and density-controlled void galaxies}. Dashes indicate no fit was obtained.}
\tabletypesize{\scriptsize}
\label{tab:ch5:lumfuncparams}
\tablehead{
\colhead{z} & \colhead{$\phi_*$} & \colhead{$M_*$} & \colhead{$\alpha$} & \colhead{$\phi_*$} & \colhead{$M_*$} & \colhead{$\alpha$}
}
\startdata
\multicolumn{4}{c}{{non-void galaxies}} & \multicolumn{3}{c}{{all void galaxies}} \\ \hline
0.0 & $0.019\pm0.001$ & $-20.36\pm0.13$ & $-0.46\pm0.18$ & $0.016\pm0.001$ & $-20.51\pm0.09$ & $-0.66\pm0.08$ \\
0.1 & $0.019\pm0.001$ & $-20.54\pm0.07$ & $-0.51\pm0.08$ & $0.014\pm0.001$ & $-20.77\pm0.09$ & $-0.80\pm0.06$ \\
0.2 & $0.018\pm0.001$ & $-20.69\pm0.07$ & $-0.55\pm0.08$ & $0.014\pm0.001$ & $-20.91\pm0.07$ & $-0.82\pm0.04$ \\
0.3 & $0.017\pm0.001$ & $-20.88\pm0.05$ & $-0.65\pm0.06$ & $0.016\pm0.001$ & $-20.91\pm0.06$ & $-0.67\pm0.06$\\
0.4 & $0.016\pm0.001$ & $-21.06\pm0.05$ & $-0.72\pm0.06$ & $0.015\pm0.001$ & $-21.05\pm0.05$ & $-0.73\pm0.05$\\
0.5 & $0.015\pm0.001$ & $-21.26\pm0.05$ & $-0.84\pm0.04$ & $0.014\pm0.001$ & $-21.25\pm0.05$ & $-0.83\pm0.04$\\
0.7 & $0.014\pm0.001$ & $-21.47\pm0.04$ & $-0.89\pm0.03$ & $0.013\pm0.001$ & $-21.49\pm0.06$ & $-0.90\pm0.05$ \\
1.0 & -- & -- & -- & -- & -- & -- \\
1.5 & -- & -- & -- & -- & -- & -- \\
2.0 & -- & -- & -- & -- & -- & -- \\
3.0 & -- & -- & -- & -- & -- & -- \\ \hline
\multicolumn{4}{c}{{radius-controlled galaxies}} & \multicolumn{3}{c}{{density-controlled galaxies}} \\ \hline
0.0 & $0.007\pm0.001$ & $-20.09\pm0.14$ & $-0.21\pm0.23$ & $0.004\pm0.001$ & $-19.72\pm0.12$ & $0.15\pm0.24$ \\
0.1 & $0.007\pm0.001$ & $-20.36\pm0.09$ & $-0.43\pm0.14$ & $0.004\pm0.001$ & $-20.02\pm0.13$ & $-0.14\pm0.23$ \\
0.2 & $0.007\pm0.001$ & $-20.57\pm0.09$ & $-0.51\pm0.12$ & $0.004\pm0.001$ & $-20.40\pm0.11$ & $-0.62\pm0.14$ \\
0.3 & $0.007\pm0.001$ & $-20.78\pm0.06$ & $-0.66\pm0.07$ & $0.004\pm0.001$ & $-20.58\pm0.13$ & $-0.76\pm0.15$ \\
0.4 & $0.006\pm0.001$ & $-20.94\pm0.06$ & $-0.75\pm0.06$  & $0.003\pm0.001$ & $-20.92\pm0.14$ & $-0.97\pm0.13$ \\
0.5 & $0.005\pm0.001$ & $-21.12\pm0.05$ & $-0.83\pm0.05$& $0.004\pm0.001$ & $-20.72\pm0.16$ & $-0.74\pm0.18$  \\
0.7 & $0.005\pm0.001$ & $-21.34\pm0.06$ & $-0.94\pm0.05$& $0.003\pm0.001$ & $-21.24\pm0.16$ & $-1.05\pm0.12$ \\
1.0 & -- & -- & -- & -- & -- & -- \\
1.5 & -- & -- & -- & -- & -- & -- \\
2.0 & -- & -- & -- & -- & -- & -- \\
3.0 & -- & -- & -- & -- & -- & -- \\ \hline
\enddata
\end{deluxetable*}

\subsection{Size Distributions}
\label{sec:ch5:sizes}

Figure~\ref{fig:ch5:galradmosaic} shows the number of galaxies as a function of radius for the non-void (circles), {all void (purple),} radius-controlled ({green}), {and density-controlled (orange) galaxies,} from which it is clear that, compared to the other two populations, there are few large radius-controlled {and density-controlled} galaxies at all redshifts. For instance, there are no radius- {or density-}controlled galaxies larger than $10h^{-1}$kpc prior to redshift $z=1.0$, and there are no {density-}controlled galaxies with radii larger than {$13h^{-1}$kpc} at redshift $z=0.1$. In addition, the void and non-void galaxies show relatively similar distributions at both small and large scales at all redshifts.


Median galaxy radii as a function of redshift are shown in the top panel of Figure~\ref{fig:ch5:galradmedians}. The bottom panel of Figure~\ref{fig:ch5:galradmedians} shows the ratios between the median galaxy radii of non-void galaxies to those of the other {three} populations. There is a drop in the median radii at redshifts $z\leq0.7$ due to our definition of $R_{\rm gal,min}$ (see \S~\ref{sec:ch5:galcuts}.) The median radii of the void and non-void populations are indistinguishable from one another across time. Compared to the non-void galaxies, the radius-controlled galaxies are, on average, $3.5\pm0.5\%$ smaller at redshifts $z\leq0.4${, while galaxies interior to the shell-crossing surface are $8.1\pm1.2\%$ smaller at $z=0.0$.}. 



\subsection{Luminosity Functions}
\label{sec:ch5:lumfuncs}

Next, we present luminosity functions (i.e., the total number of galaxies per magnitude bin per unit volume) for our {four} galaxy populations. Figure~\ref{fig:ch5:lumfuncvoidgal} shows the luminosity functions for all void galaxies {(purple)}, where the total volumes in significant \texttt{ZOBOV} voids in each snapshot were used to normalize these luminosity functions. This corresponds to $29-32\%$ of the volume of \texttt{TNG300} (see Table~1 in \citealt{curtis2025}). For $z\geq1.0$, we are unable to fit the luminosity functions by Schechter functions \citep{schechter1976} due to the small number of resolved galaxies in these snapshots. Similarly, due to the falloff in the number of the faintest galaxies, we only fit the functions down to the faintest absolute magnitudes for which the luminosity functions are still increasing. Upon closer inspection, we find that galaxies with magnitudes fainter than these cutoffs have radii between $R_{\rm gal,min}$ and $5h^{-1}$kpc.

The $z<1.0$ snapshots show good agreement between the number of bright galaxies and the best-fit Schechter luminosity function \citep{schechter1976}, with only the $z=0.3 - 0.7$ snapshots showing a slight overabundance compared to the Schechter function. 
At fainter magnitudes, there is a rapid falloff in the faint end of the distributions for all of these redshifts. This differs from a Schechter luminosity function, which has a single slope to the faint end of its distribution. 

Figures~\ref{fig:ch5:lumfuncvoidgal} {also} shows the luminosity functions and best-fit Schechter functions for the non-void {(blue)}, radius-controlled{ (green), and density-controlled (orange)} galaxies. The normalizing volumes for the luminosity functions of the non-void galaxies are the volumes in each snapshot that were not part of a \texttt{ZOBOV} void or a cluster (the latter of which comprised $\sim900h^{-3}\: \rm Mpc^3$ in each snapshot), {while} the volumes used for the luminosity functions of the radius and {density-controlled} galaxies were calculated using $0.8R_{\rm eff}$ of all voids {and the radii of the underdense spheres that meet the $\delta<-0.8$ density criterion ($\sim 10\%$ of the simulation box), respectively}. Much like before, the luminosity functions at redshifts $z\geq1.0$ cannot be fit by a Schechter function, and for $z<1.0$, to account for the falloff of the number of galaxies at the faint end, we only fit the functions down to the faintest absolute magnitudes for which the luminosity functions are still increasing. These luminosity functions again show that \texttt{TNG300}, compared to Schechter luminosity functions at redshifts $z<1.0$, slightly over-produced the brightest galaxies and under-produced the faintest galaxies, as was the case in Figure~\ref{fig:ch5:lumfuncvoidgal}. 

Table~\ref{tab:ch5:lumfuncparams} shows the parameters for the best-fit Schechter luminosity functions for each snapshots, where the dashes indicate where we were unable to properly fit a Schechter luminosity function. Compared to the non-void galaxies (void galaxies) at redshifts $\leq0.2$, the radius-controlled galaxies have values of $\phi_*$ that are lower by $0.012\pm0.002$ ($0.008\pm0.002$) on average, values of $M_*$ that are $0.19\pm0.20$ ($0.39\pm0.19$) magnitudes dimmer on average, and values of $\alpha$ that are smaller by a factor of $0.12\pm0.28$ ($-0.38\pm0.22$) on average. At redshifts $\geq0.3$, the radius-controlled galaxies have values of $\phi_*$ that are lower by $0.010\pm0.002$ ($0.009\pm0.002$), values of $M_*$ that are $0.12\pm0.11$ ($0.13\pm0.11$) magnitudes dimmer on average, and values of $\alpha$ that are smaller by a factor of $0.02\pm0.11$ ($0.01\pm0.11$).

{The density-controlled galaxies show larger deviations from the non-void and all void galaxy populations, again indicating a shift in galaxy properties interior to the surface of shell-crossing. For instance, compared to the non-void galaxies (void galaxies) at redshifts $\leq0.2$, the density-controlled galaxies have values of $\phi_*$ that are lower by $0.015\pm0.002$ ($0.011\pm0.002$) on average, values of $M_*$ that are $0.48\pm0.21$ ($-0.68\pm0.20$) magnitudes dimmer on average, and values of $\alpha$ that are smaller by a factor of $0.30\pm0.32$ ($-0.56\pm0.26$) on average, and, at redshifts $\geq0.3$, the radius-controlled galaxies have values of $\phi_*$ that are lower by $0.012\pm0.002$ ($0.011\pm0.002$), values of $M_*$ that are $-0.30\pm0.20$ ($-0.31\pm0.20$) magnitudes dimmer on average, and values of $\alpha$ that are smaller by a factor of $0.11\pm0.19$ ($0.10\pm0.20$).}

\subsection{Mass Functions}
\label{sec:ch5:massfuncs}

\begin{figure*}[!t]
\centering
    \includegraphics[width=0.9\textwidth]{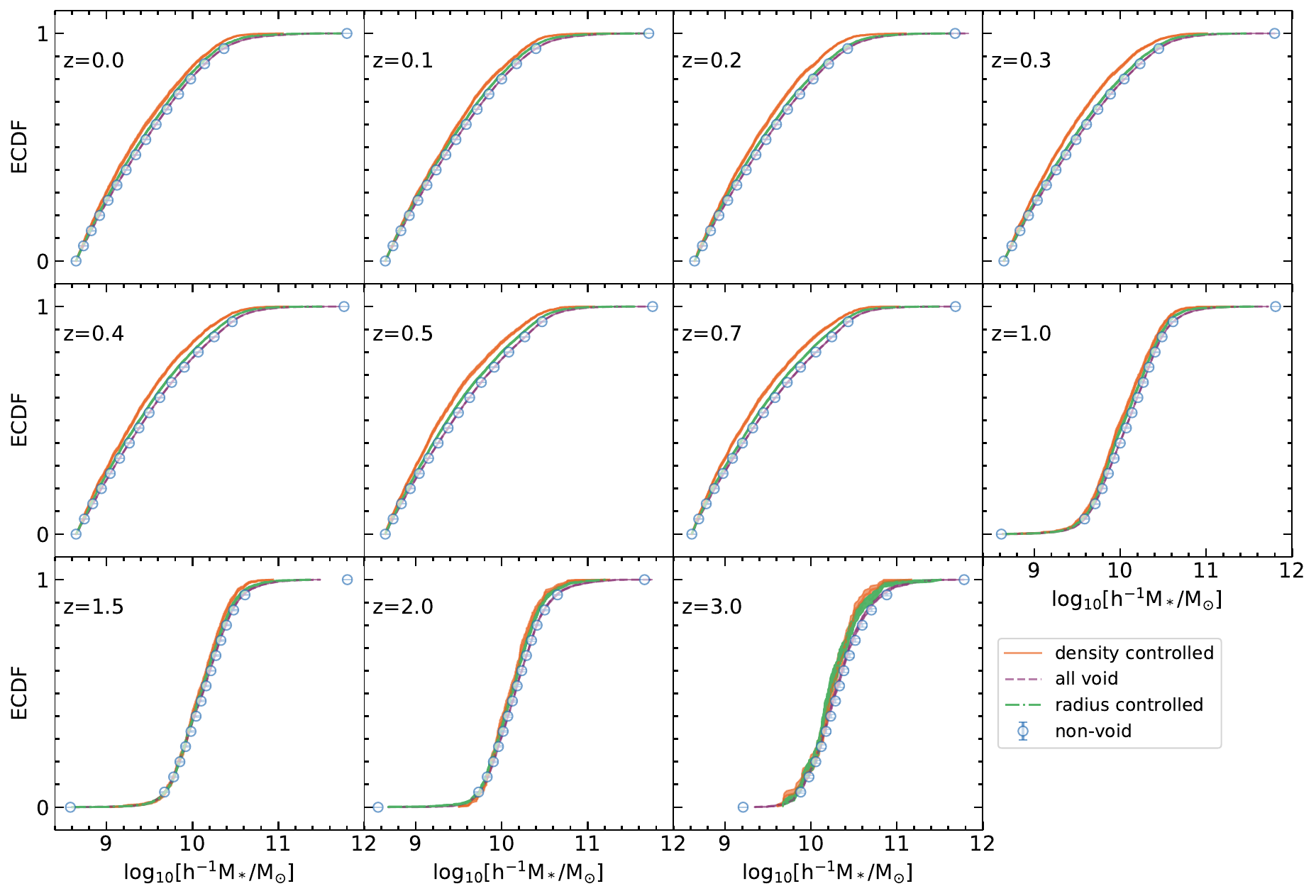}
    \caption{{Same as Figure~\ref{fig:ch5:colormag} except for stellar mass.}}
    \label{fig:ch5:Ms_mosaic_allvoid_nv}
\end{figure*}

\begin{figure}[!h]
    \centering
    \includegraphics[width=0.47\textwidth]{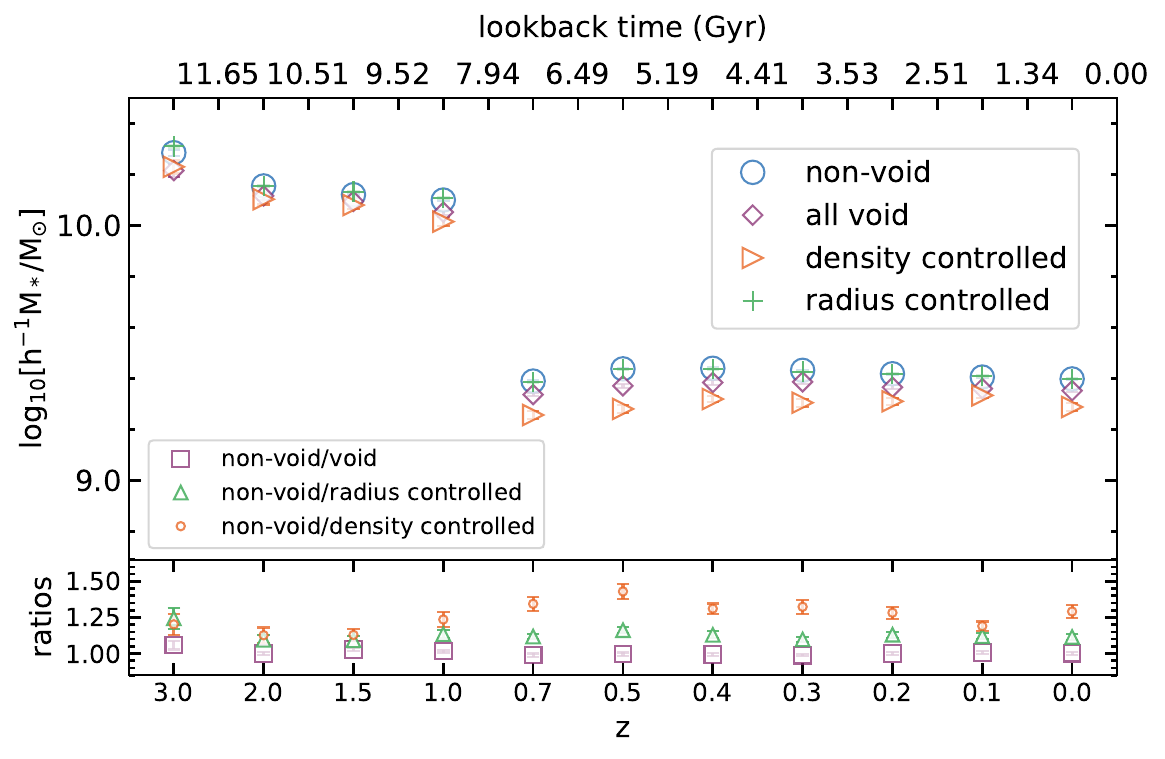}
    \caption{Same as Figure~\ref{fig:ch5:colormagmedians}, but showing the median stellar mass as a function of redshift.} 
    \label{fig:ch5:median_Ms_over_z}
\end{figure}


Figure~\ref{fig:ch5:Ms_mosaic_allvoid_nv} shows ECDFs for the stellar masses of{ non-void (circles), all void (purple), density-controlled (orange), and radius-controlled galaxies (green). The shaded regions show $1\sigma$ standard deviations in the data.}


All four galaxy populations have stellar mass ECDFs that evolve similarly across cosmic time.
For redshifts $z<1.0$, most of the subhalos in the simulation have stellar masses between $10^{10}-10^{11}h^{-1}M_{\rm \odot}$, but 
by $z=1.0$, a population of low-mass galaxies with stellar masses $\lesssim10^{9}h^{-1}M_{\rm \odot}$ emerges due to the evolving resolution cut discussed in Section~\ref{sec:ch5:galcuts}. 
From $z=0.4$ to $z=0.0$, the low-mass tails of the stellar mass ECDFs are less pronounced than they are in the $z=0.5$ and $z=0.7$ snapshots.


The median stellar masses of each galaxy population as a function of redshift are shown in Figure~\ref{fig:ch5:median_Ms_over_z}. The top panel shows the median stellar masses for the non-void (circles), void (diamonds), radius-controlled (crosses), {and density-controlled galaxies (triangles)}. The bottom panel shows the ratios of the corresponding medians between the non-void and void (squares), radius-controlled (triangles){, and density-controlled (circles) galaxies}. 

\begin{figure*}[]
    \centering
    \includegraphics[width=0.9\textwidth]{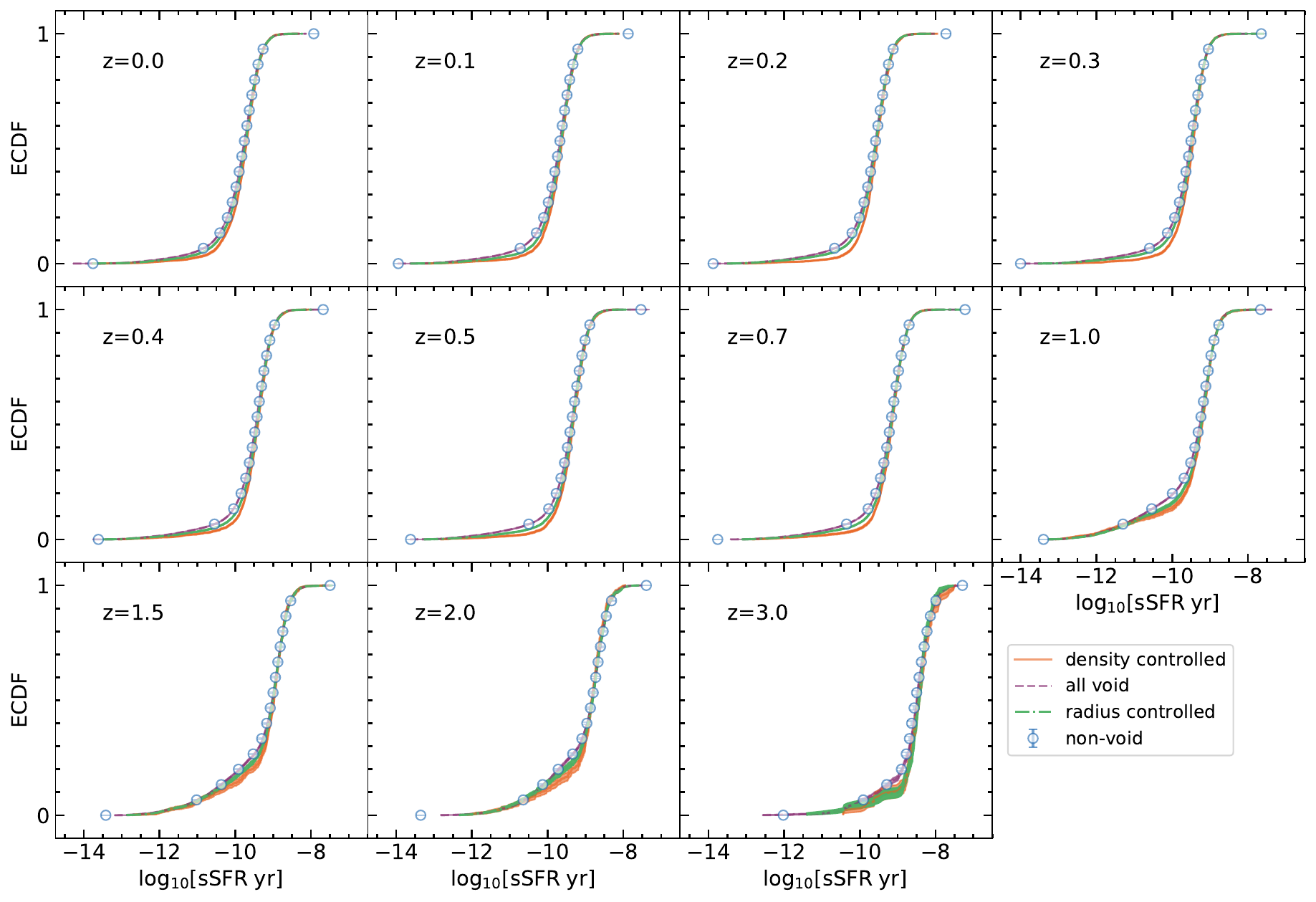}
    \caption{Same as Figure~\ref{fig:ch5:colormag} except specfic star formation rate.}
    \label{fig:ch5:ssfr_mosaic_allvoid_nv}
\end{figure*}

There is a drop in the median stellar masses of all {four} galaxy populations at $z=0.7$, which can be attributed to our definition of $R_{\rm gal,min}$ (see \S~\ref{sec:ch5:galcuts}). Within the formal error bars, the median stellar masses of the all void galaxies are indistinguishable from those of the non-void galaxies. Compared to the non-void galaxies, the radius-controlled galaxies are $\sim13.5\pm2.4\%$ less massive at redshifts $3.0>z\geq1.0$ and $\sim12.0\pm2.0\%$ less massive at redshifts $z<0.7$. {Interior to the shell-crossing surface at redshifts $<1.0$, the density-controlled galaxies show a stronger deviation from those in the most dense environments, where, compared to the non-void galaxies, they are $43.0\pm5.7\%$ less massive at $z=0.5$ and $29.1\pm3.9\%$ less massive at $z=0.0$. We note that the fact that the density-controlled galaxies reach a maximum deviation at $z=0.5$ might be  due to the sporadic late-time major mergers that have been reported for void galaxies \citep{rodriguesmedrano2024}. However, tracking merger trees is the beyond the scope of this study.}



\subsection{sSFRs}
\label{sec:ch5:ssfr}

\begin{figure}[]
    \centering
    \includegraphics[width=0.47\textwidth]{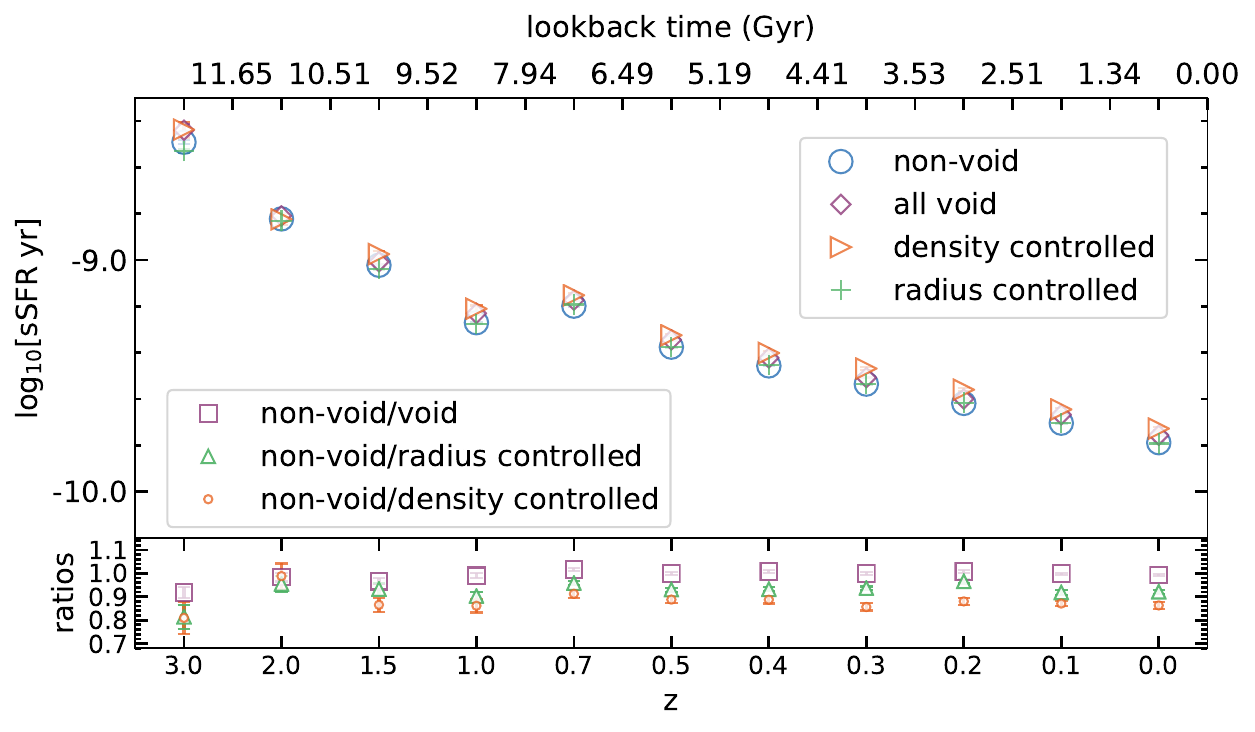}
    \caption{Same as Figure~\ref{fig:ch5:colormagmedians}, but showing the median sSFR as a function of redshift.}
    \label{fig:ch5:median_ssfr_overz}
\end{figure}

Figure~\ref{fig:ch5:ssfr_mosaic_allvoid_nv} shows ECDFs for sSFRs of the {non-void (circles), all void (purple), density-controlled (orange), and radius-controlled (green) galaxies. Shaded regions show $1\sigma$ standard deviations in the data}. From these figures, it is clear that there are a few differences between the evolution of the sSFR ECDFs of each population. The $z=3.0$ to $z=1.5$ snapshots contain a sizeable population of galaxies with very low sSFRs ($10^{-11}-10^{-9}\rm yr^{-1}$). Despite this, star formation is still more prevalent in the higher redshift snapshots than in other snapshots, in agreement with observational studies of the cosmic star formation history \citep{madau2014}. At low redshifts, most of the galaxies have sSFRs between $10^{-9.5}-10^{-8.5}\rm yr^{-1}$, and the ECDFs change very little from snapshot to snapshot.

The top panel of Figure~\ref{fig:ch5:median_ssfr_overz} shows the median sSFR of {non-void (circles), all void (diamonds), density-controlled (triangles), and radius-controlled (crosses) galaxies,} while the bottom panel of Figure~\ref{fig:ch5:median_ssfr_overz} shows the ratios between the non-void population and the other {three} void galaxy populations. For the epochs that we studied, star formation peaks at $z=3.0$ for each galaxy population. The median sSFRs for all void galaxies are equal to those of the non-void galaxies at all redshifts, but compared to the non-void galaxies, the radius-controlled galaxies have median sSFRs that are $\sim7.0\pm0.9\%$ higher. {Galaxies interior to the shell-crossing region are much more actively star forming, where, compared to the non-void galaxies, they are $13.5\pm3.1\%$ more actively star forming at $z=1.5$ and $13.7\pm1.6\%$ more actively star forming at $z=0.0$.} 



\subsection{Stellar and Gas Chemical Abundance Ratios}
\label{sec:ch5:metals}

\figsetstart
\figsetnum{12}
\figsettitle{Stellar and gaseous chemical abundance ratios.}

\figsetgrpstart
\figsetgrpnum{figurenumber.1}
\figsetgrptitle{Stellar chemical abundance ratios at z=0.0}
\figsetplot{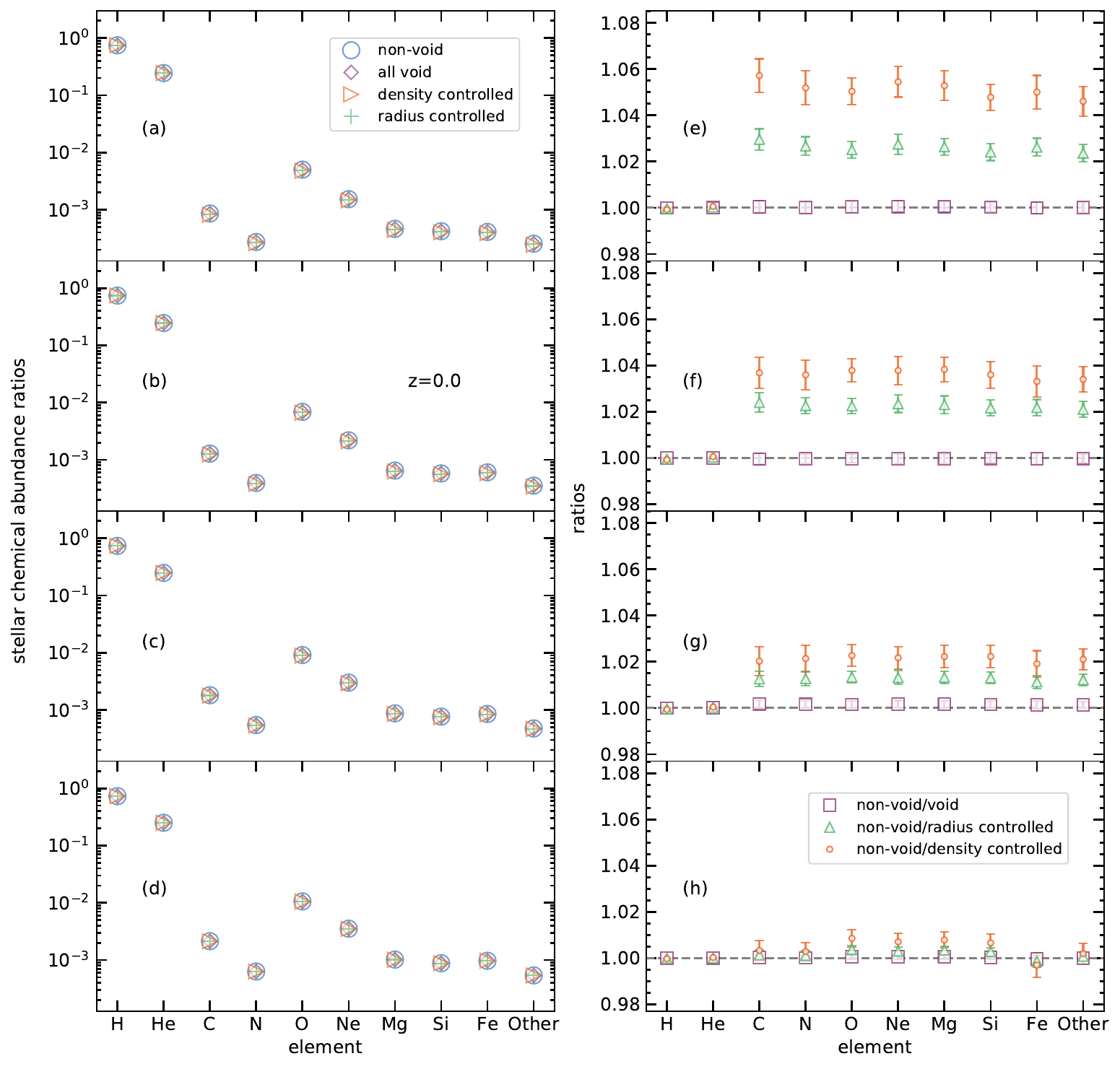}
\figsetgrpnote{{{\it Left:} Average stellar chemical abundance ratios for non-void (circles){, all void (diamonds), density-controlled (triangles), and radius-controlled (crosses)} galaxies in the $z=0.0$ snapshot. {\it Right:} Ratios between the stellar abundances of the non-void and void {(squares), non-void and radius-controlled (triangles), and non-void and density-controlled (circles)} galaxies from the left panels. Galaxies with stellar masses $10^{8.5} h^{-1} M_\odot \le M_* < 10^{9.0}h^{-1}$ $ M_\odot$: panels (a) and (e). Galaxies with stellar masses $10^{9.0} h^{-1} M_\odot \le M_* < 10^{9.5}h^{-1}$ $ M_\odot$: panels (b) and (f). Galaxies with stellar masses $10^{9.5} h^{-1} M_\odot \le M_* < 10^{10.0}h^{-1}$ $ M_\odot$: panels (c) and (g). Galaxies with stellar masses $10^{10.0} h^{-1} M_\odot \le M_* < 10^{10.5}h^{-1}$ $ M_\odot$: panels (d) and (h). {The complete Figure Set (22 images) is available in the online journal, showing stellar and gaseous chemical abundance ratios for all 11 snapshots. Data are omitted when there were no galaxies, stellar particles, or gas cells in that group.}}}
\figsetgrpend

\begin{figure*}[!htbp]
    \centering
    \digitalasset
    \includegraphics[width=0.7\textwidth, keepaspectratio]{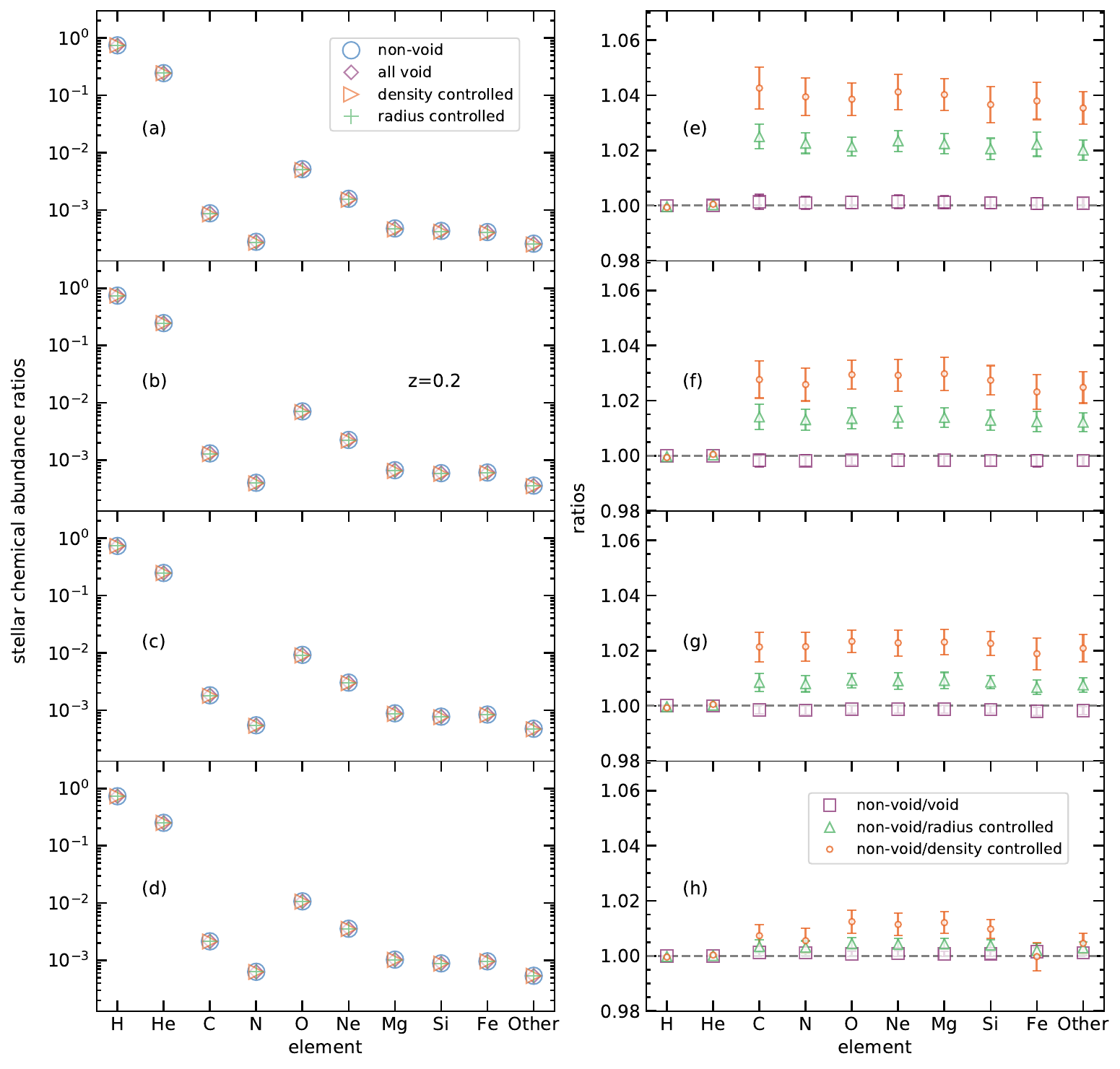}
    \caption{{\it Left:} Average stellar chemical abundance ratios for non-void (circles){, all void (diamonds), density-controlled (triangles), and radius-controlled (crosses)} galaxies in the $z=0.2$ snapshot. {\it Right:} Ratios between the stellar abundances of the non-void and void {(squares), non-void and radius-controlled (triangles), and non-void and density-controlled (circles)} galaxies from the left panels. Galaxies with stellar masses $10^{8.5} h^{-1} M_\odot \le M_* < 10^{9.0}h^{-1}$ $ M_\odot$: panels (a) and (e). Galaxies with stellar masses $10^{9.0} h^{-1} M_\odot \le M_* < 10^{9.5}h^{-1}$ $ M_\odot$: panels (b) and (f). Galaxies with stellar masses $10^{9.5} h^{-1} M_\odot \le M_* < 10^{10.0}h^{-1}$ $ M_\odot$: panels (c) and (g). Galaxies with stellar masses $10^{10.0} h^{-1} M_\odot \le M_* < 10^{10.5}h^{-1}$ $ M_\odot$: panels (d) and (h). {The complete Figure Set (22 images) is available in the online journal, showing stellar and gaseous chemical abundance ratios for all 11 snapshots. In the complete Figure Set, data are omitted when there were no galaxies, stellar particles, or gas cells in that group.}}
    \label{fig:ch5:gas_met_frac_ex_z02}
\end{figure*}

\figsetgrpstart
\figsetgrpnum{figurenumber.2}
\figsetgrptitle{Stellar chemical abundance ratios at z=0.1}
\figsetplot{star_metal_fracs_z00.pdf}
\figsetgrpnote{Same as Figure~\ref{fig:ch5:gas_met_frac_ex_z02} except for galaxies at $z=0.1$.}
\figsetgrpend

\figsetgrpstart
\figsetgrpnum{figurenumber.2}
\figsetgrptitle{Stellar chemical abundance ratios at z=0.1}
\figsetplot{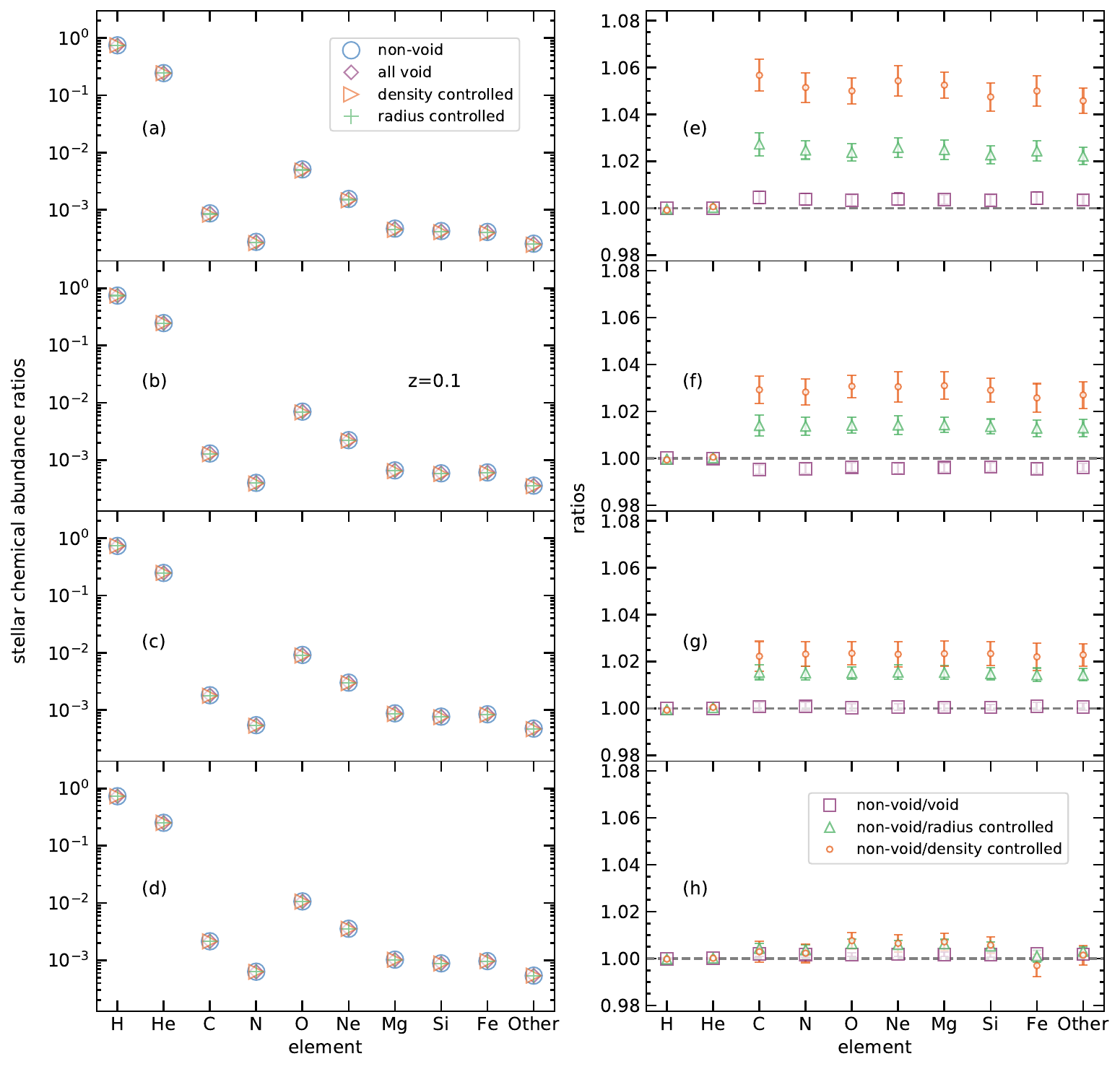}
\figsetgrpnote{Same as Figure~\ref{fig:ch5:gas_met_frac_ex_z02} except for galaxies at $z=0.1$.}
\figsetgrpend

\figsetgrpstart
\figsetgrpnum{figurenumber.3}
\figsetgrptitle{Stellar chemical abundance ratios at z=0.2}
\figsetplot{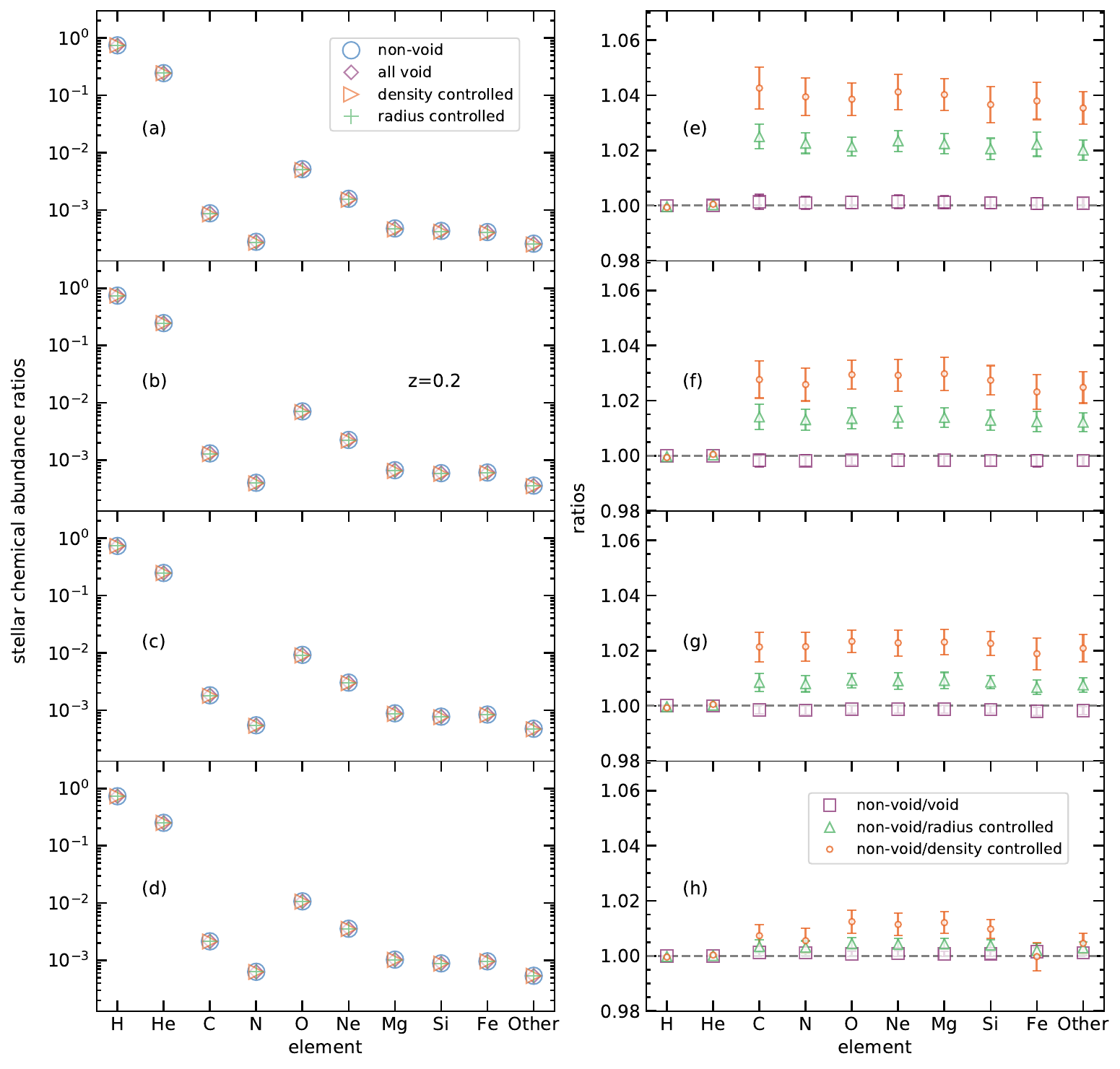}
\figsetgrpnote{Same as Figure~\ref{fig:ch5:gas_met_frac_ex_z02} except for galaxies at $z=0.2$.}
\figsetgrpend

\figsetgrpstart
\figsetgrpnum{figurenumber.4}
\figsetgrptitle{Stellar chemical abundance ratios at z=0.3}
\figsetplot{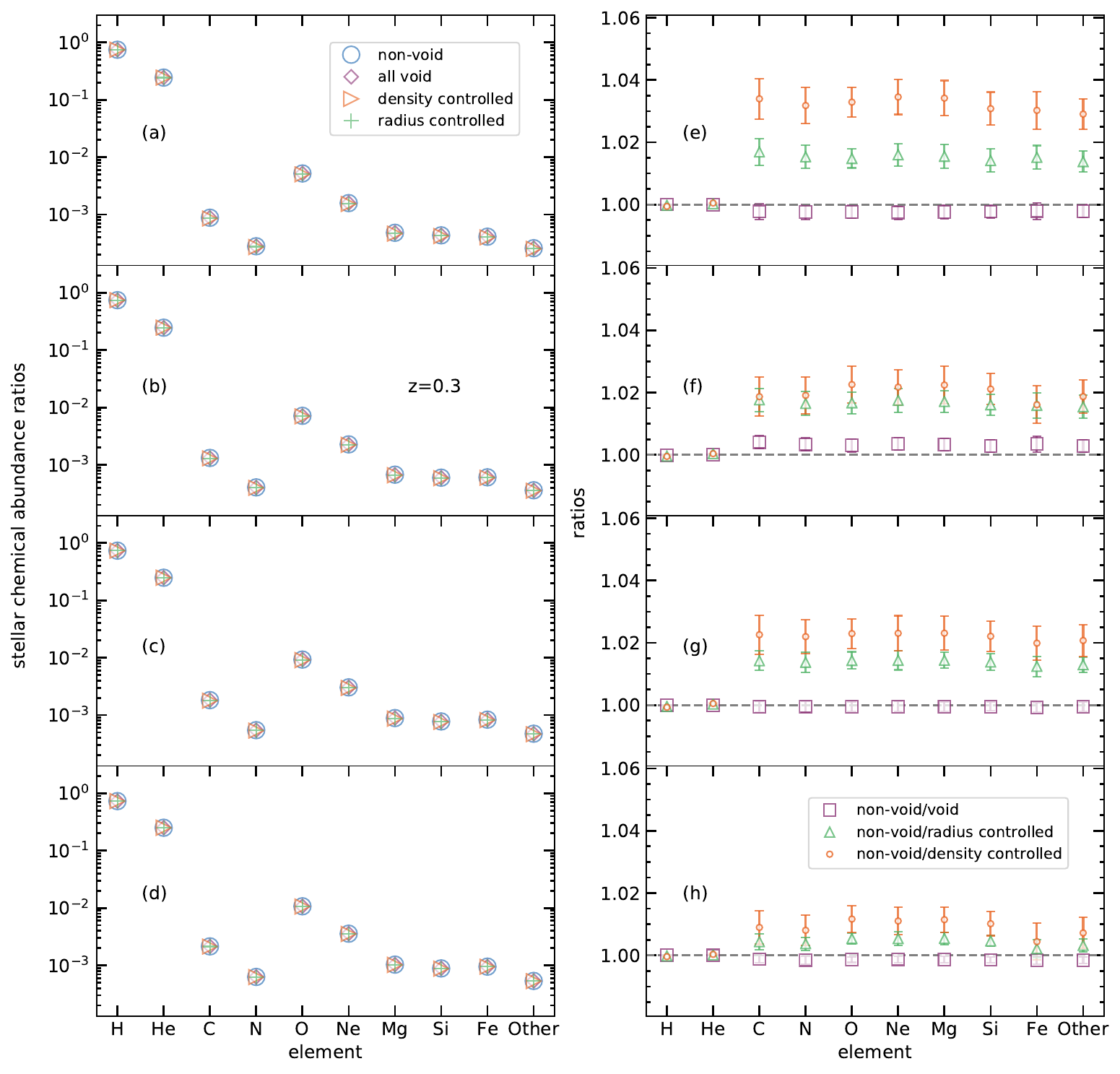}
\figsetgrpnote{Same as Figure~\ref{fig:ch5:gas_met_frac_ex_z02} except for galaxies at $z=0.3$.}
\figsetgrpend

\figsetgrpstart
\figsetgrpnum{figurenumber.5}
\figsetgrptitle{Stellar chemical abundance ratios at z=0.4}
\figsetplot{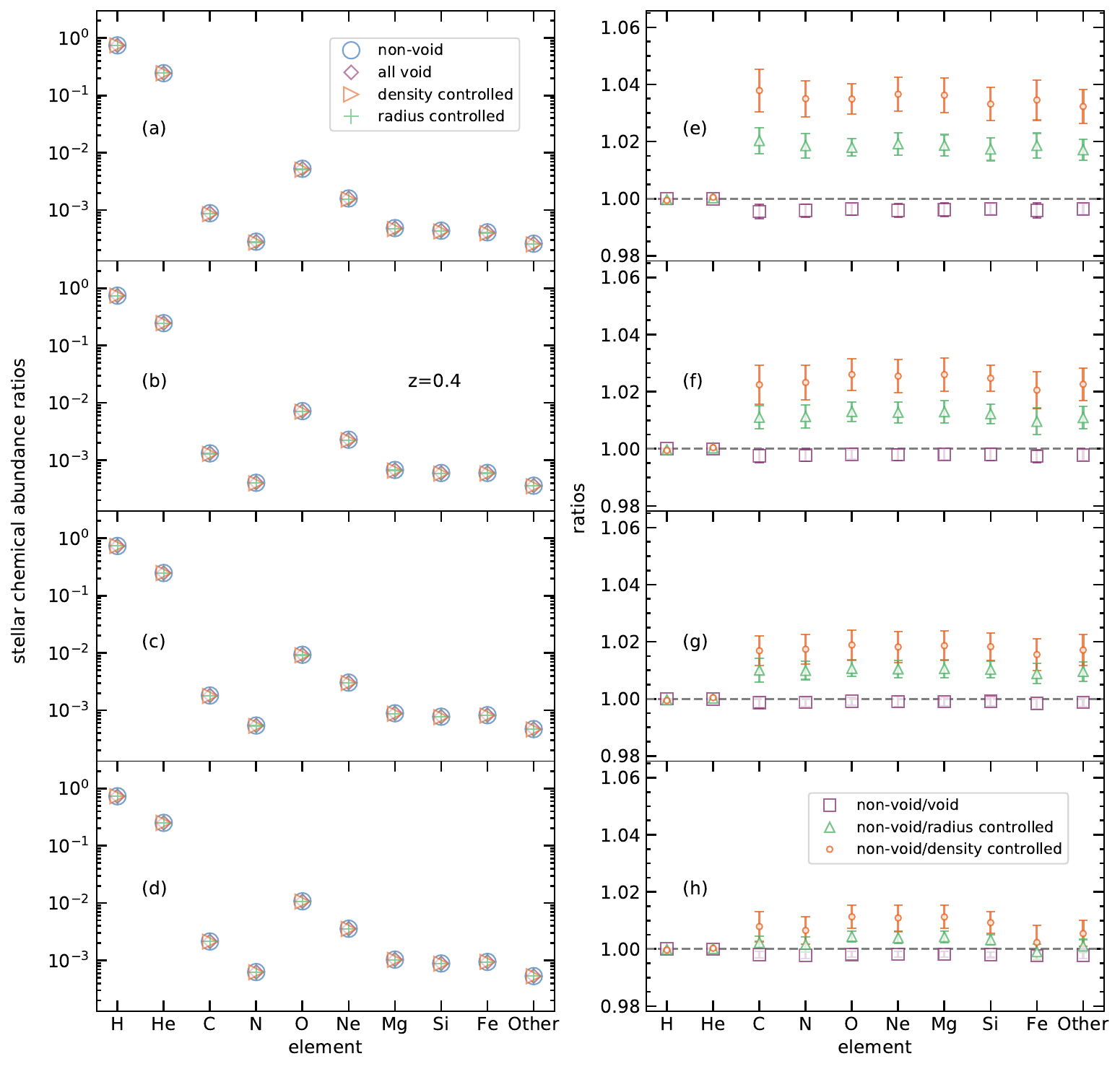}
\figsetgrpnote{Same as Figure~\ref{fig:ch5:gas_met_frac_ex_z02} except for galaxies at $z=0.4$.}
\figsetgrpend

\figsetgrpstart
\figsetgrpnum{figurenumber.6}
\figsetgrptitle{Stellar chemical abundance ratios at z=0.5}
\figsetplot{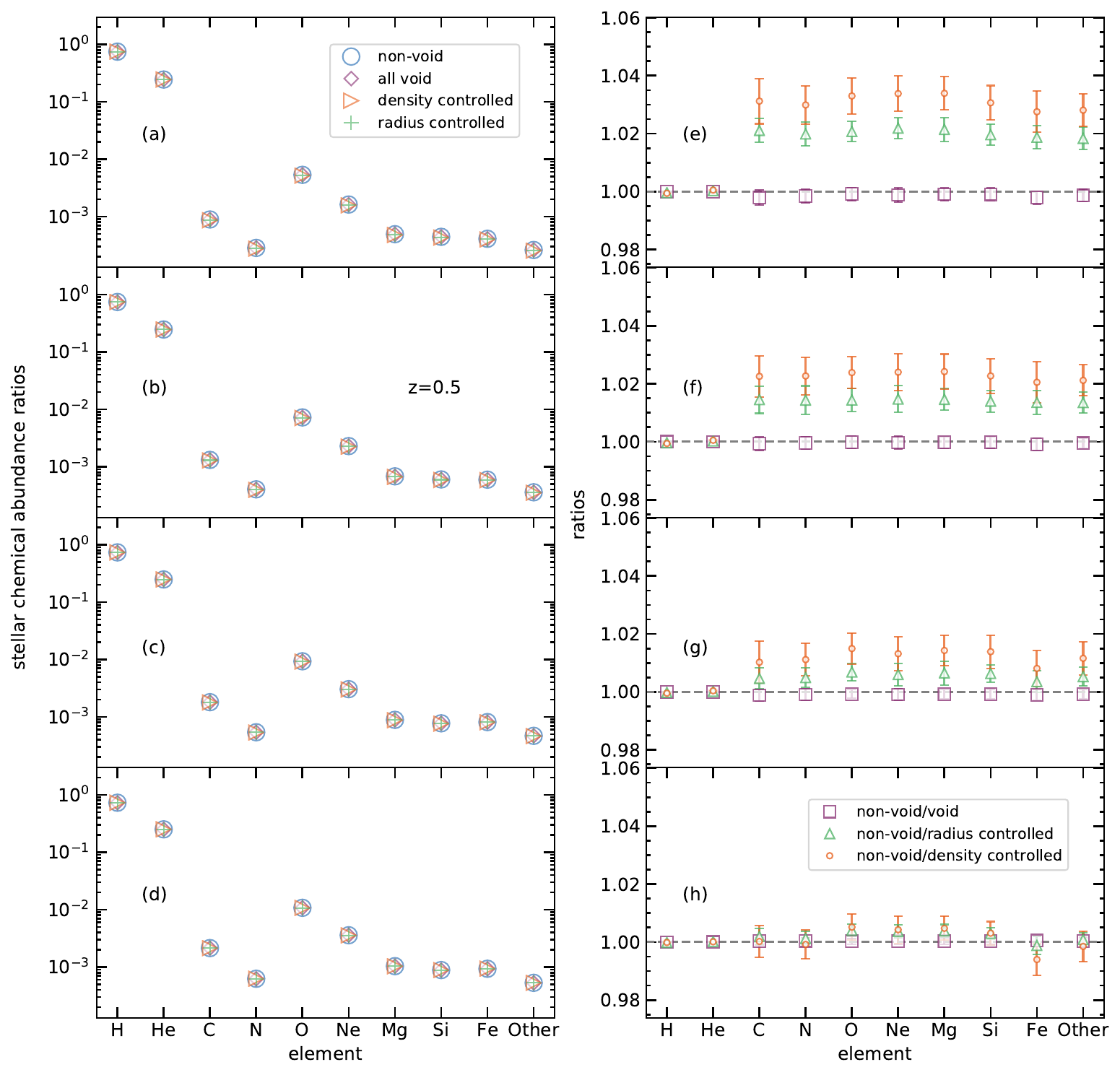}
\figsetgrpnote{Same as Figure~\ref{fig:ch5:gas_met_frac_ex_z02} except for galaxies at $z=0.5$.}
\figsetgrpend

\figsetgrpstart
\figsetgrpnum{figurenumber.7}
\figsetgrptitle{Stellar chemical abundance ratios at z=0.7}
\figsetplot{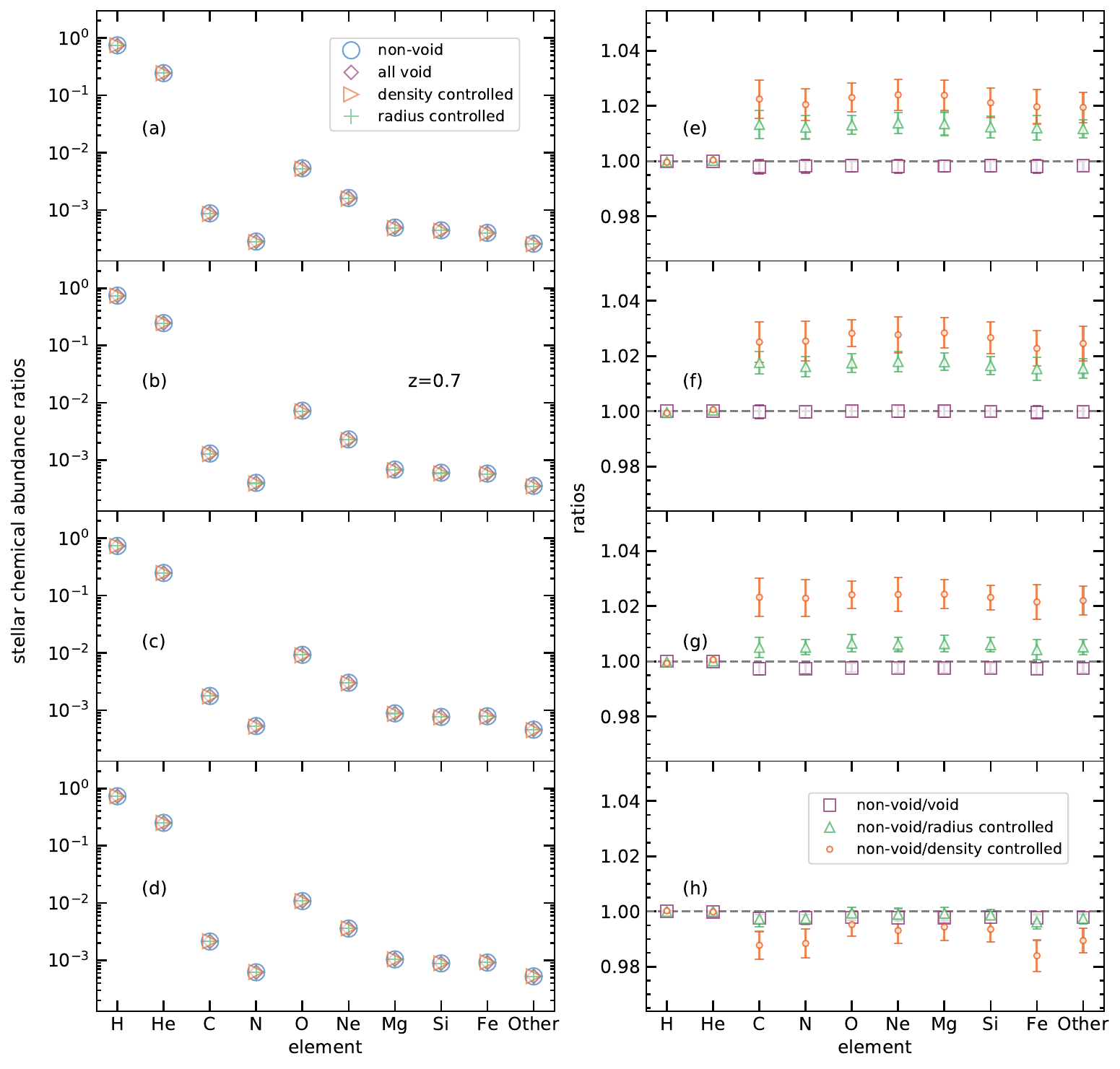}
\figsetgrpnote{Same as Figure~\ref{fig:ch5:gas_met_frac_ex_z02} except for galaxies at $z=0.7$.}
\figsetgrpend

\figsetgrpstart
\figsetgrpnum{figurenumber.8}
\figsetgrptitle{Stellar chemical abundance ratios at z=1.0}
\figsetplot{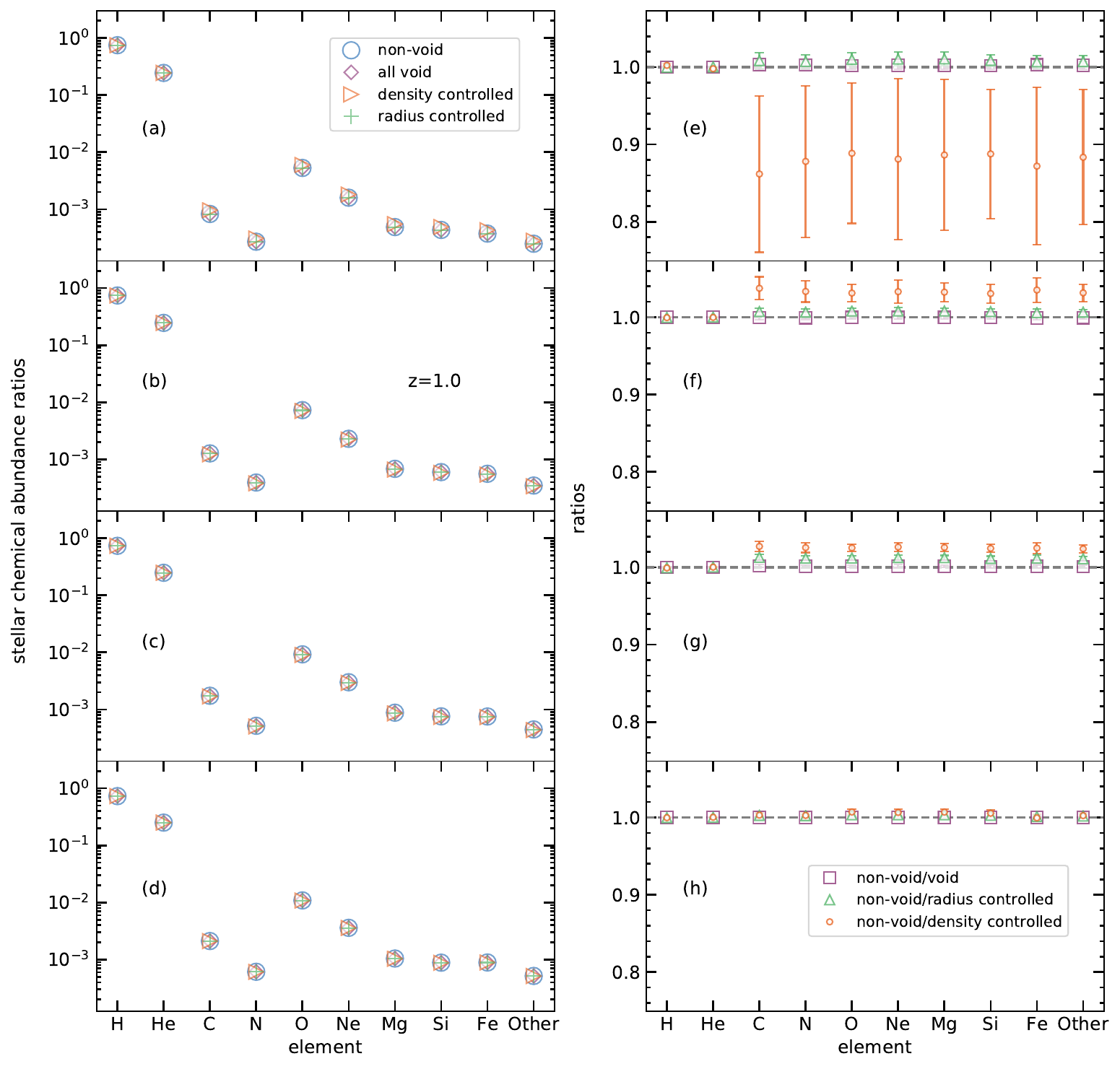}
\figsetgrpnote{Same as Figure~\ref{fig:ch5:gas_met_frac_ex_z02} except for galaxies at $z=1.0$.}
\figsetgrpend

\figsetgrpstart
\figsetgrpnum{figurenumber.9}
\figsetgrptitle{Stellar chemical abundance ratios at z=1.5}
\figsetplot{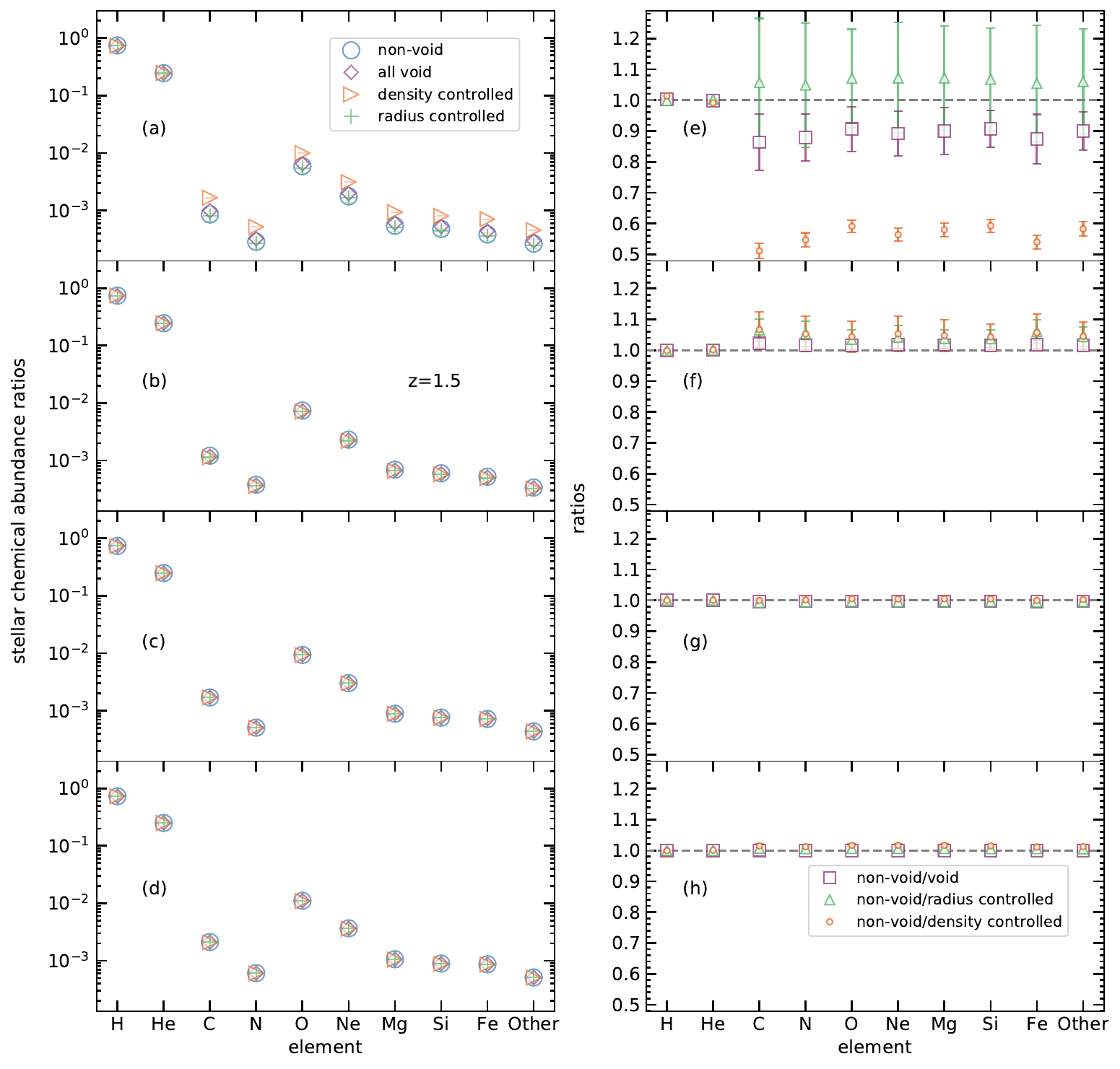}
\figsetgrpnote{Same as Figure~\ref{fig:ch5:gas_met_frac_ex_z02} except for galaxies at $z=1.5$.}
\figsetgrpend

\figsetgrpstart
\figsetgrpnum{figurenumber.10}
\figsetgrptitle{Stellar chemical abundance ratios at z=2.0}
\figsetplot{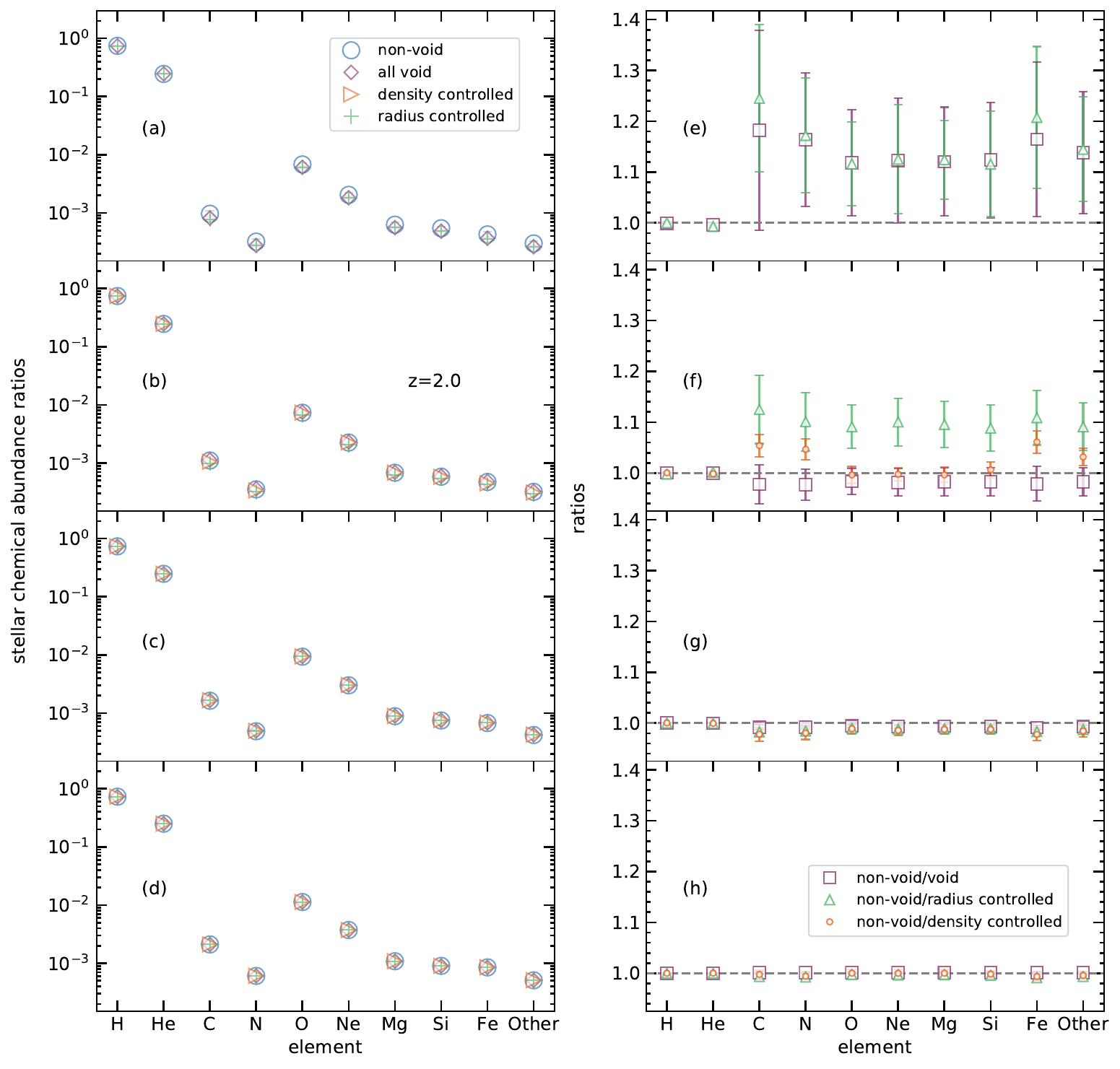}
\figsetgrpnote{Same as Figure~\ref{fig:ch5:gas_met_frac_ex_z02} except for galaxies at $z=2.0$.}
\figsetgrpend

\figsetgrpstart
\figsetgrpnum{figurenumber.11}
\figsetgrptitle{Stellar chemical abundance ratios at z=3.0}
\figsetplot{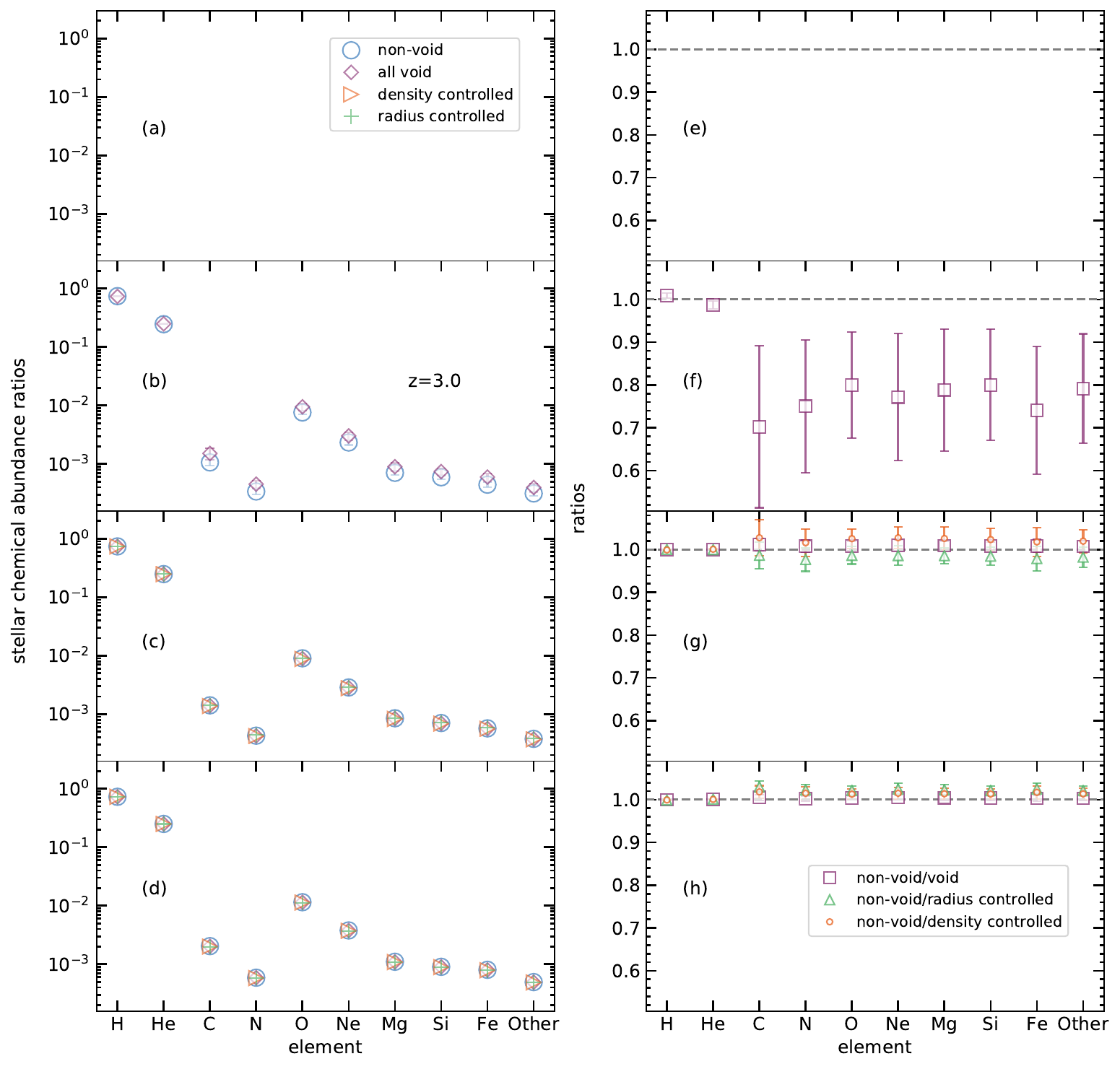}
\figsetgrpnote{Same as Figure~\ref{fig:ch5:gas_met_frac_ex_z02} except for galaxies at $z=3.0$.}
\figsetgrpend

\figsetgrpstart
\figsetgrpnum{figurenumber.12}
\figsetgrptitle{Gaseous chemical abundance ratios at $z=0.0$}
\figsetplot{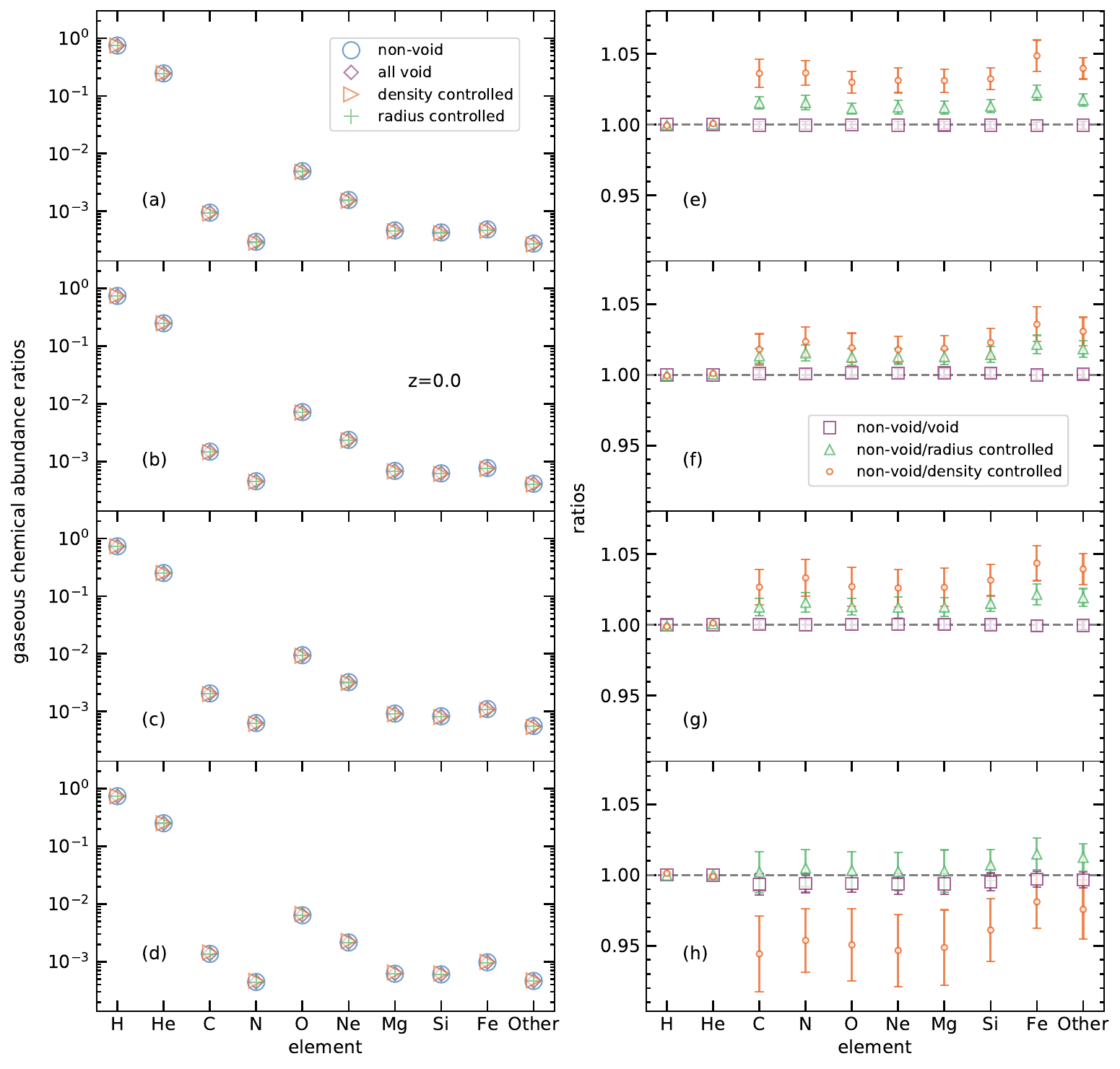}
\figsetgrpnote{Same as Figure~\ref{fig:ch5:gas_met_frac_ex_z02} except for gaseous chemical abundances ratios at $z=0.0$.}
\figsetgrpend

\figsetgrpstart
\figsetgrpnum{figurenumber.13}
\figsetgrptitle{Gaseous chemical abundance ratios at $z=0.1$}
\figsetplot{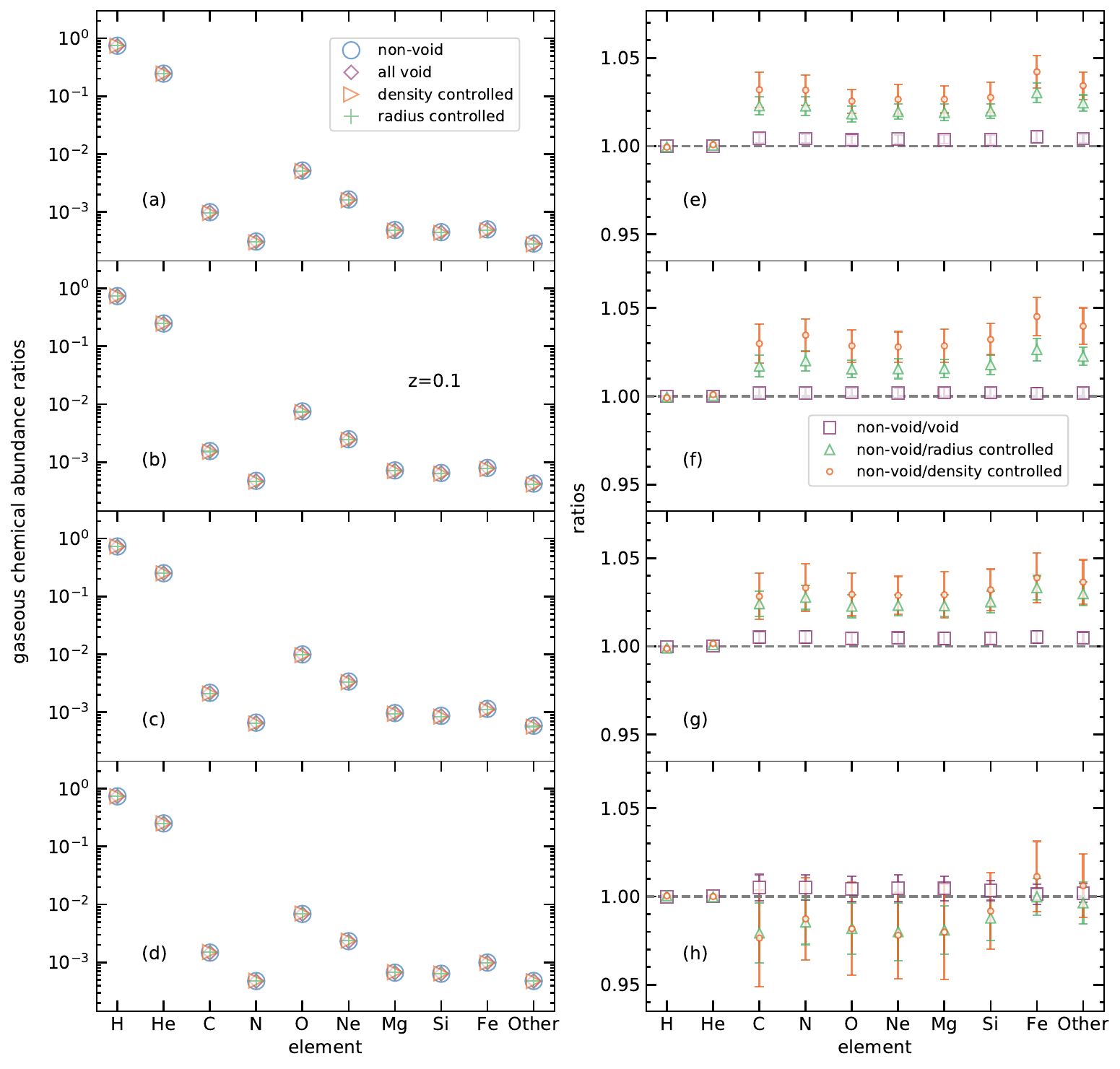}
\figsetgrpnote{Same as Figure~\ref{fig:ch5:gas_met_frac_ex_z02} except for gaseous chemical abundances ratios at $z=0.1$.}
\figsetgrpend

\figsetgrpstart
\figsetgrpnum{figurenumber.14}
\figsetgrptitle{Gaseous chemical abundance ratios at $z=0.2$}
\figsetplot{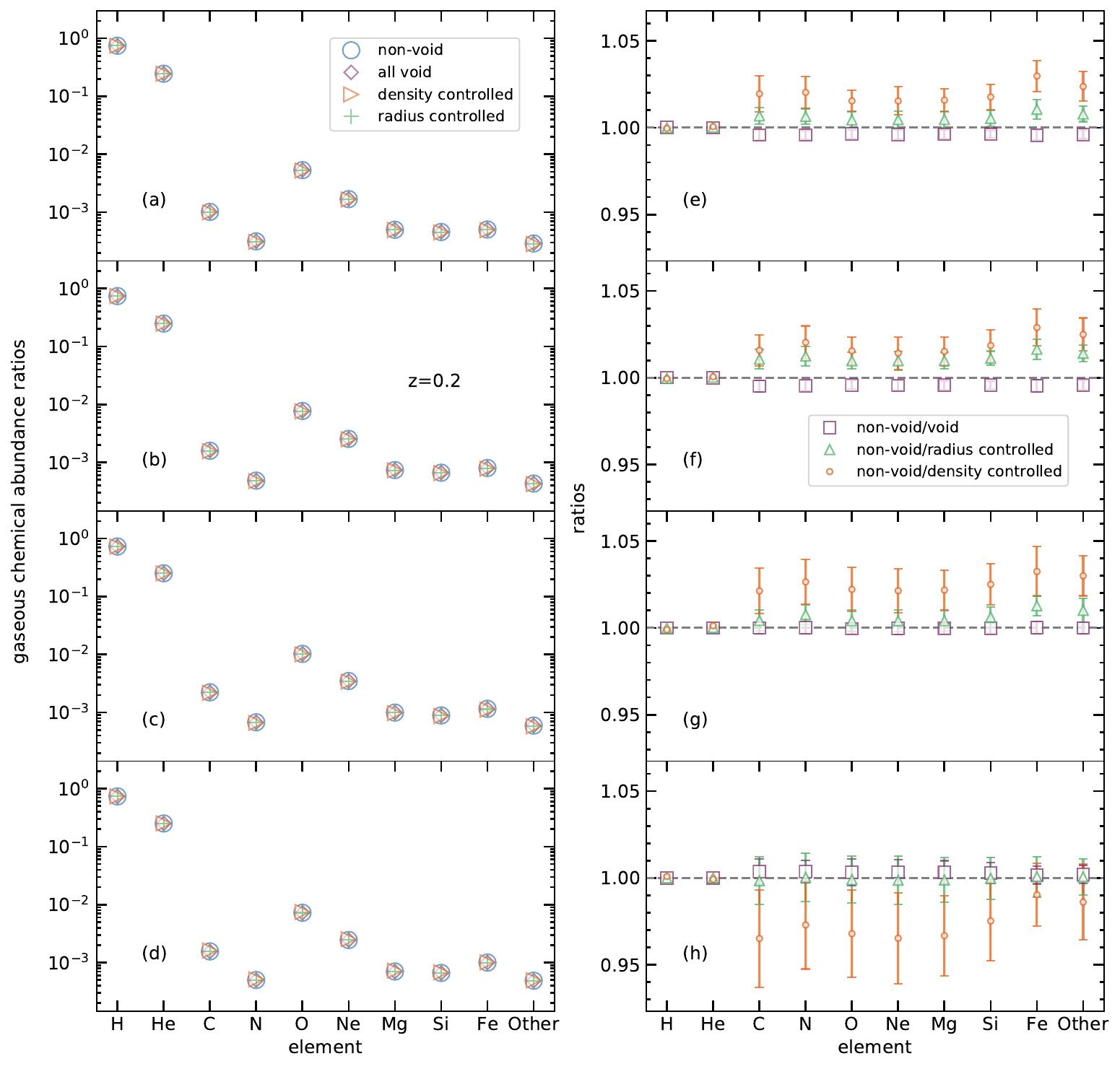}
\figsetgrpnote{Same as Figure~\ref{fig:ch5:gas_met_frac_ex_z02} except for gaseous chemical abundances ratios at $z=0.2$.}
\figsetgrpend

\figsetgrpstart
\figsetgrpnum{figurenumber.15}
\figsetgrptitle{Gaseous chemical abundance ratios at $z=0.3$}
\figsetplot{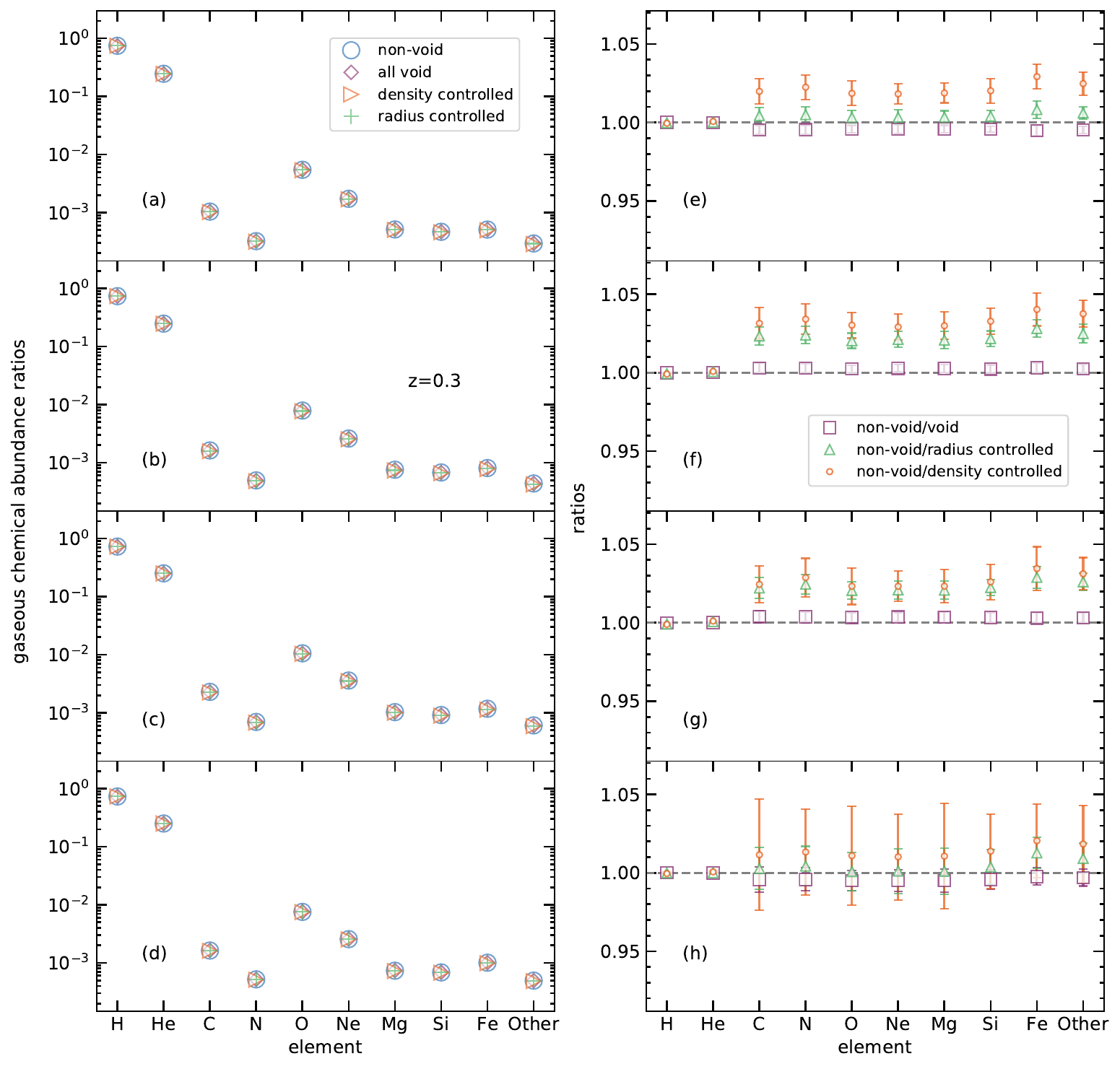}
\figsetgrpnote{Same as Figure~\ref{fig:ch5:gas_met_frac_ex_z02} except for gaseous chemical abundances ratios at $z=0.3$.}
\figsetgrpend

\figsetgrpstart
\figsetgrpnum{figurenumber.16}
\figsetgrptitle{Gaseous chemical abundance ratios at $z=0.4$}
\figsetplot{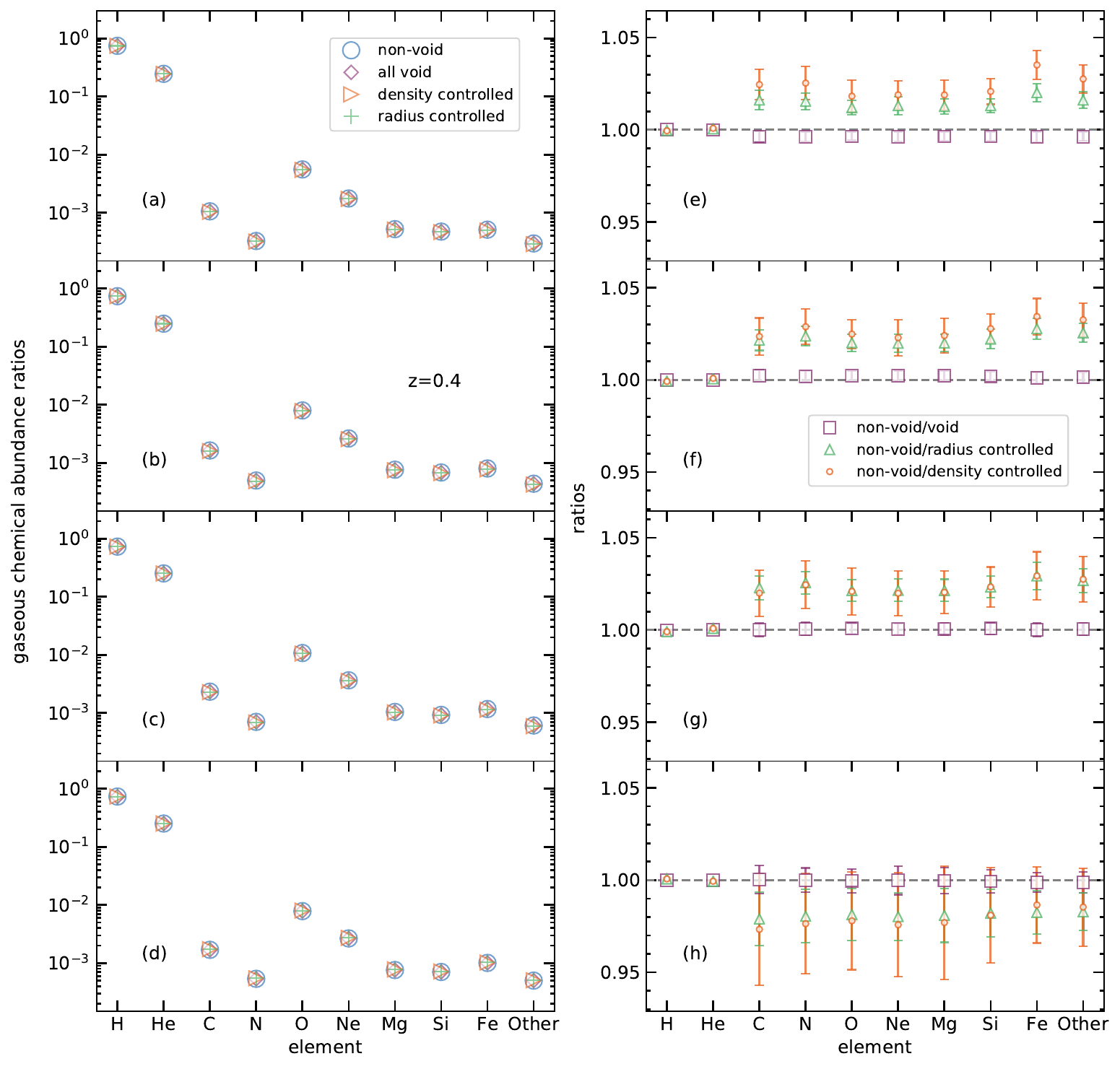}
\figsetgrpnote{Same as Figure~\ref{fig:ch5:gas_met_frac_ex_z02} except for gaseous chemical abundances ratios at $z=0.4$.}
\figsetgrpend

\figsetgrpstart
\figsetgrpnum{figurenumber.17}
\figsetgrptitle{Gaseous chemical abundance ratios at $z=0.5$}
\figsetplot{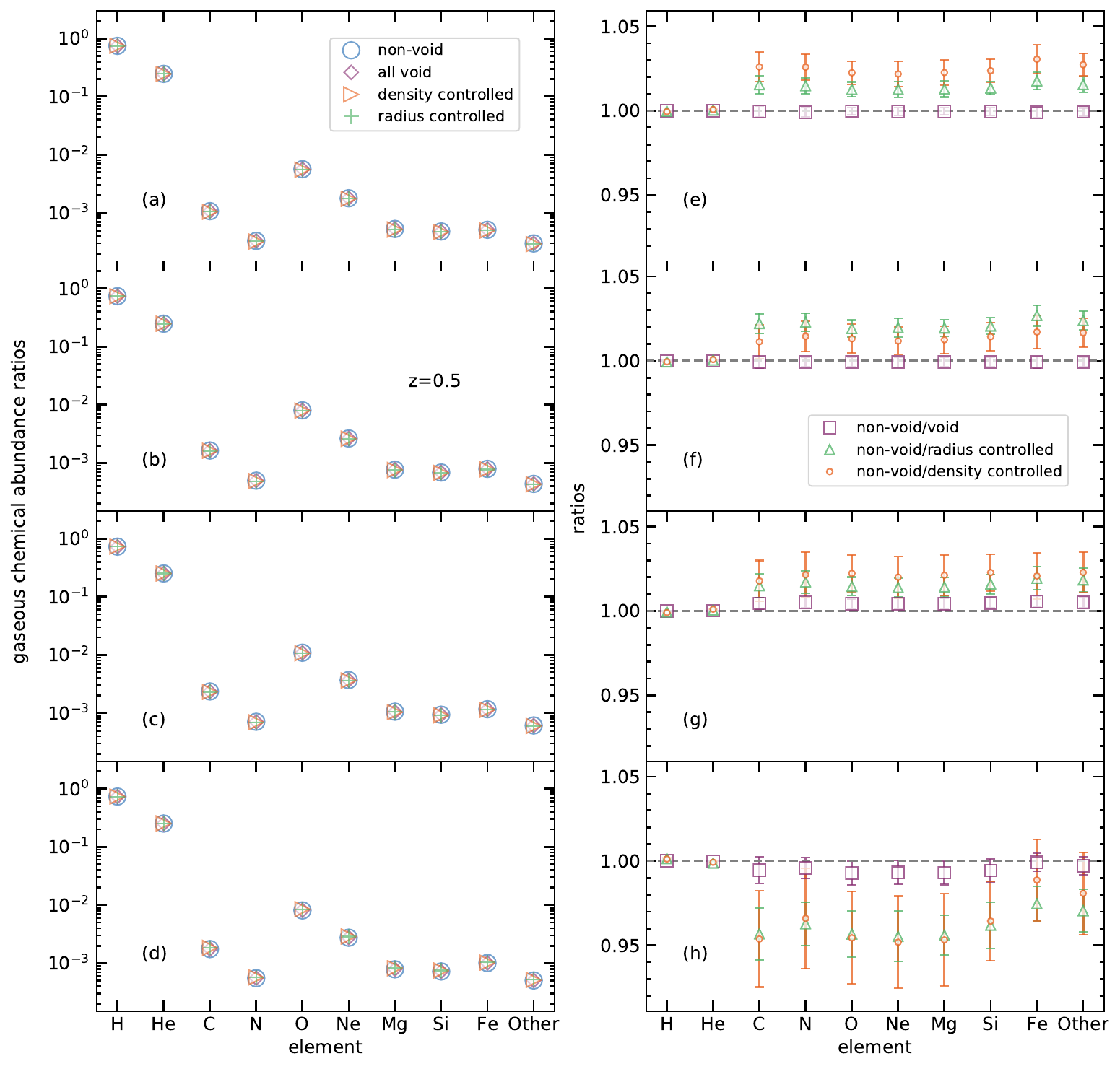}
\figsetgrpnote{Same as Figure~\ref{fig:ch5:gas_met_frac_ex_z02} except for gaseous chemical abundances ratios at $z=0.5$.}
\figsetgrpend

\figsetgrpstart
\figsetgrpnum{figurenumber.18}
\figsetgrptitle{Gaseous chemical abundance ratios at $z=0.7$}
\figsetplot{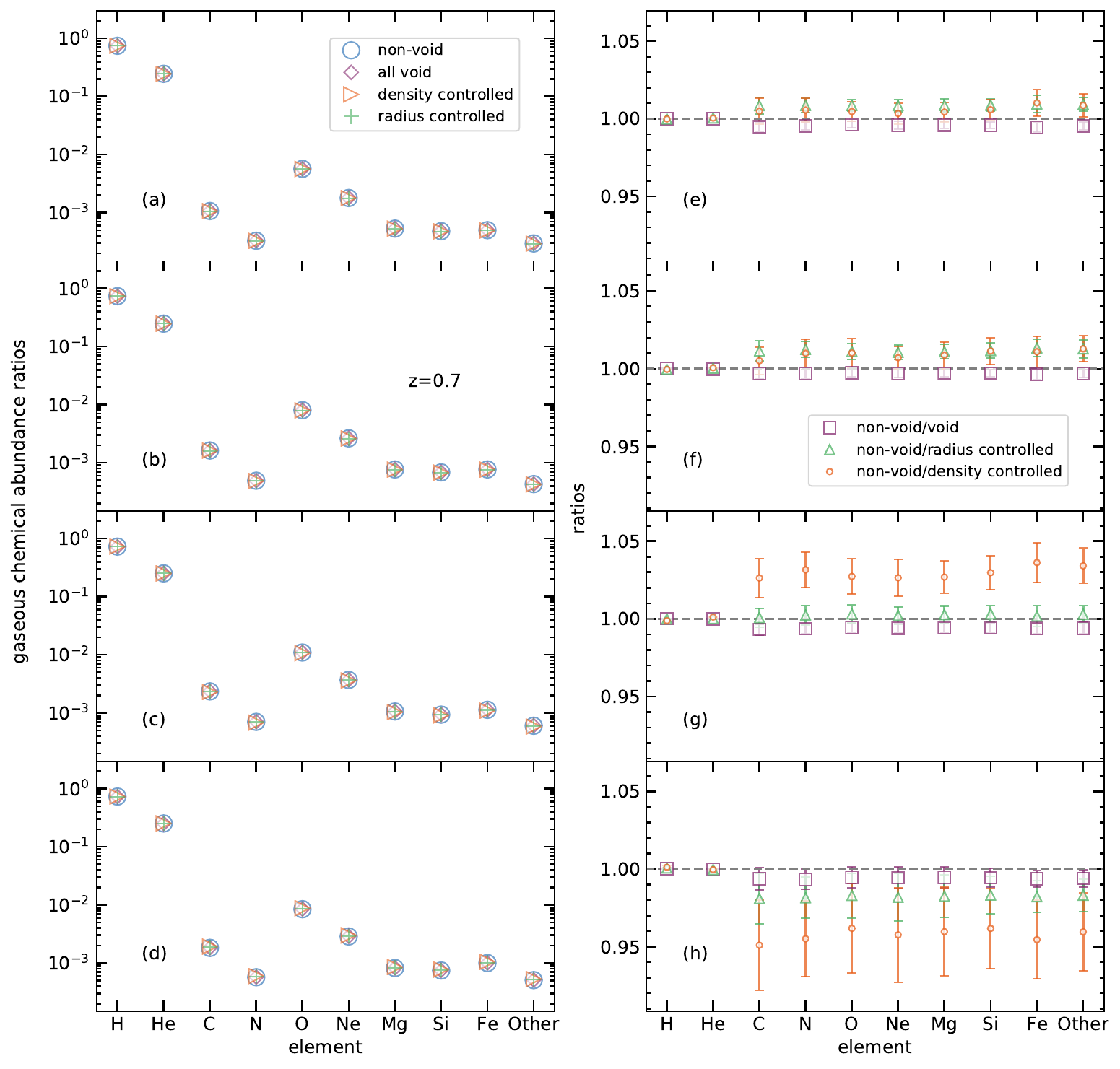}
\figsetgrpnote{Same as Figure~\ref{fig:ch5:gas_met_frac_ex_z02} except for gaseous chemical abundances ratios at $z=0.7$.}
\figsetgrpend

\figsetgrpstart
\figsetgrpnum{figurenumber.19}
\figsetgrptitle{Gaseous chemical abundance ratios at $z=1.0$}
\figsetplot{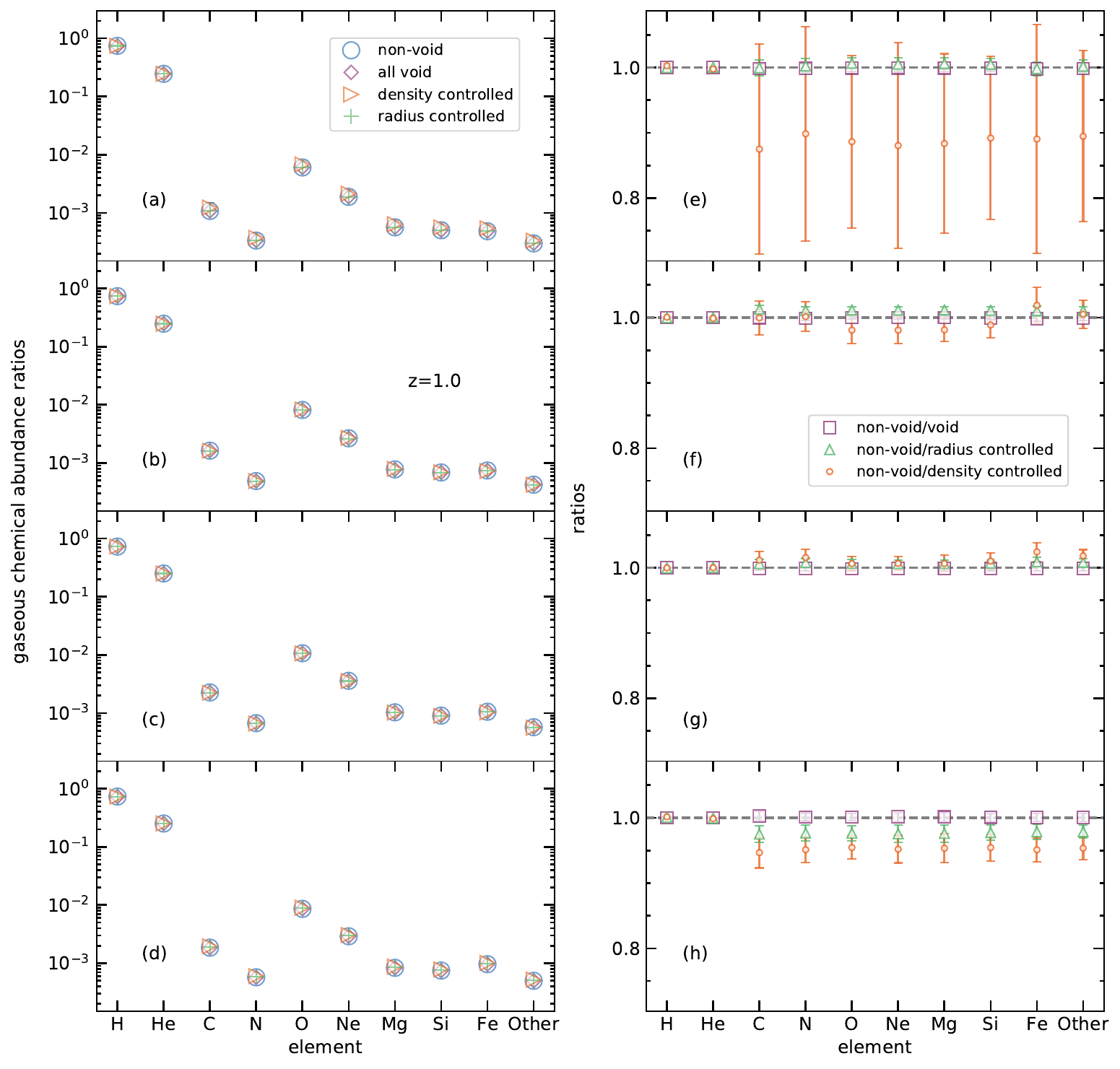}
\figsetgrpnote{Same as Figure~\ref{fig:ch5:gas_met_frac_ex_z02} except for gaseous chemical abundances ratios at $z=1.0$.}
\figsetgrpend

\figsetgrpstart
\figsetgrpnum{figurenumber.20}
\figsetgrptitle{Gaseous chemical abundance ratios at $z=1.5$}
\figsetplot{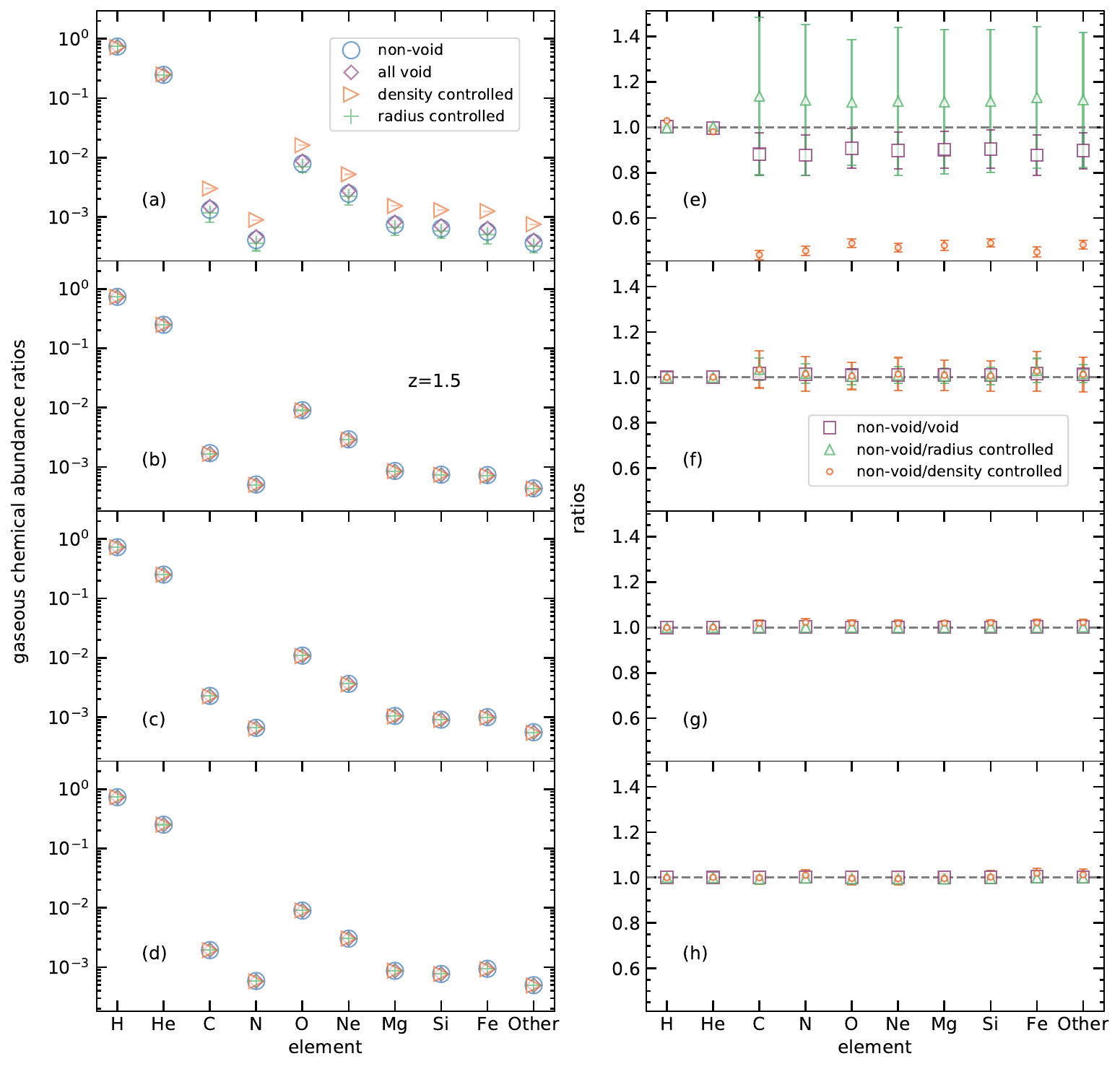}
\figsetgrpnote{Same as Figure~\ref{fig:ch5:gas_met_frac_ex_z02} except for gaseous chemical abundances ratios at $z=1.5$.}
\figsetgrpend

\figsetgrpstart
\figsetgrpnum{figurenumber.21}
\figsetgrptitle{Gaseous chemical abundance ratios at $z=3.0$}
\figsetplot{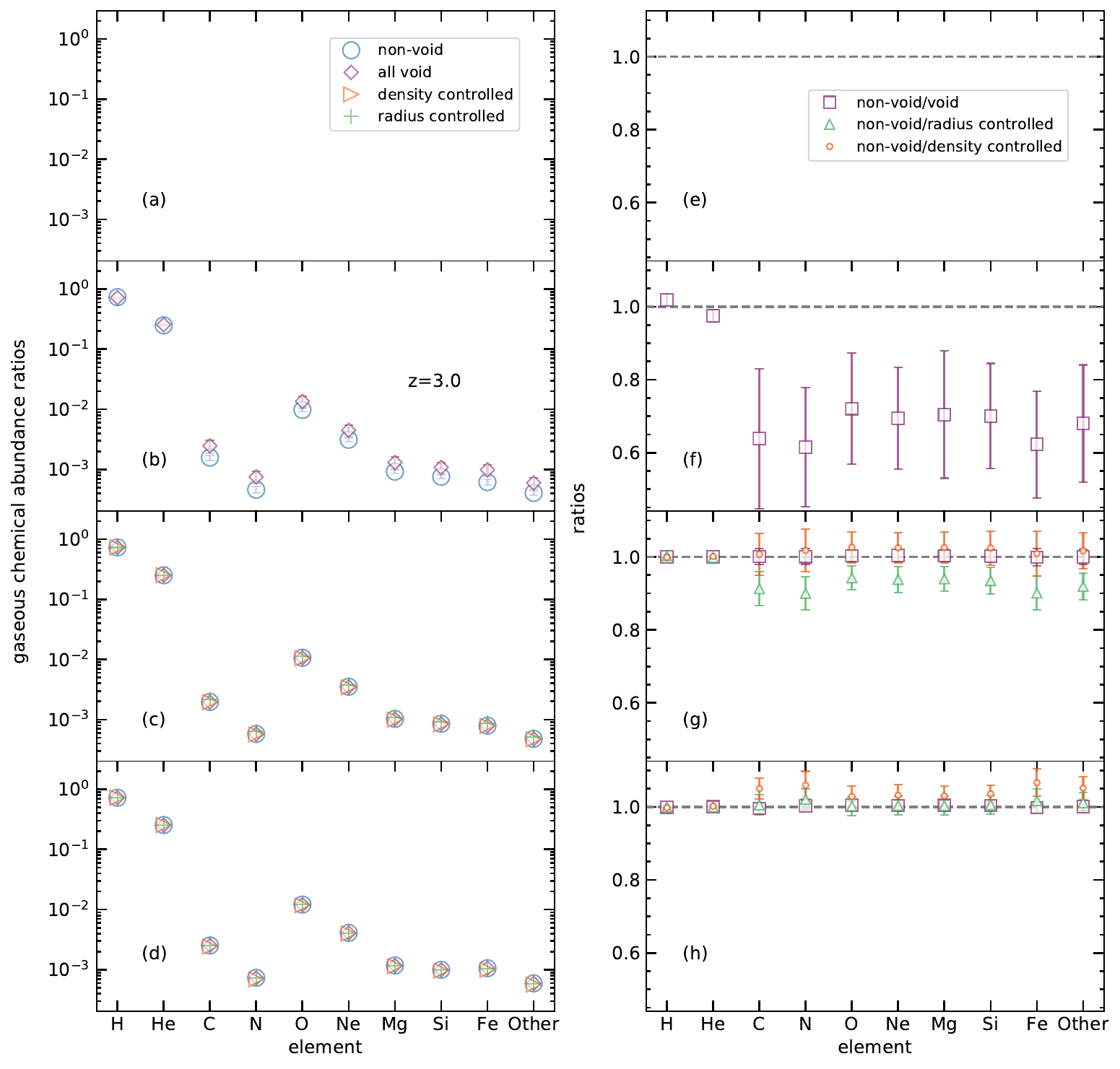}
\figsetgrpnote{Same as Figure~\ref{fig:ch5:gas_met_frac_ex_z02} except for gaseous chemical abundances ratios at $z=3.0$.}
\figsetgrpend

\figsetgrpstart
\figsetgrpnum{figurenumber.22}
\figsetgrptitle{Gaseous chemical abundance ratios at $z=3.0$}
\figsetplot{gas_metal_fracs_z30.pdf}
\figsetgrpnote{Same as Figure~\ref{fig:ch5:gas_met_frac_ex_z02} except for gaseous chemical abundances ratios at $z=3.0$.}
\figsetgrpend

\figsetend

\begin{deluxetable*}{c c c c | c c c | c c c | c c c}
\tabletypesize{\scriptsize}  
\tablecaption{Total number of active and inactive galaxies as well as the fraction of non-void (left), void (center {left}), and {radius-controlled} ({center }right){, and density-controlled (right)} galaxies that are in their active state for each redshift snapshot. \label{tab:ch5:agncounts}}
\tablehead{
 & \multicolumn{3}{c}{non-void galaxies} & \multicolumn{3}{c}{all void galaxies} & \multicolumn{3}{c}{radius-controlled galaxies} & \multicolumn{3}{c}{density-controlled galaxies} \\
 \colhead{z} & \colhead{$N_{\rm active}$} & \colhead{$N_{\rm inactive}$} & \colhead{fraction} 
 & \colhead{$N_{\rm active}$} & \colhead{$N_{\rm inactive}$} & \colhead{fraction} 
 & \colhead{$N_{\rm active}$} & \colhead{$N_{\rm inactive}$} & \colhead{fraction} 
 & \colhead{$N_{\rm active}$} & \colhead{$N_{\rm inactive}$} & \colhead{fraction} 
}
\startdata
0.0 & 7956 & 65777 & $0.108\pm0.001$ & 4637 & 39216 & $0.106\pm0.001$ & 912 & 7187 & $0.113\pm0.003$ & 1880 & 1637 & $0.535\pm0.008$ \\
0.1 & 9618 & 67248 & $0.125\pm0.001$ & 5087 & 36675 & $0.122\pm0.002$ & 988 & 6637 & $0.130\pm0.004$ & 1948 & 1670 & $0.538\pm0.008$ \\
0.2 & 10684 & 67929 & $0.136\pm0.001$ & 5447 & 34759 & $0.135\pm0.002$ & 1097 & 6582 & $0.143\pm0.004$ & 1797 & 1513 & $0.543\pm0.009$ \\
0.3 & 10959 & 63077 & $0.148\pm0.001$ & 5360 & 31060 & $0.147\pm0.002$ & 1106 & 6019 & $0.155\pm0.004$ & 1662 & 1387 & $0.545\pm0.009$ \\
0.4 & 13356 & 66658 & $0.167\pm0.001$ & 6091 & 30986 & $0.164\pm0.002$ & 1207 & 5633 & $0.176\pm0.005$ & 1717 & 1395 & $0.552\pm0.009$\\
0.5 & 14877 & 67277 & $0.181\pm0.001$ & 6002 & 27246 & $0.181\pm0.002$ & 1214 & 4930 & $0.198\pm0.005$ & 1557 & 1225 & $0.560\pm0.009$ \\
0.7 & 16676 & 58790 & $0.221\pm0.002$ & 8030 & 27410 & $0.227\pm0.002$ & 1629 & 5119 & $0.241\pm0.005$ & 1621 & 1201 & $0.574\pm0.009$ \\
1.0 & 6519 & 30937 & $0.174\pm0.002$ & 3976 & 18559 & $0.176\pm0.002$ & 829 & 3398 & $0.196\pm0.006$ & 1248 & 1002 & $0.555\pm0.010$\\
1.5 & 6609 & 19982 & $0.249\pm0.003$ & 2825 & 8335 & $0.253\pm0.004$ & 540 & 1508 & $0.264\pm0.010$ & 705 & 513 & $0.579\pm0.014$\\
2.0 & 5948 & 9123 & $0.395\pm0.004$ & 2210 & 3317 & $0.400\pm0.007$ & 400 & 601 & $0.400\pm0.015$ & 353 & 216 & $0.620\pm0.020$ \\
3.0 & 2325 & 1140 & $0.671\pm0.008$ & 848 & 377 & $0.692\pm0.013$ & 133 & 48 & $0.735\pm0.033$ & 129 & 33 & $0.796\pm0.032$ \\
\enddata
\end{deluxetable*}

In this section, we present the median stellar and gas chemical abundance ratios for various chemical elements in each galaxy population across cosmic time. To do this, we perform the same analyses as in \cite{curtis2024}. That is, we first divide each galaxy population into four mass bins, the boundaries of which are $M_*=10^{8.5}h^{-1}M_\odot$, $10^{9.0}h^{-1}M_\odot$, $10^{9.5}h^{-1}M_\odot$, $10^{10.0}h^{-1}M_\odot$, and $10^{10.5}h^{-1}M_\odot$. In each mass bin, we then determine the average stellar and gas chemical abundance ratios of stellar particles and gas cells in galaxies. {We refer to these mass groups as (a), (b), (c), and (d), respectively (see Figure Set~\ref{fig:ch5:gas_met_frac_ex_z02}).}

Figure Set~\ref{fig:ch5:gas_met_frac_ex_z02} shows an example {of the stellar chemical abundance ratios} from the $z=0.2$ snapshot. The left panels show the stellar chemical abundance ratios for the {non-void (circles), all void (diamonds), density-controlled (triangles), and interior void (crosses) galaxies.} The right panels show the ratios between the points in the corresponding panel to the left. As in \cite{curtis2024}, the mass bins increase from top to bottom such that the lowest stellar mass bin is shown in panels (a) and (e) while the highest stellar mass bin is shown in panels (d) and (h). 

In this example, we find that the radius-controlled galaxies in the first three stellar mass bins have stellar chemical abundance ratios that are lower than those of non-void galaxies by $\sim 1.5-2.5\pm0.7\%$, but in the fourth stellar mass bin, the radius-controlled galaxies have stellar chemical abundance ratios that are roughly equal to those of non-void galaxies. {The density-controlled galaxies show larger deviations from the non-void galaxies, being $\sim3.5-4.0\pm1\%$ metal enriched in mass group (a) and $\sim2.0-2.5\pm0.8\%$ less metal enriched in mass groups (b) and (c), and $\sim0.5-1.0\pm0.5\%$ less metal enriched in mass group (d). The complete sample of all void galaxies has gas and stellar chemical abundance ratios that are similar to those of the non-void galaxies. These results highlight the importance for controlling for local matter density when constructing void galaxy studies. As we restrict our void galaxy samples to regions with high galaxy number densities (all void galaxies), intermediate densities (radius-controlled galaxies), and very low densities (density-controlled galaxies; see Figure~\ref{fig:underdensities}), the increasing ratios in panels (e), (f), (g), and (h) of Figure~\ref{fig:ch5:gas_met_frac_ex_z02} indicate that the shell-crossing region plays an important role in regulating the evolution of baryonic material in and around voids.} 


{The online version of this article contains a complete Figure Set showing the stellar and gaseous chemical abundance ratios for all populations, mass bins, and redshift bins (22 images). In this Figure Set, data are omitted when there were no galaxies, stellar particles, or gas cells in a particular group. This primarily occurs at high redshifts and for density-controlled galaxies in mass groups (a) and (b), where there are few galaxies present in our samples.} For the most part, the density and radius-controlled galaxies in mass bins {(a), (b), and (c)} have stellar and gas chemical abundance ratios that are slightly lower than those of non-void galaxies. On the other hand, there are only marginal differences between the most massive galaxies in mass group (d), {suggesting that the most massive galaxies still have time to chemically enrich their gas regardless of their location with the LSS}. 

{The differences between the gaseous chemical abundance ratios are less distinct and exhibit larger spreads in their distributions.} For example, compared to the non-void galaxies at $z=0.4$, the N/H ratio in gas cells in the {density-controlled} galaxies decreases from $1.025\pm0.010$ in mass group (a) to $0.980\pm0.025$ in mass group (d). There are entries for which {lower mass groups} have slightly higher chemical abundance ratios than {more massive groups}, such as at $z=0.3$, which shows that, {compared to the gas cells in the non-void galaxies, the N/H ratio in gas cells within the density-controlled galaxies is $\sim2.0\pm1.0\%$ lower in mass group (a) and $\sim3.2\pm1\%$ lower in mass group (b).} While these ratios change across cosmic time, there does not appear to be any noticeable trend in the data as a function of redshift.


\section{Active Galactic Nuclei}
\label{sec:ch5:AGN}

\figsetstart
\figsetnum{13}
\figsettitle{Supermassive black hole mass vs. galaxy stellar mass relations}

\figsetgrpstart
\figsetgrpnum{figurenumber.1}
\figsetgrptitle{SMBH masses vs. galaxy stellar masses at $z=0.0$}
\figsetplot{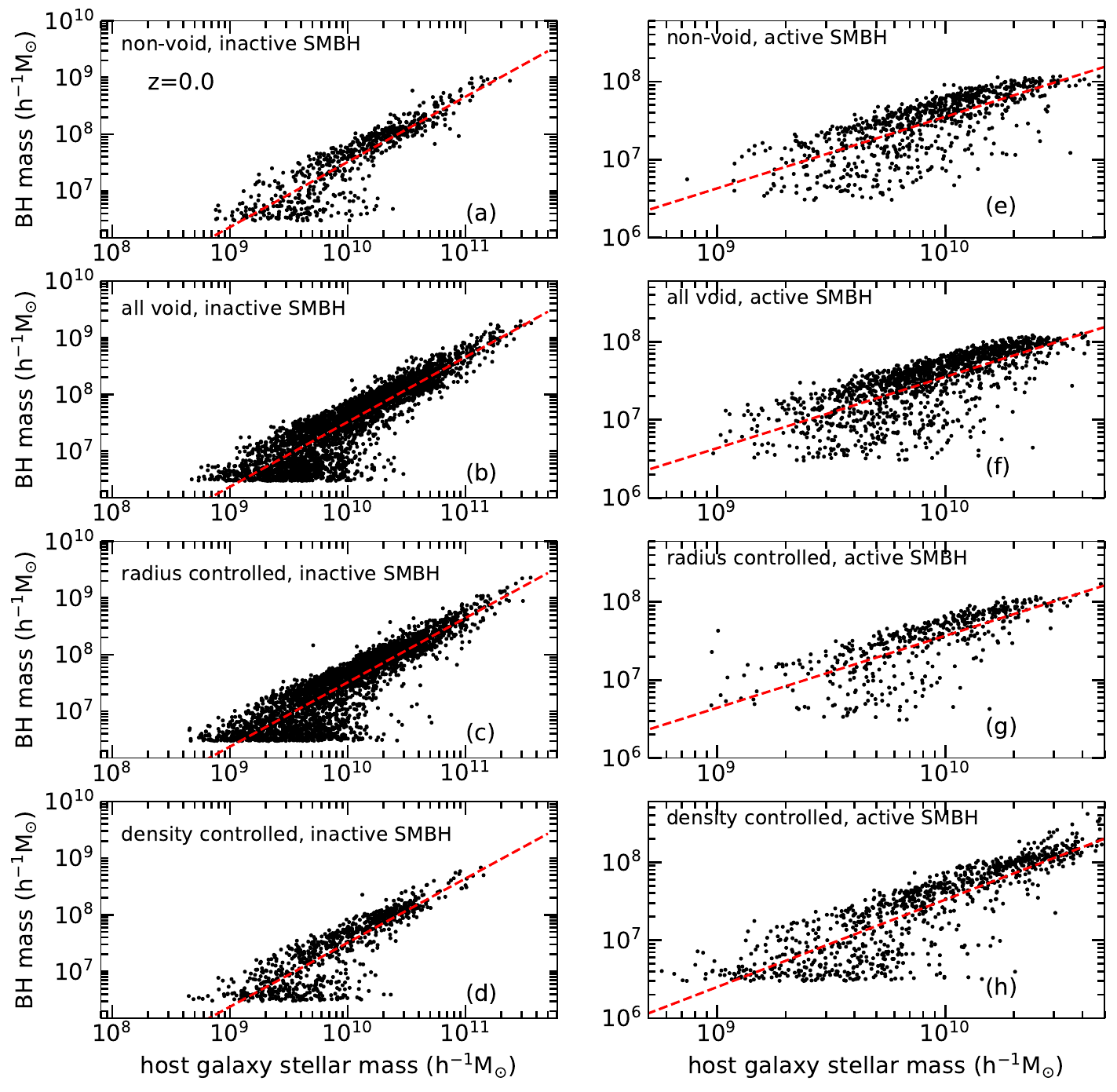}
\figsetgrpnote{Black points: Supermassive black hole (SMBH) mass vs. galaxy stellar mass in the $z=0.0$ snapshot of \texttt{TNG300}. {Top:} non-void galaxies. {Middle:} all void galaxies.  {Bottom:} radius-controlled galaxies.  {Left:} inactive galaxies {Right:} active galaxies.  Dashed red lines: Best-fitting relationship between SMBH mass and galaxy stellar mass. For clarity of the figure, $5-30\%$ of the data points were randomly selected and plotted.}
\figsetgrpend

\begin{figure*}[]
    \digitalasset
    \centering
    \includegraphics[width=0.7\textwidth]{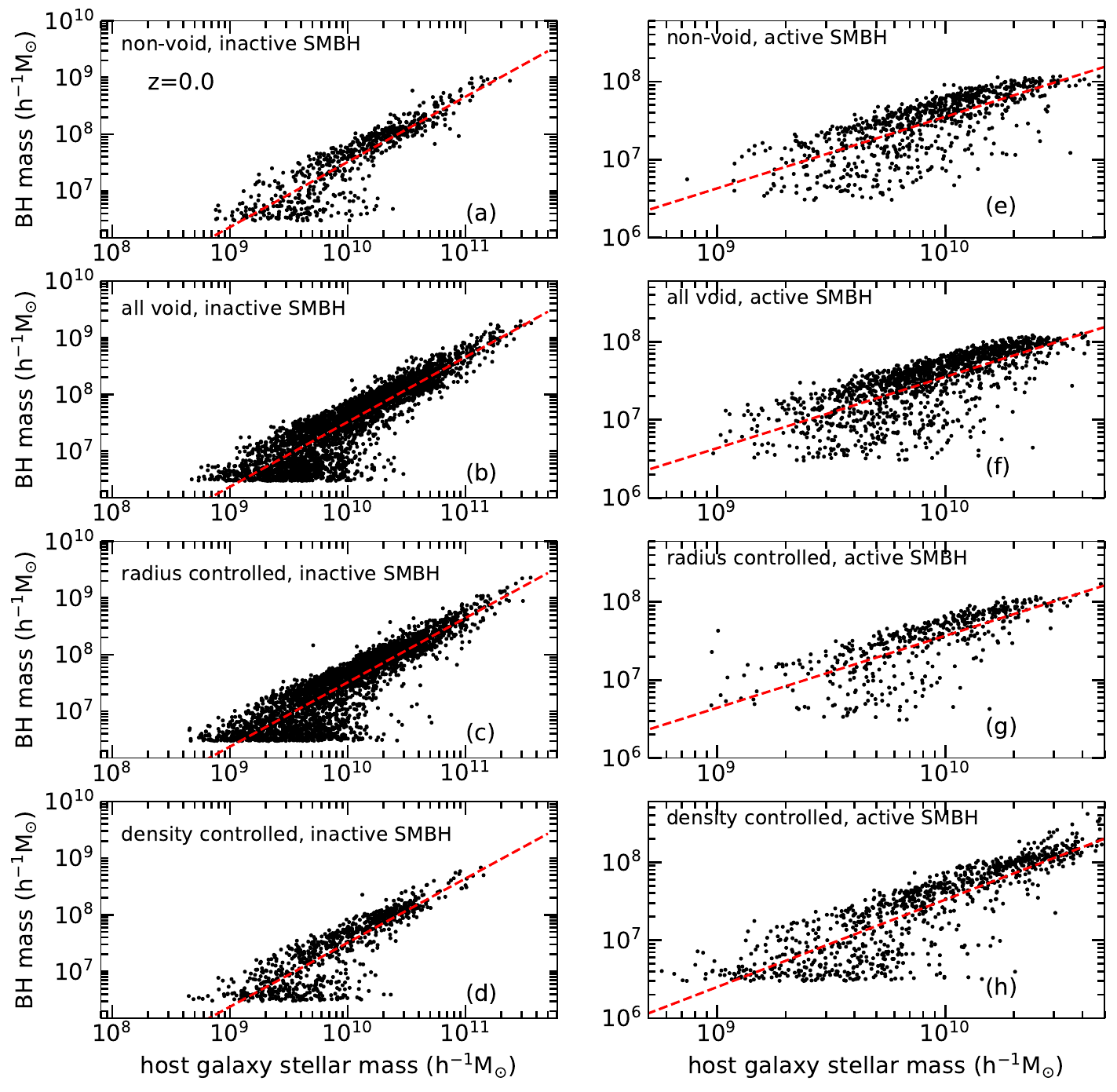}
    \caption{Black points: {Representative sample of} supermassive black hole (SMBH) mass vs. galaxy stellar mass in the $z=0.0$ snapshot of \texttt{TNG300}. {First row:} non-void galaxies. {Second row:} all void galaxies.  {Third row:} radius-controlled galaxies. {Fourth row:} {density-controlled galaxies.}  {Left:} inactive galaxies {Right:} active galaxies.  Dashed red lines: Best-fitting relationship between SMBH mass and galaxy stellar mass. For clarity of the figure, $5-30\%$ of the data points were randomly selected and plotted. {The complete Figure Set (11 images) is available in the online journal.}}
    \label{fig:ch5:z00MbhMs}
\end{figure*}

\figsetgrpstart
\figsetgrpnum{figurenumber.2}
\figsetgrptitle{SMBH masses vs. galaxy stellar masses at $z=0.1$}
\figsetplot{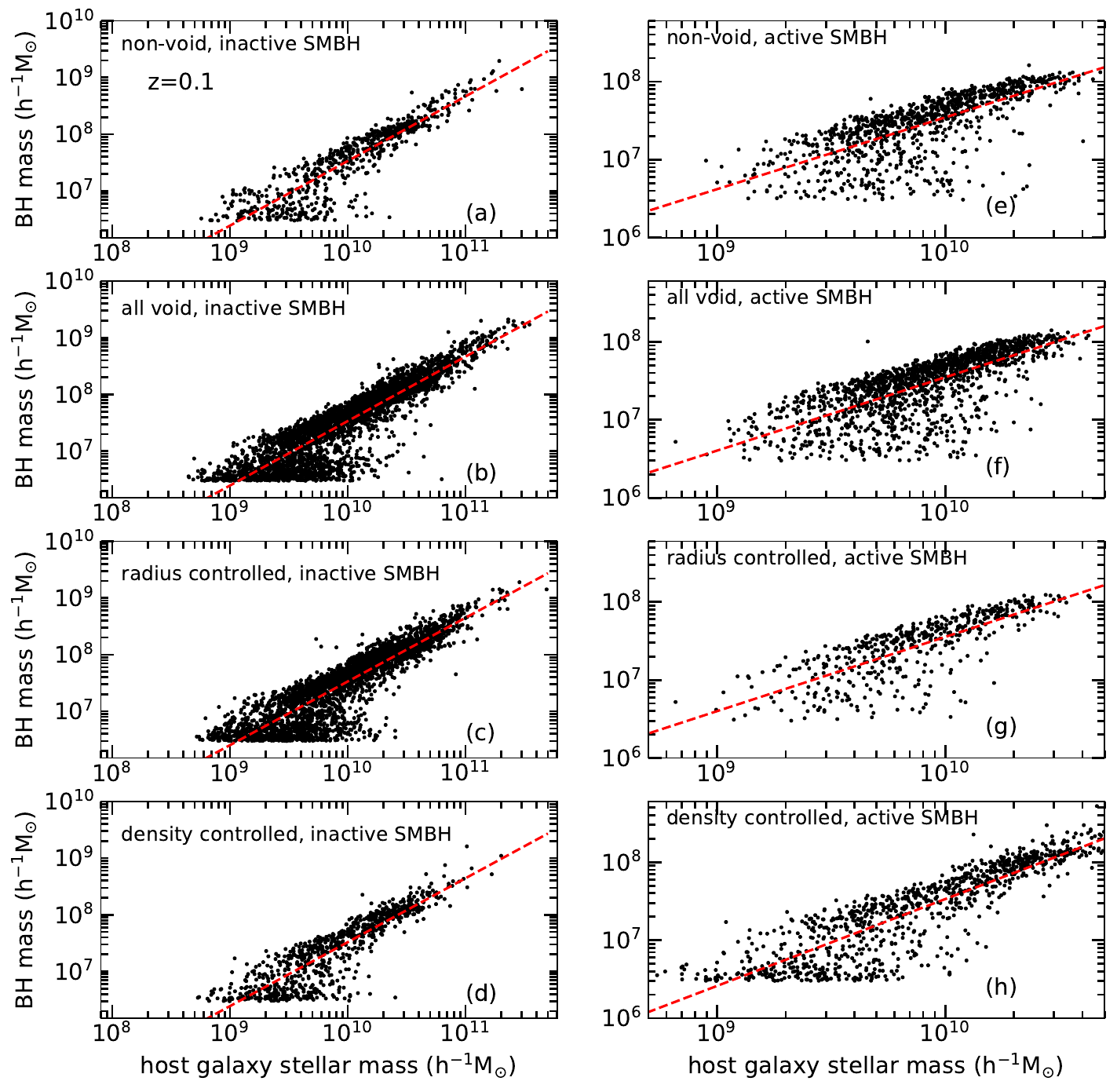}
\figsetgrpnote{Same as Figure~\ref{fig:ch5:z00MbhMs} except for galaxies at $z=0.1$.}
\figsetgrpend

\figsetgrpstart
\figsetgrpnum{figurenumber.3}
\figsetgrptitle{SMBH masses vs. galaxy stellar masses at $z=0.2$}
\figsetplot{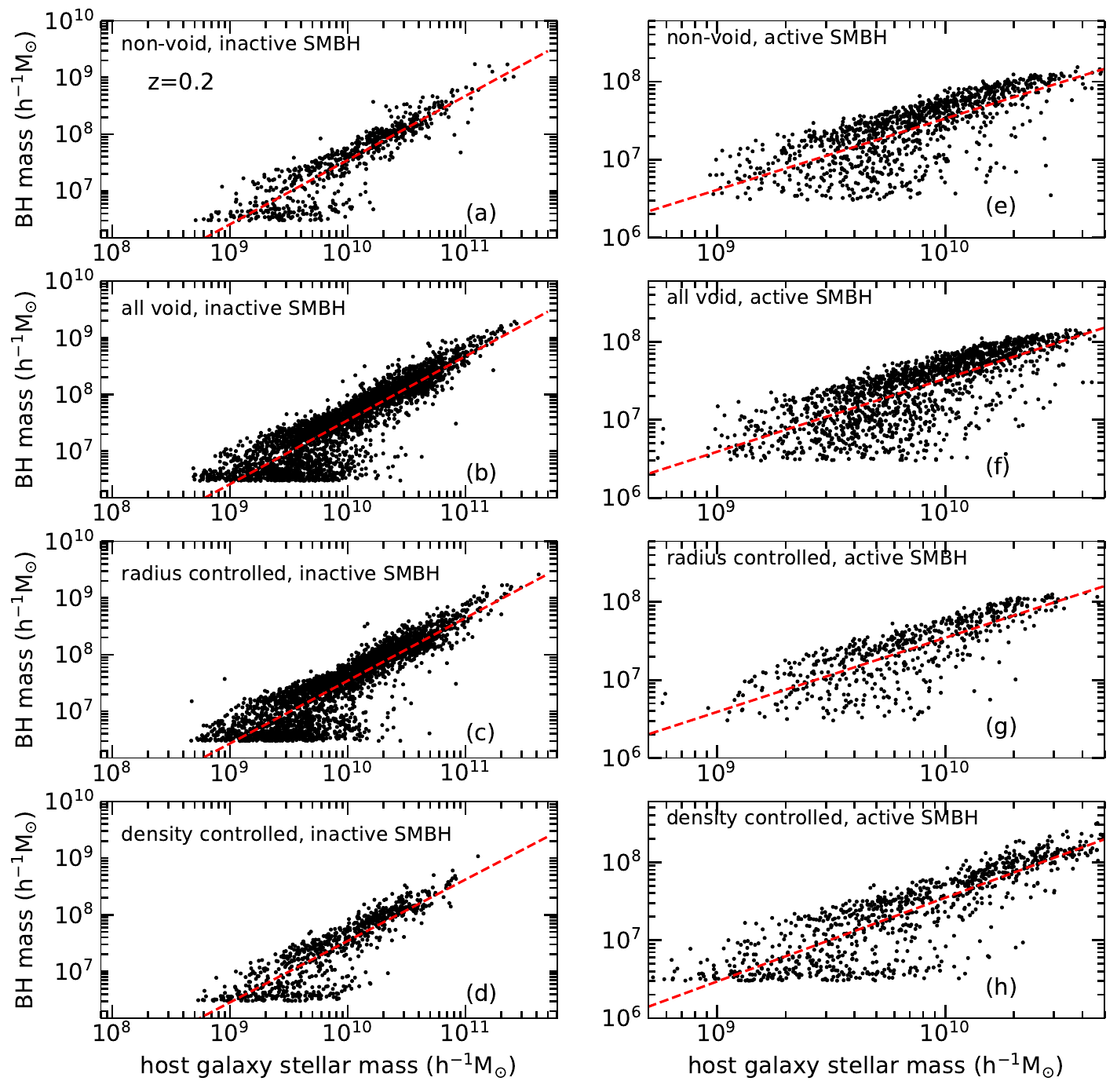}
\figsetgrpnote{Same as Figure~\ref{fig:ch5:z00MbhMs} except for galaxies at $z=0.2$.}
\figsetgrpend

\figsetgrpstart
\figsetgrpnum{figurenumber.4}
\figsetgrptitle{SMBH masses vs. galaxy stellar masses at $z=0.3$}
\figsetplot{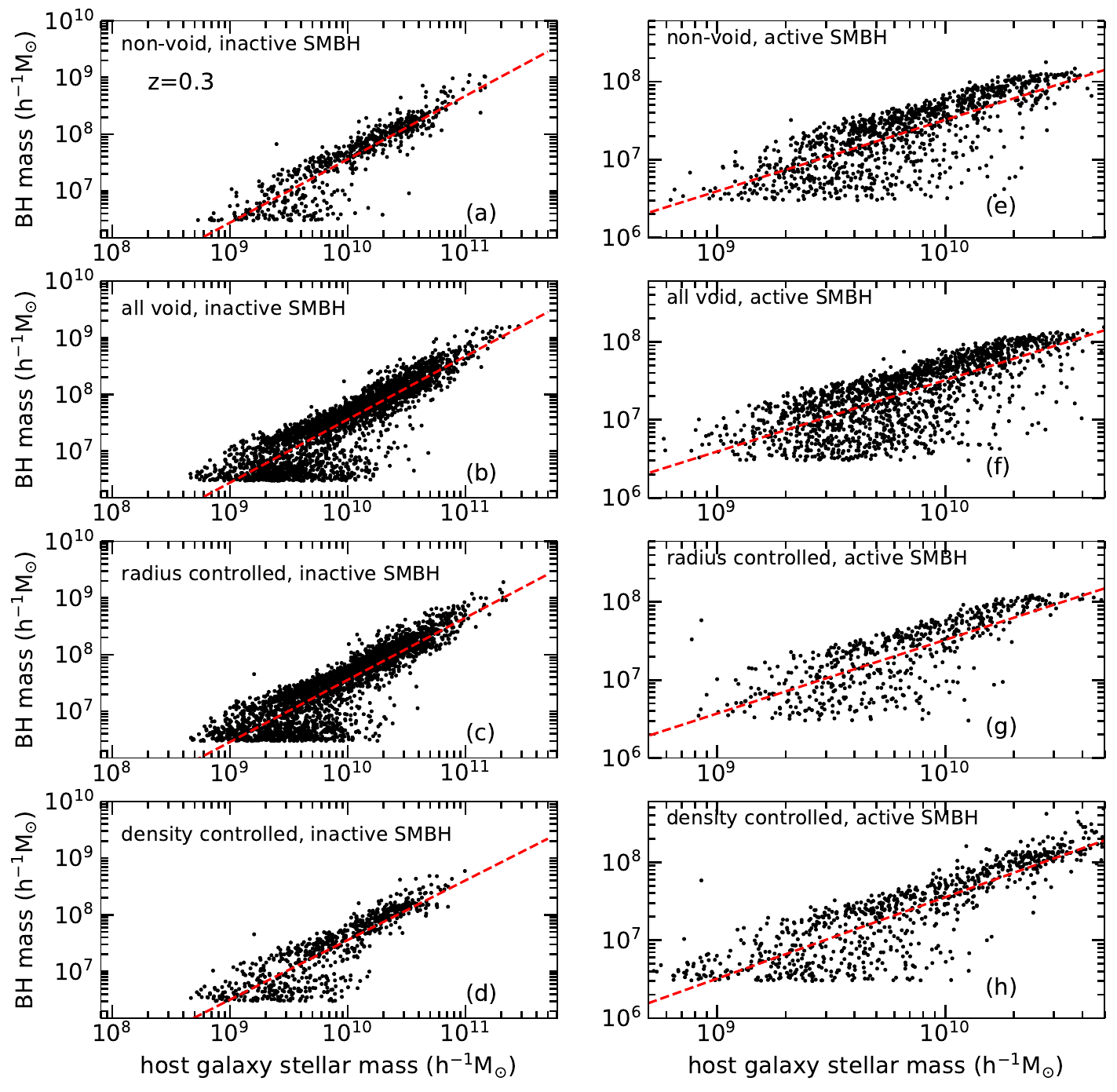}
\figsetgrpnote{Same as Figure~\ref{fig:ch5:z00MbhMs} except for galaxies at $z=0.3$.}
\figsetgrpend

\figsetgrpstart
\figsetgrpnum{figurenumber.5}
\figsetgrptitle{SMBH masses vs. galaxy stellar masses at $z=0.4$}
\figsetplot{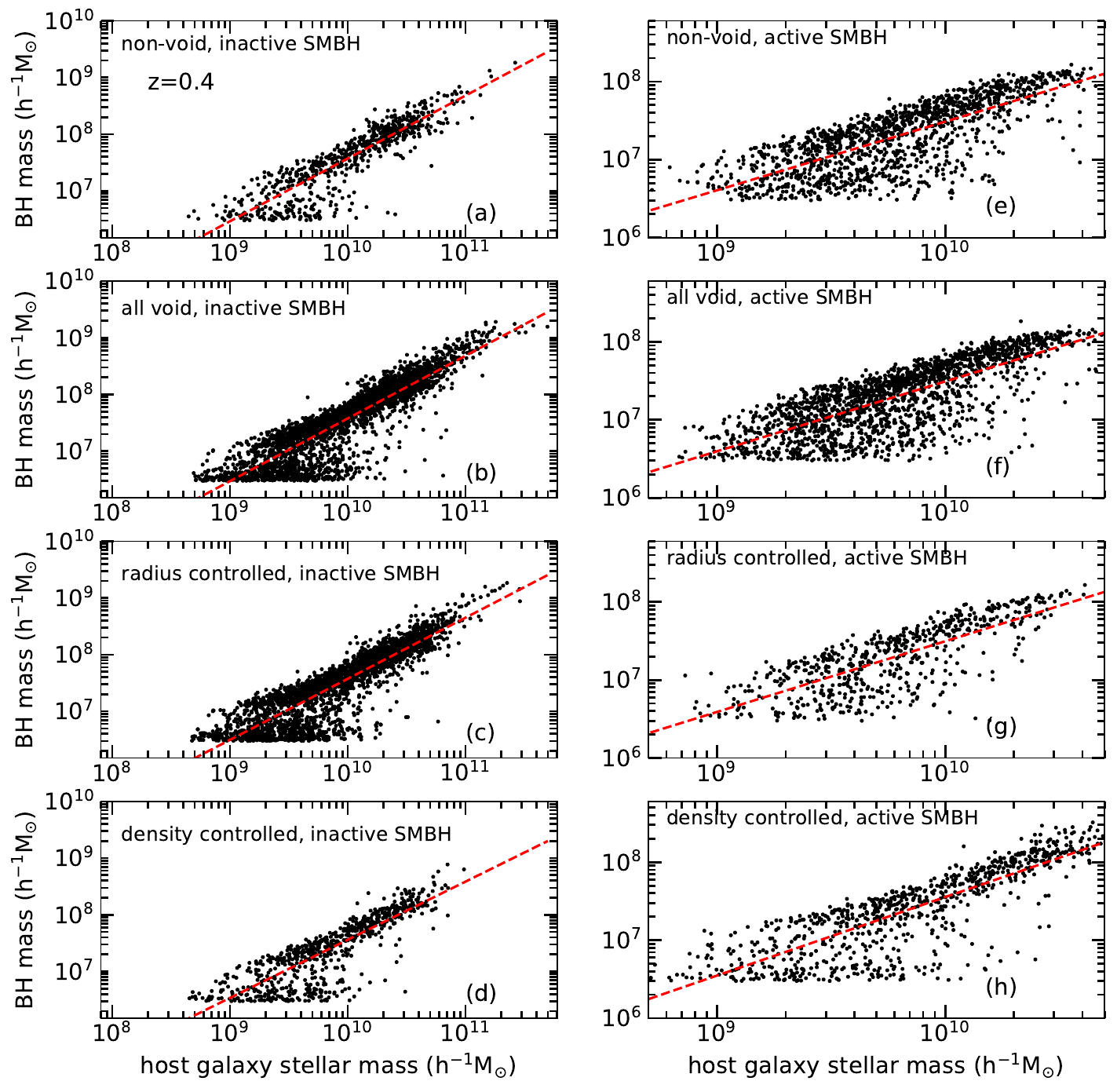}
\figsetgrpnote{Same as Figure~\ref{fig:ch5:z00MbhMs} except for galaxies at $z=0.4$.}
\figsetgrpend

\figsetgrpstart
\figsetgrpnum{figurenumber.6}
\figsetgrptitle{SMBH masses vs. galaxy stellar masses at $z=0.5$}
\figsetplot{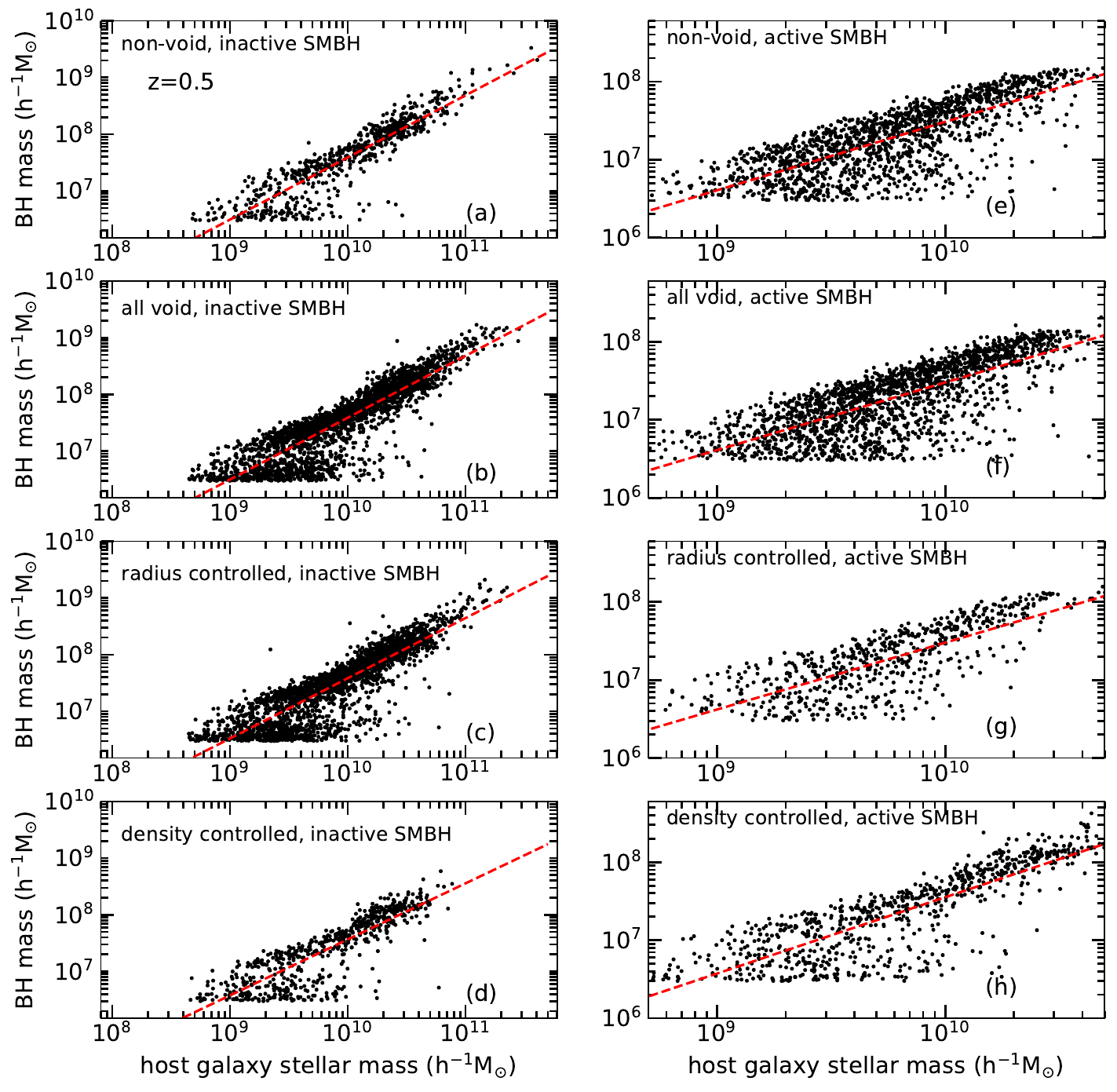}
\figsetgrpnote{Same as Figure~\ref{fig:ch5:z00MbhMs} except for galaxies at $z=0.5$.}
\figsetgrpend

\figsetgrpstart
\figsetgrpnum{figurenumber.7}
\figsetgrptitle{SMBH masses vs. galaxy stellar masses at $z=0.7$}
\figsetplot{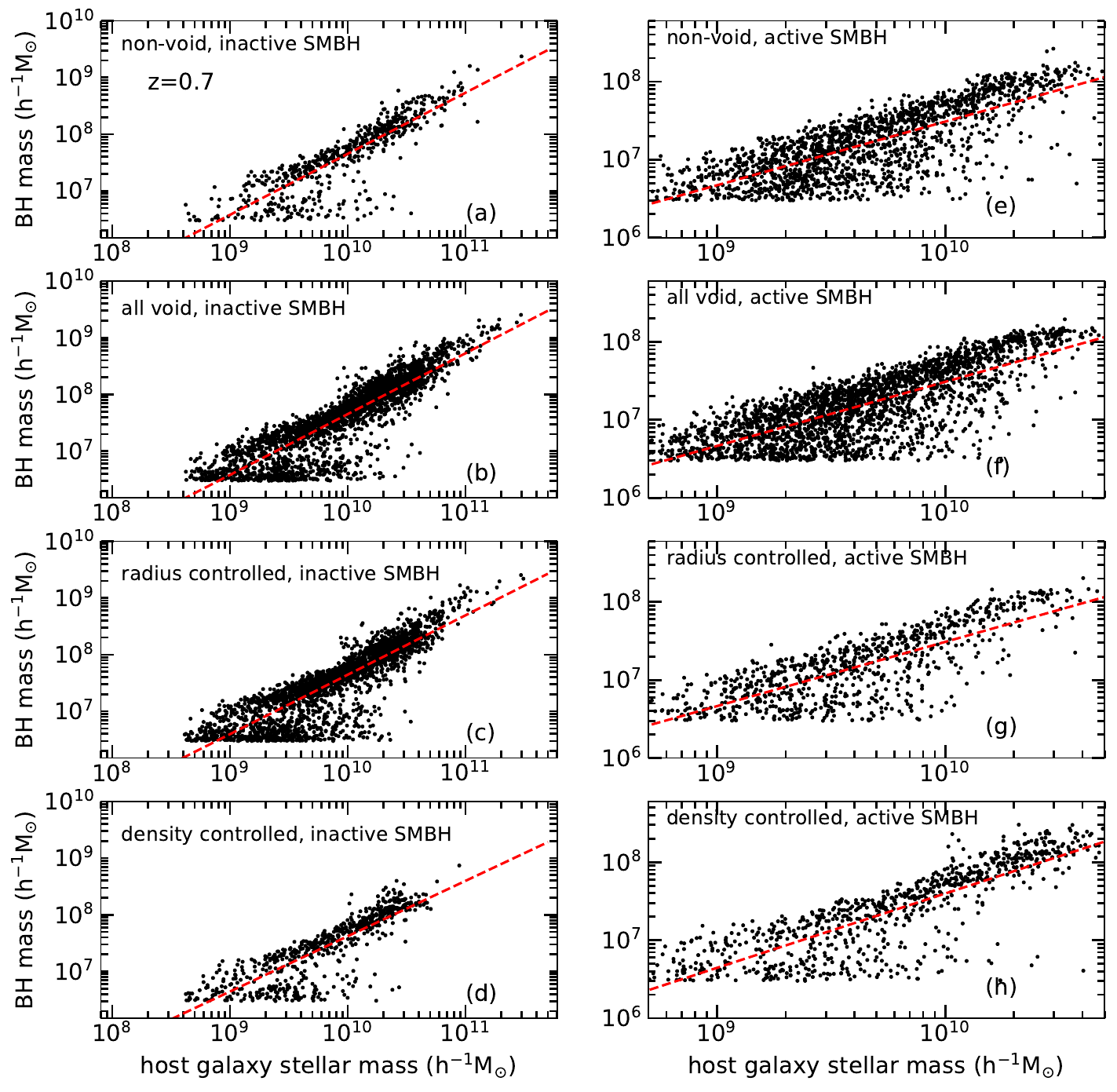}
\figsetgrpnote{Same as Figure~\ref{fig:ch5:z00MbhMs} except for galaxies at $z=0.7$.}
\figsetgrpend

\figsetgrpstart
\figsetgrpnum{figurenumber.8}
\figsetgrptitle{SMBH masses vs. galaxy stellar masses at $z=1.0$}
\figsetplot{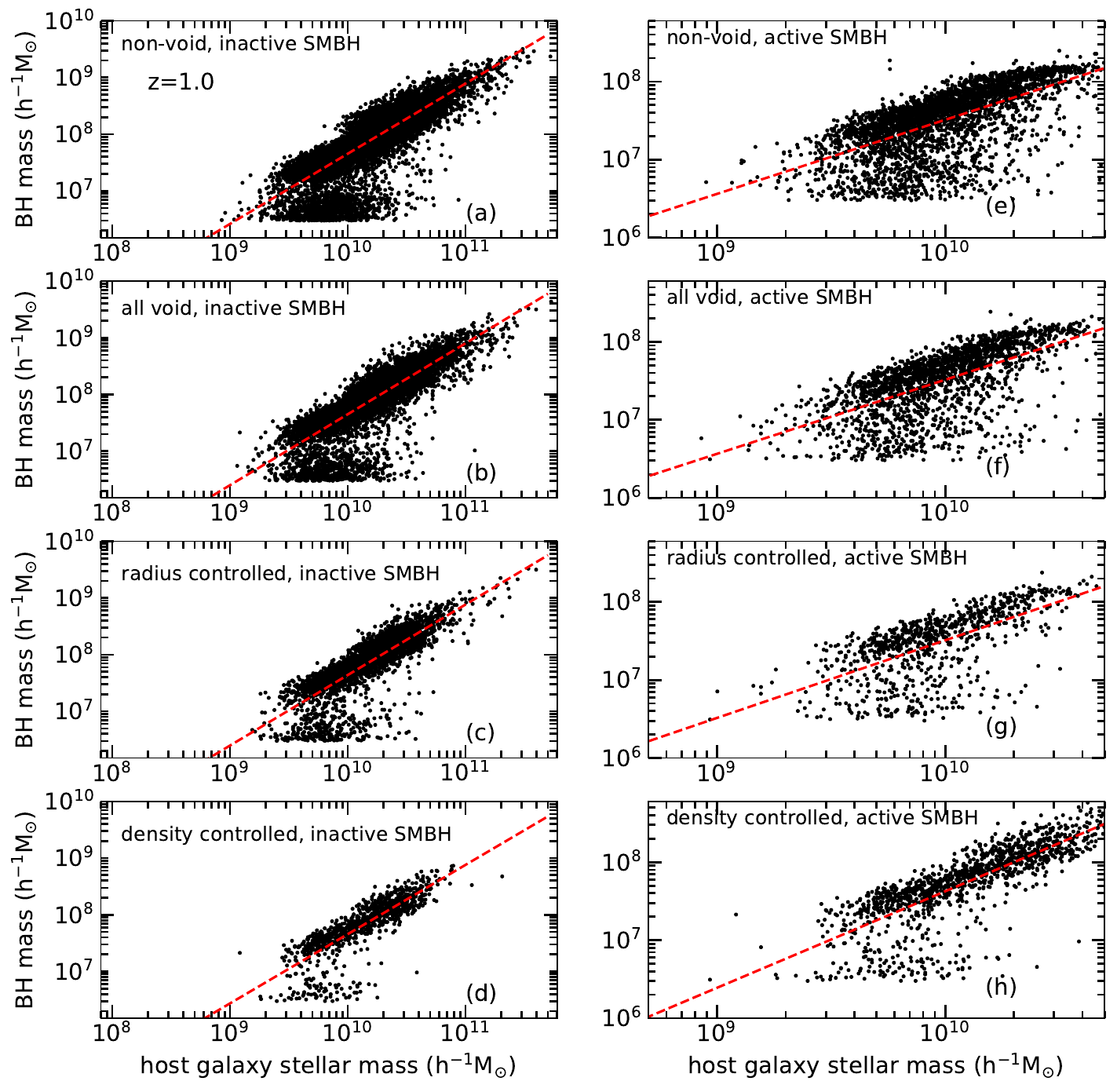}
\figsetgrpnote{Same as Figure~\ref{fig:ch5:z00MbhMs} except for galaxies at $z=1.0$.}
\figsetgrpend

\figsetgrpstart
\figsetgrpnum{figurenumber.9}
\figsetgrptitle{SMBH masses vs. galaxy stellar masses at $z=1.5$}
\figsetplot{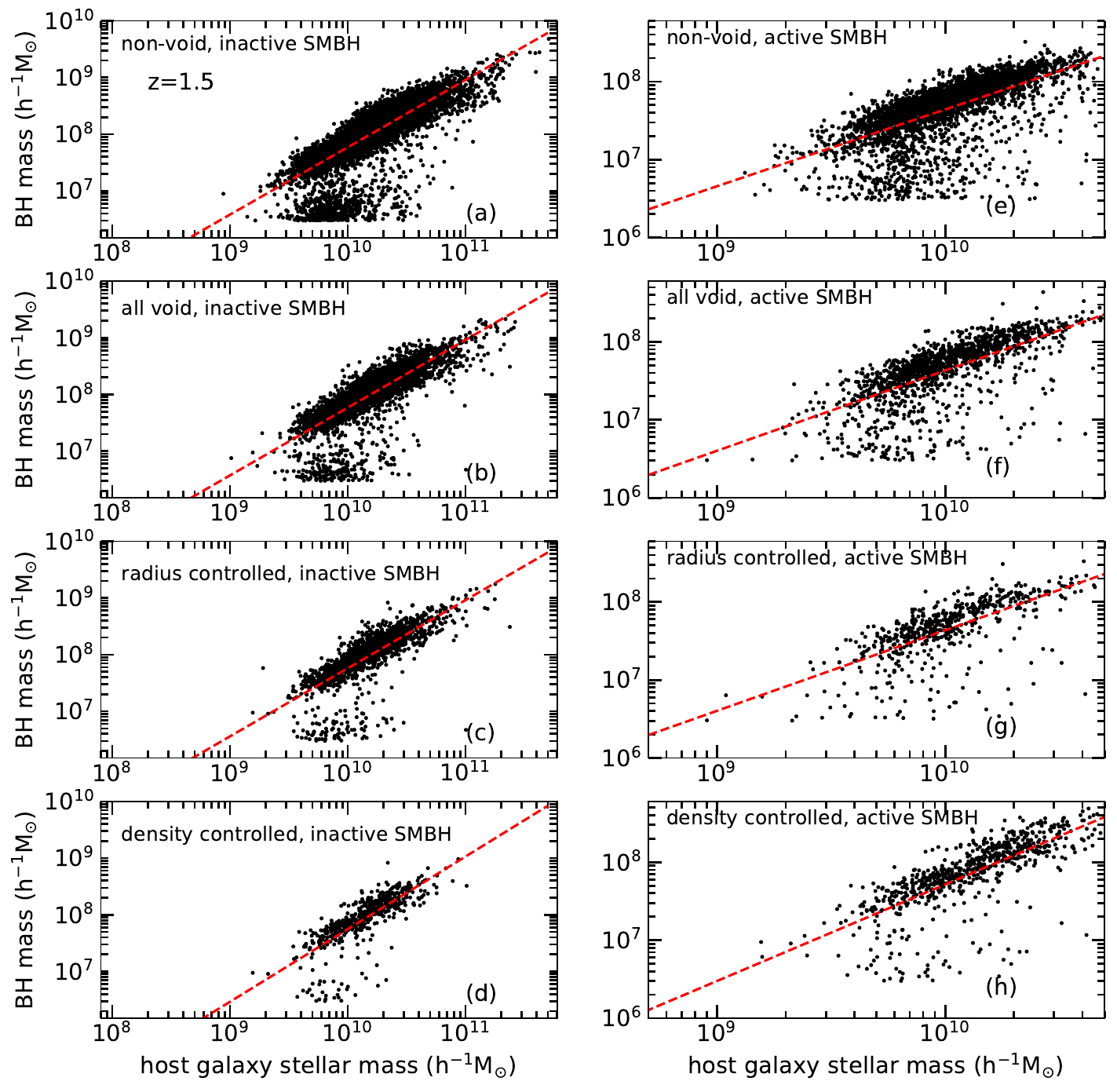}
\figsetgrpnote{Same as Figure~\ref{fig:ch5:z00MbhMs} except for galaxies at $z=1.5$.}
\figsetgrpend

\figsetgrpstart
\figsetgrpnum{figurenumber.10}
\figsetgrptitle{SMBH masses vs. galaxy stellar masses at $z=2.0$}
\figsetplot{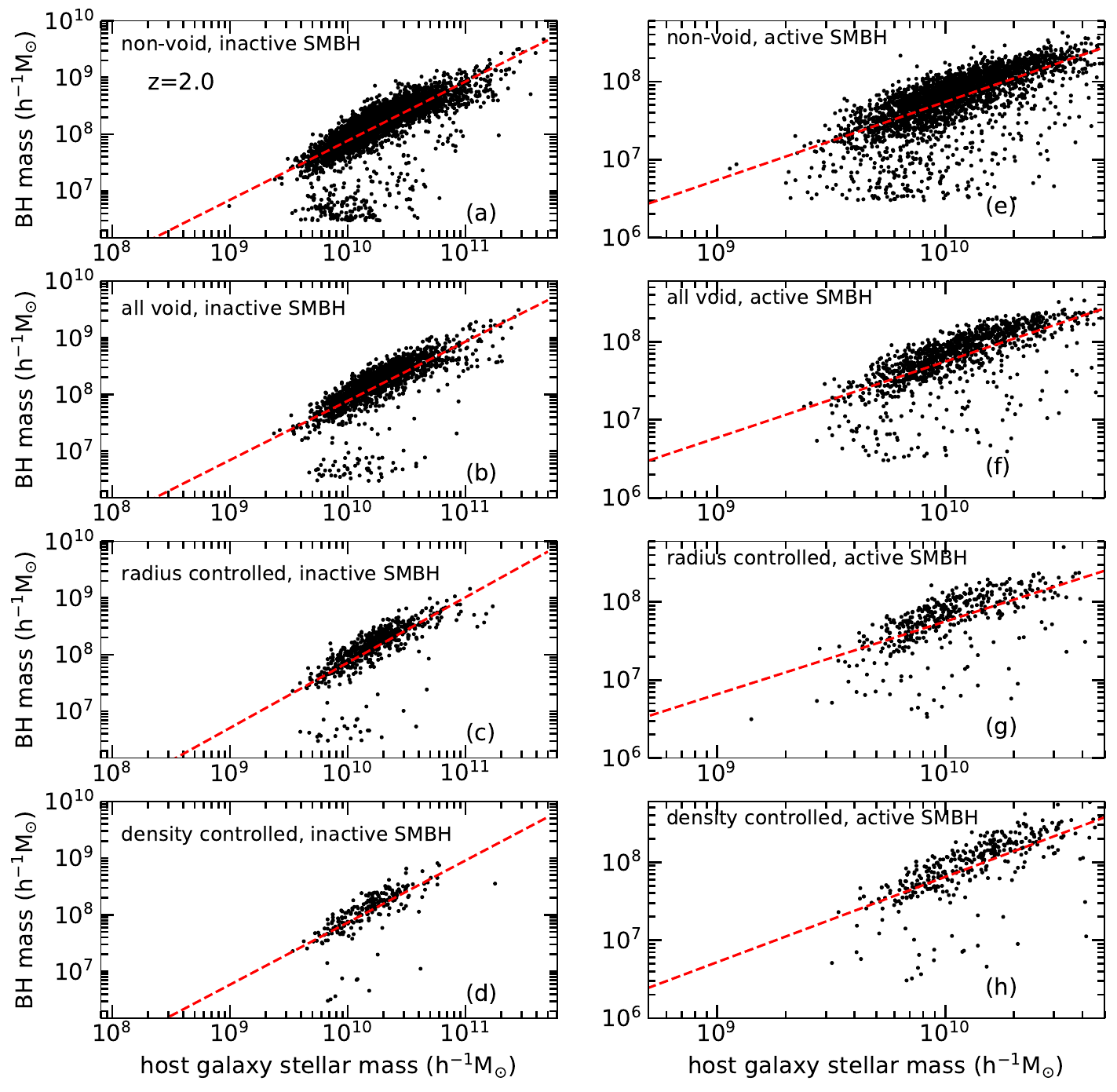}
\figsetgrpnote{Same as Figure~\ref{fig:ch5:z00MbhMs} except for galaxies at $z=2.0$.}
\figsetgrpend

\figsetgrpstart
\figsetgrpnum{figurenumber.11}
\figsetgrptitle{SMBH masses vs. galaxy stellar masses at $z=3.0$}
\figsetplot{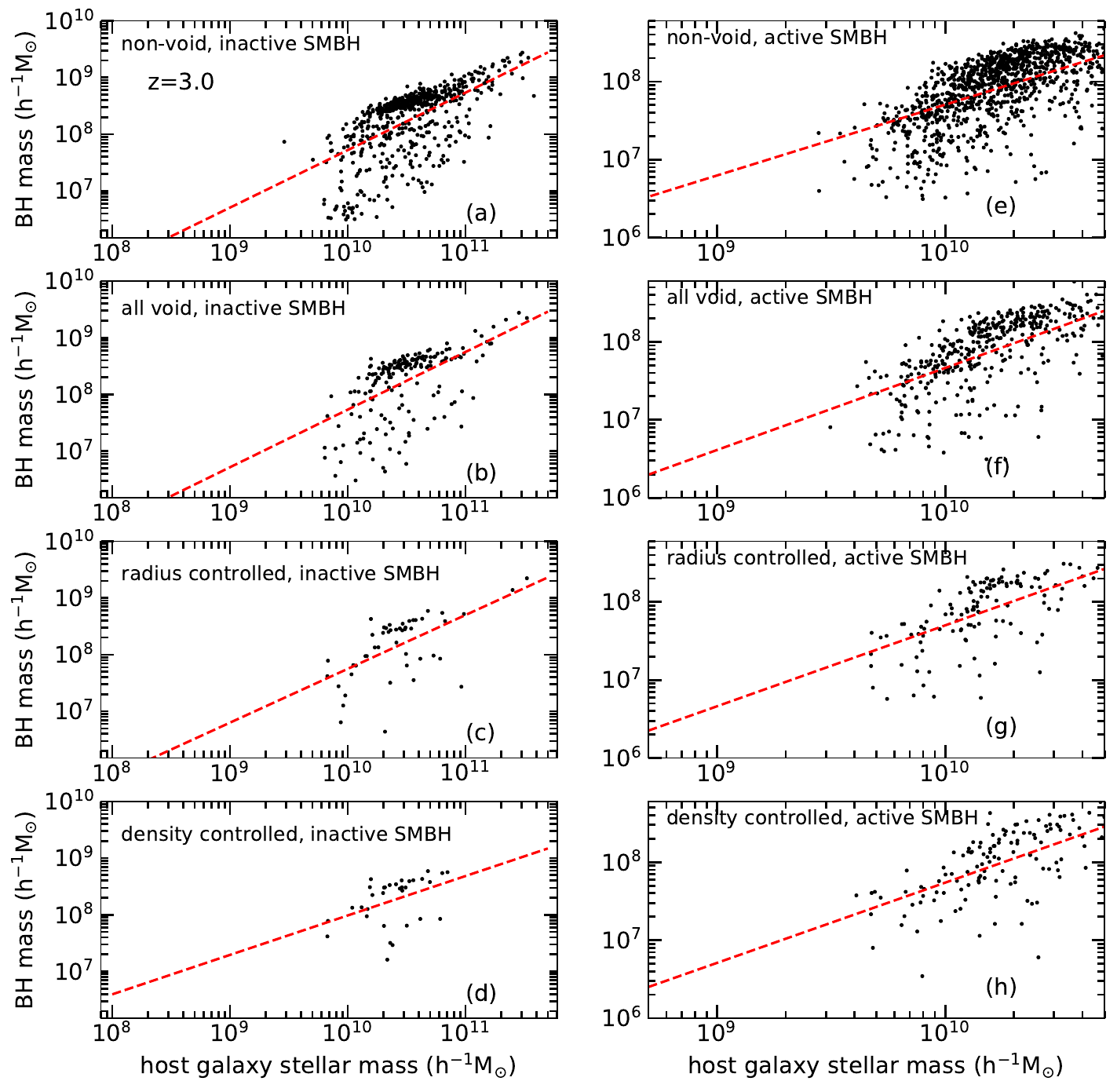}
\figsetgrpnote{Same as Figure~\ref{fig:ch5:z00MbhMs} except for galaxies at $z=3.0$.}
\figsetgrpend

\figsetend

In this Section, we investigate the abundance of AGN within our void and non-void galaxy catalogs. Here we define our AGN and inactive galaxy populations in the same manner as \cite{curtis2024}. That is, galaxies are considered to be in their ``active'' state if the ratio of Bondi accretion rate to Eddington accretion rate is $\geq 0.05$ \citep{weinberger2017}. Table~\ref{tab:ch5:agncounts} shows the total number of active and inactive galaxies in each galaxy population for each redshift. For all populations, the AGN fraction peaks at early redshifts. In fact, at $z=3.0$, there are more galaxies with AGN than without AGN in all {four} populations. These ratios decrease steadily over cosmic time. For example, $17-19\%$ of galaxies contain AGN at $z=1.0$, whereas only $\sim10-11\%$ of galaxies contain AGN at $z=0.0$. 

{From Table~\ref{tab:ch5:agncounts}}, it is clear that the AGN fractions for the {all void} and non-void galaxies are consistent with each other across cosmic time. However, for redshifts $z<1.0$, the radius-controlled galaxies have a somewhat larger AGN fraction than the other two populations, reaching a maximum of $24.1\pm0.5\%$ at $z=0.7$. {Inside the shell-crossing surface, a majority of the density-controlled galaxies are consistently in their active state, with AGN fractions of $53.5\pm0.8\%$ at redshift $z=0.0$, $57.4\pm0.9\%$ at $z=0.7$, and $79.6\pm3.2\%$ at $z=3.0$. This indicates a very strong correlation between AGN activity and location within underdense environments. Compared to the non-void galaxies, the AGN fractions of the density (radius) controlled galaxies are $79.8\pm0.4\%$ ($4.4\pm2.7\%$) higher at $z=0.0$, $61.5\pm0.7\%$ ($8.3\pm2.1\%$) higher at $z=0.7$, and $15.7\pm3.5\%$ ($8.7\pm4.2\%$) higher at $z=3.0$. That is, AGN fraction increases as we select for galaxies in the most underdense regions of voids that are far from the void ridges.} 


Next, we explore the $M_{*}-M_{\rm SMBH}$ relation for void and non-void galaxies. Figure {Set}~\ref{fig:ch5:z00MbhMs} show{s} the $M_{\rm{SMBH}}-M_*$ relationships for the $z=0.0$ to $z=3.0$ snapshots. The left panels of these figures show the relationships for galaxies with inactive SMBH, while the right panels show active SMBH. Results for non-void galaxies, all void galaxies, radius-controlled galaxies, {and density-controlled galaxies} are shown in the {first, second, third, and fourth} rows, respectively. Each panel shows a randomly selected sample of $5-50\%$ of the original data for the void and non-void populations, and $50-100\%$ of the data for the radius {and density} controlled galaxies. The red dashed lines in each panel indicate best-fitting lines corresponding to

\begin{equation}
    \log_{10}[h^{-1}M_{\rm{SMBH}}/M_\odot] = \alpha + \beta \times \log_{10}[h^{-1}M_{\rm{*}}/M_\odot] \: ,
    \label{eq:ch5:msmbhms}
\end{equation}

\noindent where $\alpha$ and $\beta$ were derived using a linear least squares fitting routine for each set of data. We list the slopes and y-intercepts for each fit ($\beta$ and $\alpha$ in Equation~\ref{eq:ch5:msmbhms}) {as well as the root-mean-square deviations between each fit and the corresponding in Tables~\ref{tab:ch5:agnslopes}}. {The online version of this article contains a complete Figure Set showing the $M_*-M_{\rm SMBH}$ relation for all populations and redshift bins (11 images).}

\begin{deluxetable*}{c c c | c c | c c | c c}
\tabletypesize{\scriptsize}
\tablecaption{{Slopes, intercepts, and root-mean square deviations of the best-fit lines to the $M_{\rm SMBH}-M_*$ planes of non-void, all void, radius-controlled void, and density-controlled void galaxies.}}
\label{tab:ch5:agnslopes}
\tablehead{
 & \multicolumn{2}{c}{non-void galaxies}
 & \multicolumn{2}{c}{all void galaxies}
 & \multicolumn{2}{c}{radius-controlled galaxies} 
 & \multicolumn{2}{c}{density-controlled galaxies} \\
 \colhead{$z$}
 & \colhead{$\beta_{\rm active}$} & \colhead{$\beta_{\rm inactive}$}
 & \colhead{$\beta_{\rm active}$} & \colhead{$\beta_{\rm inactive}$}
 & \colhead{$\beta_{\rm active}$} & \colhead{$\beta_{\rm inactive}$}
 & \colhead{$\beta_{\rm active}$} & \colhead{$\beta_{\rm inactive}$}
}
\startdata
0.0 & $0.917\pm0.010$ & $1.149\pm0.002$ & $0.913\pm0.013$ & $1.143\pm0.003$ & $0.923\pm0.028$ & $1.134\pm0.007$ & $1.119\pm0.015$ & $1.310\pm0.016$ \\
0.1 & $0.920\pm0.009$ & $1.140\pm0.002$ & $0.940\pm0.018$ & $1.137\pm0.003$ & $0.951\pm0.026$ & $1.123\pm0.007$ & $1.112\pm0.014$ & $1.127\pm0.015$ \\
0.2 & $0.914\pm0.008$ & $1.132\pm0.002$ & $0.936\pm0.011$ & $1.127\pm0.003$ & $0.950\pm0.022$ & $1.107\pm0.007$ & $1.073\pm0.015$ & $1.083\pm0.016$ \\
0.3 & $0.915\pm0.007$ & $1.118\pm0.002$ & $0.915\pm0.011$ & $1.115\pm0.003$ & $0.944\pm0.024$ & $1.096\pm0.070$ & $1.043\pm0.015$ & $1.052\pm0.016$ \\
0.4 & $0.879\pm0.007$ & $1.108\pm0.002$ & $0.894\pm0.010$ & $1.106\pm0.003$ & $0.906\pm0.218$ & $1.077\pm0.008$ & $1.006\pm0.015$ & $1.025\pm0.016$ \\
0.5 & $0.877\pm0.006$ & $1.095\pm0.002$ & $0.866\pm0.009$ & $1.090\pm0.003$ & $0.856\pm0.021$ & $1.062\pm0.008$ & $0.977\pm0.016$ & $0.988\pm0.017$ \\
0.7 & $0.815\pm0.005$ & $1.077\pm0.002$ & $0.821\pm0.008$ & $1.072\pm0.003$ & $0.823\pm0.165$ & $1.048\pm0.008$ & $0.952\pm0.016$ & $0.976\pm0.018$ \\
1.0 & $0.949\pm0.015$ & $1.238\pm0.005$ & $0.949\pm0.019$ & $1.251\pm0.006$ & $0.994\pm0.043$ & $1.243\pm0.016$ & $1.236\pm0.027$ & $1.221\pm0.028$ \\
1.5 & $0.983\pm0.014$ & $1.188\pm0.007$ & $1.029\pm0.022$ & $1.195\pm0.010$ & $1.032\pm0.050$ & $1.202\pm0.027$ & $1.237\pm0.044$ & $1.280\pm0.048$ \\
2.0 & $1.002\pm0.015$ & $1.043\pm0.010$ & $0.976\pm0.024$ & $1.045\pm0.017$ & $0.931\pm0.060$ & $1.152\pm0.046$ & $1.092\pm0.062$ & $1.095\pm0.078$ \\
3.0 & $0.907\pm0.024$ & $1.014\pm0.043$ & $1.051\pm0.043$ & $1.016\pm0.073$ & $1.037\pm0.114$ & $0.949\pm0.187$ & $1.028\pm0.111$ & $0.695\pm0.274$ \\ \hline
 $z$ & $\alpha_{\rm active}$ & $\alpha_{\rm inactive}$ & $\alpha_{\rm active}$ & $\alpha_{\rm inactive}$ & $\alpha_{\rm active}$ & $\alpha_{\rm inactive}$ & $\alpha_{\rm active}$ & $\alpha_{\rm inactive}$ \\ 
 0.0 & $-1.62\pm0.10$ & $-3.98\pm0.02$ & $-1.58\pm0.13$ & $-3.92\pm0.03$ & $-1.67\pm0.27$ & $-3.83\pm0.07$ & $-3.67\pm0.15$ & $-3.80\pm0.16$  \\
 0.1 & $-1.66\pm0.09$ & $-3.87\pm0.02$ & $-1.85\pm0.12$ & $-3.84\pm0.03$ & $-1.96\pm0.25$ & $-3.70\pm0.07$ & $-3.60\pm0.14$ & $-3.76\pm0.15$ \\
 0.2 & $-1.61\pm0.08$ & $-3.77\pm0.02$ & $-1.83\pm0.11$ & $-3.73\pm0.03$ & $-1.96\pm0.22$ & $-3.53\pm0.07$ & $-3.18\pm0.14$ & $-3.29\pm0.16$ \\
 0.3 & $-1.64\pm0.07$ & $-3.62\pm0.02$ & $-1.64\pm0.10$ & $-3.59\pm0.03$ & $-1.93\pm0.23$ & $-3.40\pm0.07$ & $-2.88\pm0.15$ & $-2.96\pm0.16$ \\
 0.4 & $-1.31\pm0.07$ & $-3.51\pm0.02$ & $-1.45\pm0.10$ & $-3.48\pm0.03$ & $-1.56\pm0.21$ & $-3.20\pm0.08$ & $-2.50\pm0.15$ & $-2.69\pm0.16$ \\
 0.5 & $-1.29\pm0.06$ & $-3.36\pm0.02$ & $-1.18\pm0.09$ & $-3.31\pm0.03$ & $-1.09\pm0.20$ & $-3.03\pm0.08$ & $-2.22\pm0.16$ & $-2.31\pm0.17$ \\
 0.7 & $-0.66\pm0.05$ & $-3.11\pm0.02$ & $-0.72\pm0.07$ & $-3.07\pm0.03$ & $-0.74\pm0.16$ & $-2.83\pm0.08$ & $-1.92\pm0.15$ & $-2.14\pm0.18$ \\
 1.0 & $-1.98\pm0.15$ & $-4.72\pm0.05$ & $-1.97\pm0.19$ & $-4.86\pm0.07$ & $-2.43\pm0.42$ & $-4.79\pm0.16$ & $-4.73\pm0.27$ & $-4.55\pm0.28$ \\
 1.5 & $-2.19\pm0.14$ & $-4.10\pm0.07$ & $-2.65\pm0.22$ & $-4.18\pm0.11$ & $-2.68\pm0.50$ & $-4.25\pm0.27$ & $-.466\pm0.44$ & $-5.06\pm0.49$ \\
 2.0 & $-2.28\pm0.15$ & $-2.55\pm0.10$ & $-2.01\pm0.24$ & $-2.56\pm0.18$ & $-1.56\pm0.60$ & $-3.66\pm0.47$ & $-3.11\pm0.63$ & $-3.09\pm0.79$ \\
 3.0 & $-1.36\pm0.25$ & $-2.42\pm0.45$ & $-2.84\pm0.44$ & $-2.43\pm0.77$ & $-2.67\pm1.16$ & $-1.74\pm1.95$ & $-2.54\pm1.14$ & $1.04\pm2.85$ \\
 \hline
$z$ & $\rm RMSD_{\rm \rm active}$ & $\rm RMSD_{\rm \rm inactive}$ & $\rm RMSD_{\rm \rm active}$ & $\rm RMSD_{\rm \rm inactive}$ & $\rm RMSD_{\rm \rm active}$ & $\rm RMSD_{\rm \rm inactive}$ & $\rm RMSD_{\rm \rm active}$ & $\rm RMSD_{\rm \rm inactive}$ \\ 
0.0 & $0.2660$ & $0.285$ & $0.266$ & $0.282$ & $0.262$ & $0.281$ & $0.288$ & $0.292$ \\
0.1 & $0.271$ & $0.286$ & $0.271$ & $0.284$ & $0.255$ & $0.281$ & $0.280$ & $0.284$ \\
0.2 & $0.275$ & $0.286$ & $0.275$ & $0.285$ & $0.255$ & $0.289$ & $0.282$ & $0.289$ \\
0.3 & $0.281$ & $0.288$ & $0.281$ & $0.286$ & $0.287$ & $0.284$ & $0.289$ & $0.291$ \\
0.4 & $0.289$ & $0.291$ & $0.289$ & $0.288$ & $0.282$ & $0.290$ & $0.294$ & $0.296$ \\
0.5 & $0.283$ & $0.292$ & $0.283$ & $0.292$ & $0.289$ & $0.292$ & $0.302$ & $0.305$ \\
0.7 & $0.285$ & $0.296$ & $0.285$ & $0.297$ & $0.277$ & $0.303$ & $0.309$ & $0.311$ \\
1.0 & $0.328$ & $0.289$ & $0.328$ & $0.291$ & $0.331$ & $0.301$ & $0.298$ & $0.279$ \\
1.5 & $0.292$ & $0.287$ & $0.293$ & $0.295$ & $0.297$ & $0.302$ & $0.309$ & $0.290$ \\
2.0 & $0.286$ & $0.296$ & $0.286$ & $0.302$ & $0.288$ & $0.303$ & $0.300$ & $0.291$ \\
3.0 & $0.346$ & $0.478$ & $0.349$ & $0.463$ & $0.339$ & $0.446$ & $0.345$ & $0.374$ \\
\hline
\enddata
\end{deluxetable*}

We find differences between the best-fitting values of $\alpha$ and $\beta$ for active and inactive populations. Specifically, we find that, compared to inactive galaxies, active galaxies have shallower slopes for redshifts $z\leq1.5$. At these redshifts, the average differences between $\beta_{\rm inactive}$ and $\beta_{\rm active}$ for the non-void, all void, and radius-controlled, and density-controlled galaxies are $0.202\pm0.004$, $0.183\pm0.005$, $0.168\pm0.009${, and $-0.018\pm0.018$} (corresponding to average slope ratios of $1.223\pm0.017$, $1.200\pm0.025$, $1.182\pm0.058${, and $0.982\pm0.055$}), respectively. In all cases, these results indicate that the manner in which SMBHs grow in mass compared to their host galaxies depends on whether that galaxy is active or inactive. {Compared to those derived for the non-void galaxies, the density-controlled populations have steeper values of $\beta_{\rm active}$ and $\beta_{\rm inactive}$ (e.g., $\beta_{\rm active}=0.917\pm0.010$ for the non-void galaxies vs. $1.119\pm0.015$ for the density-controlled galaxies at $z=0.0$), which implies that SMBHs in the most underdense environments are growing faster relative to their host galaxies than those in denser regions are.}

\section{Discussion}
\label{sec:ch5:discussion}

{When all \texttt{ZOBOV} members are used as a void galaxy population}, we find only minor differences between \texttt{TNG300} void and non-void field galaxies. {However, we show that we can recover the expected population statistics of void galaxies through careful cuts to our original void galaxy catalogs. For example,} differences between void and non-void galaxies are more pronounced when only the galaxies that are interior to a sphere of radius $0.8R_{\rm eff}$ are used to define the void galaxy population (i.e., the ``radius-controlled galaxies''). {Although radius cuts have previously been proposed as a way to recover void galaxy properties (see, e.g., \citealt{Veyrat2023}; \citealt{Zaidouni2025}), Figure~\ref{fig:underdensities} shows that }
this method is not foolproof, since \texttt{ZOBOV} voids are not necessarily spherical. By relying on the effective radius of the void, we may still be including galaxies within the ridges of non-spherical voids. Regardless, this method of selecting galaxies in a sphere around a void center is comparable to a {SVF} and thus provides a reasonable approximation for galaxies that are located within the inner regions of voids. {Instead, we define an interior sample of void galaxies as those that lie within spheres of controlled underdensity contrasts that can be more directly compared to SVF-based analyses. }

For redshifts $z\geq1.0$, we find that, compared to non-void galaxies, radius-controlled and {density} controlled galaxies have similar $(g-r)$ colors, stellar masses, star formation rates, and chemical abundance ratios but have smaller radii. Compared to the non-void galaxies at redshifts $z\leq0.7$, the {density-controlled galaxies are $\sim10.6\pm0.9\%$ bluer, $\sim8.1\pm1.2\%$ smaller, $\sim29.1\pm3.9\%$ less massive, $\sim13.7\pm1.6\%$ more star-forming, and have stellar particles and gas cells that are $\sim5\pm1\%$ less metal-enriched.} {Many of these differences are large enough that past and ongoing population-level studies of void galaxies will be able to measure them (e.g., \citealt{dominguezgomez2022}). There is also slight evidence for a time evolution in the properties of the density-controlled galaxies. For example, compared to the mass function of the non-void galaxies, the density-controlled galaxies reach a maximum deviation at $z\sim0.5$ where they are $43.0\pm5.7\%$ less massive on average, which supports the claim made by \cite{rodriguesmedrano2024} who find that void galaxies in \texttt{TNG300} experience mergers at later times.}


The apparent differences between the properties of all void galaxies, radius-controlled galaxies, {and density-controlled galaxies} indicate that void galaxy properties in \texttt{TNG300} likely vary somewhat as a function of void-centric distance (i.e., the distance between a galaxy and the center of a void.) For instance, the redder colors of the ``all void galaxies'' population indicate that, compared to radius-controlled and density-controlled galaxies, galaxies near void ridges experience more environmental factors that affect their stellar populations than galaxies. {Indeed, differences between galaxy populations are most pronounced interior to the shell-crossing surface (i.e., the density-controlled galaxies) than they are in the other populations where we are still including in our void galaxy populations those near the void ridges where densities are intermediate to high.} However, determining void galaxy properties specifically as a function of void-centric distance is beyond the scope of this work. 


The fact that we are only considering snapshots of voids may bias our conclusions. 
For instance, large amounts of material can flow into and out of voids (see, e.g., \citealt{tully2016}; \citealt{courtois2023}). Thus, some of the galaxies that we identified near the ridges of voids might not have originated within the voids themselves. Further study is necessary to investigate how cosmic flow patterns in \texttt{TNG300} are affected by the ridges of voids and how this might alter the global properties that are obtained for void galaxies.

Recently, \cite{rodriguesmedrano2024} used the \texttt{Popcorn} {\citep{Paz2023} void finding algorithm} to study the evolutionary pathways of \texttt{TNG300} void galaxies, and they report differences between the physical properties of void and non-void galaxies across cosmic time. Compared to \texttt{TNG300} non-void galaxies at $z=0.0$, they report that \texttt{TNG300} void galaxies have higher star formation rates, smaller stellar-to-halo-mass ratios, higher gas metallicities, and lower stellar metallicities. Compared to the properties of the {density}-controlled galaxies that we identified at $z=0.0$ with stellar masses between $10^{9.5}$ and $10^{10.0}h^{-1}M_{\odot}$, \cite{rodriguesmedrano2024} identify void galaxies that have similar average specific star formation rates ($\sim1.4\times10^{-10}\rm yr^{-1}$ in their study and $\sim1.2\times10^{-10}\rm yr^{-1}$ in our study), higher average gaseous chemical abundance ratios ($\sim1.280$ for their study and $\sim1.04$ in our study), and slightly lower stellar chemical abundance ratios ($\sim0.950$ in their study and $\sim1.055$ in our study).

In Section \ref{sec:ch5:AGN}, we examined the properties of \texttt{TNG300} void and non-void AGN across cosmic time. {We identify a strong correlation between underdense environments and AGN fraction, in agreement with many recent studies (e.g., \citealt{kauffmann2004}; \citealt{constantin2008}; \citealt{platenPhD}; \citealt{lopes2017}; \citealt{mishra2021}; \citealt{curtis2024}; \citealt{Aradhey2025}). At all redshifts, more than $50\%$ of all density-controlled galaxies were in their active state. Compared to the AGN fractions of the non-void galaxies, the AGN fractions of the density-controlled galaxies are $79.8\pm0.4\%$ higher at $z=0.0$ and $61.5\pm0.7\%$ greater at higher redshifts. In contrast, }
we find that the AGN fraction of radius-controlled galaxies is {only marginally} greater than those of the other two populations for redshifts. Compared to the non-void (void) galaxies, the radius-controlled galaxies have AGN fractions that are $4.6\pm2.9\%$ ($6.6\pm3.0\%$) higher at $z=0.0$, $9.0\pm2.5\%$ ($6.2\pm2.4\%$) higher at $z=0.7$, and $12.6\pm3.7\%$ ($11.4\pm3.6\%$) at $z=1.0$. {Since it is thought that AGN will evolve from active to quiescent states as gas reservoirs are depleted \citep{Ellingson1991, Best2012}, this result further shows that void galaxies within are experiencing stunted growth.}




Finally, we find slight differences between the best-fit parameters for the linear fits to the $M_{*}-M_{SMBH}$ relationships of active and inactive galaxies. Compared to inactive galaxies, active galaxies tend to have steeper slopes for the best-fit $M_{\rm{SMBH}}-M_*$ relationships. {For the non-void, all void, and radius-controlled galaxies, $\beta_{\rm active}$ and $\beta_{\rm inactive}$ are similar from population to population (e.g., $\beta_{\rm active}\sim0.9$ and $\beta_{\rm inacitve}\sim1.14$ for each population at $z=0.0$) while the density-controlled galaxies have steeper slopes (e.g., $\beta_{\rm active}\sim1.12$ and $\beta_{\rm inactive}\sim1.31$ at $z=0.0$, suggesting that SMBHs in the innermost regions of voids are growing at faster rates with regards to their host galaxies than those in more dense regions of the universe are.} The growth channels for SMBHs in \texttt{TNG300} depend on a variety of factors, such as the seeding of SMBHs into sufficiently massive halos, the modeling of AGN feedback mechanisms, and the handling of the two-phase state of nuclear activity (i.e., either active or inactive). However, determining the cause of the differing slopes that we find is beyond the scope of this work.
 
\section*{Acknowledgements}
{We are grateful to the anonymous reviewer for helpful comments and suggestions that improved the manuscript.} 
This work was partially supported by National Science Foundation grant AST-2009397. The IllustrisTNG simulations were undertaken with compute time awarded by the Gauss Centre for Supercomputing (GCS) under GCS Large-Scale Projects GCS-ILLU and GCS-DWAR on the GCS share of the supercomputer Hazel Hen at the High Performance Computing Center Stuttgart (HLRS), as well as on the machines of the Max Planck Computing and Data Facility (MPCDF) in Garching, Germany. In addition, we are pleased to acknowledge that the computational work reported on in this paper was performed on the Shared Computing Cluster which is administered by Boston University’s Research Computing Services. BM acknowledges support from Northeastern University's Future Faculty Postdoctoral Fellowship program.


\bibliography{bibliography}{}
\bibliographystyle{aasjournalv7}

\end{document}